\documentclass[11pt,a4paper]{article}
\pdfoutput=1
\usepackage{jcappub}

\usepackage[T1]{fontenc}
\usepackage[utf8]{inputenc}

\usepackage{graphicx}
\usepackage{times}%
\usepackage{natbib}
\usepackage{graphics}
\usepackage{epsfig}
\usepackage{array}
\usepackage{stfloats}
\usepackage{fixltx2e}
\usepackage{amsmath}

\newcommand{\gsim}{\,\lower2truept\hbox{${>\atop\hbox{\raise4truept\hbox{$\sim$}}}$}\,}

\newcommand{\be}{\begin{equation}}
\newcommand{\ee}{\end{equation}}
\newcommand{\bea}{\begin{eqnarray}}
\newcommand{\eea}{\end{eqnarray}}

\def\ltsima{$\; \buildrel < \over \sim \;$}
\def\simlt{\lower.5ex\hbox{\ltsima}}
\def\gtsima{$\; \buildrel > \over \sim \;$}
\def\simgt{\lower.5ex\hbox{\gtsima}}

\def\rd{{\rm d}}
\title{ Linear Perturbation constraints on Multi-coupled Dark Energy}
\author[a]{Arpine Piloyan,}
\author[b]{Valerio Marra,}
\author[c,d,e]{Marco Baldi}
\author[b]{and\\ Luca Amendola}
\affiliation[a]{Yerevan State University, Alex Manoogian 1, Yerevan 0025, Armenia;}
\affiliation[b]{Institut für Theoretische Physik, Universität Heidelberg, Philosophenweg
16, 69120 Heidelberg, Germany;}
\affiliation[c]{Dipartimento di Fisica e Astronomia, Università di Bologna, Viale
C.~Berti-Pichat 6/2, I-40127 Bologna, Italy;}
\affiliation[d]{INAF, Osservatorio Astronomico di Bologna, Viale C.~Berti-Pichat
6/2, I-40127 Bologna, Italy;}
\affiliation[e]{ INFN - Sezione di Bologna, Viale C.~Berti Pichat 6/2, I-40127 Bologna, Italy.}

\abstract{
The Multi-coupled Dark Energy (McDE) scenario has been recently proposed as a specific
example of a cosmological model characterized by a non-standard physics of the dark sector of the
universe that nevertheless gives an expansion history which does not significantly differ from the one of the standard $\Lambda $CDM model.
Thanks to a dynamical screening mechanism, in fact, the interaction between the
Dark Energy field and the Dark Matter sector is effectively suppressed at the background level
during matter domination. As a consequence, background observables cannot discriminate a McDE
cosmology from $\Lambda $CDM for a wide range of
model parameters. On the other hand, linear
perturbations are expected to provide tighter bounds due to the existence of attractive and
repulsive fifth-forces associated %to
with the dark interactions. In this work, we present the first
constraints on the McDE scenario obtained by comparing the predicted evolution of linear density
perturbations with a large compilation of recent data sets for the growth rate $f\sigma _{8}$,
including 6dFGS, LRG, BOSS, WiggleZ and VIPERS. Confirming qualitative expectations, growth rate
data provide much tighter bounds on the model parameters as compared to the extremely loose bounds
that can be obtained when only the background expansion history is considered. In particular, the
95\% confidence level on the coupling strength $|\beta |$ is reduced from $|\beta |\leq 83$ (background constraints only) to
$|\beta |\leq 0.88$ (background and linear perturbation constraints).
We also investigate how these constraints
further improve %by
when using data from future wide-field surveys
such as supernova data from LSST and growth rate data from Euclid-type missions. In this case the 95\% confidence level on the coupling further reduce to $|\beta |\leq 0.85$. Such
constraints are in any case still consistent with a scalar fifth-force of gravitational strength,
and we foresee that tighter bounds might be possibly obtained from the investigation of nonlinear
structure formation in McDE cosmologies.
}


\makeatother

\begin{document}
\maketitle

\section{Introduction}
\label{intro}

Understanding the fundamental origin of the observed accelerated expansion
of the Universe~\citep[]{Riess_etal_1998,Perlmutter_etal_1999,Schmidt_etal_1998,SNLS}
represents the driving scientific case for a large number of complex
and ambitious international initiatives planned for the next decade
of cosmological observations, including e.g.~the Baryon Oscillation Spectroscopic Survey~\citep[BOSS,][]{Ahn_etal_2013}, the Dark Energy Survey~\citep[DES,][]{DES}, the Large Synoptic Space Telescope~\citep[LSST,][]{LSST}
and the ESA satellite mission Euclid\footnote{www.euclid-ec.org}~\citep[][]{Euclid-r}.
Besides an exquisite quality of observational
data and a rigorous control of any possible systematics, such a challenging
task will also require reliable predictions of how different theoretical
scenarios might be scrutinised and possibly disentangled by efficiently
combining different observational probes. More specifically, in order
to discriminate a wide variety of Dark Energy (DE) or Modified Gravity
(MG) models from the standard cosmological constant -- presently assumed
as the fiducial scenario -- an appropriate combination of geometrical
and dynamical observables is generally required, as competing models
often feature strong degeneracies with the standard cosmological parameters when
single observational probes are considered.

In this respect, it is particularly instructive to investigate cosmological
scenarios that are practically indistinguishable from the fiducial
$\Lambda$CDM model
%through
as far as some particular observational probes are concerned, while showing characteristic
features through other observational channels. While the most widely
investigated alternatives to the cosmological constant such as {\em
Quintessence} \citep[]{Wetterich_1988,Ratra_Peebles_1988}, {\em
k-essence} \citep[]{kessence}, {\em phantom} \citep[][]{Caldwell_2002} and
{\em quintom} \citep[][]{Feng_Wang_Zhang_2005} DE models, or more complex scenarios
like {\em interacting DE} \citep[]{Wetterich_1995,Amendola_2000,Farrar2004,Baldi_2011a},
{\em the Growing Neutrino model} \citep[]{Amendola_Baldi_Wetterich_2008}
and generic MG theories \citep[as proposed e.g.~by][]{Buchdahl_1970,Starobinsky_1980,Hu_Sawicki_2007,Sotiriou_Faraoni_2010,
Dvali_Gabadadze_Porrati_2000,Nicolis_Rattazzi_Trincherini_2009} generally affect both the
background and linear perturbations evolution of the universe, some
specific realisations of these models were found to have the appealing
(and challenging) feature of sharing the same expansion history of
the $\Lambda$CDM cosmology to an extreme level of precision. Such
models -- which include e.g.~the Hu \& Sawicki realisation
of $f(R)$ theories \citep[][]{Hu_Sawicki_2007} and the Multi-coupled Dark Energy (McDE) scenario
\citep[]{Baldi_2012a} discussed in the present work -- represent
an ideal benchmark to test the predictive power of future multi-probe
observational surveys.

In particular, the McDE model was proposed as a particular realisation
of the more general framework discussed by Ref.~\cite{Brookfield_VanDeBruck_Hall_2008}
with the {aim of providing} an extension to the standard coupled Dark
Energy (cDE) scenario in terms of a multi-particle nature of the CDM
field, without introducing additional free parameters. In fact, the
McDE model discussed in the present work is characterised by a hidden
symmetry in the CDM sector associated with two distinct particle species
with opposite couplings to a single DE scalar field, thereby requiring
the same number of free parameters (a self-interaction potential slope
$\alpha$ and a coupling constant $\beta$) as the widely investigated
coupled Quintessence models. However, differently from the latter,
the CDM internal symmetry that characterises McDE scenarios has been
shown to provide a self-regulating mechanism of the effective interaction
strength, thereby very effectively suppressing the DE-CDM interaction at
the background level.

Such background screening of the DE-CDM interaction has been qualitatively
demonstrated by \citep[][]{Baldi_2012a} and subsequently investigated
in full detail by our team in \cite{Piloyan:2013mla}. In the latter paper, we
provided the first direct comparison of McDE cosmologies with real
observational data consisting of the supernova luminosities of
the publicly available Union2.1 sample \cite{Suzuki_etal_2012}.
Confirming the previous qualitative results of \cite{Baldi_2012a},
our analysis directly showed how present observational data on the
background expansion history of the Universe are fully consistent
with McDE scenarios even for very large values of the DE coupling,
up to three orders of magnitude larger than the present bounds on
the coupling for standard cDE models ($\beta \lesssim 0.1$, see e.g.~\cite{Pettorino_2013,Xia_2013}).
In fact, the
conclusion of our previous paper was that the background dynamics
could hardly offer any significant constraint on the McDE model and
anticipated that more severe constraints could be obtained by working
out the behavior of perturbations.

This paper is devoted to such a task. Our primary goal is to derive observational constraints
on the main parameters of the McDE model based on the latest available data on the
growth of linear density perturbations, with the aim to significantly improve the extremely loose bounds
derived through background observables, thereby reducing the viable parameter space.
This task is also particularly relevant in view of the further extension of the investigation
of the McDE scenario to the nonlinear regime of structure formation by means of dedicated N-body simulations. As the latter are in general quite computationally expensive, especially
for large values of the coupling, constraining the viable region of the model's parameter space through linear observables will avoid wasting precious computational resources.
To this end, we will first  derive the full
set of linear perturbation equations in the McDE model and we will analytically solve them
for the simplified cases of the background critical points in phase space.
Then, we will integrate numerically the equations to obtain the full solution
and check the analytical results. With the full numerical evolution of the linear perturbation growth at hand, we will
finally compare the predictions obtained for different choices of model parameters with our sample of observational
data by performing a detailed sampling of the parameter space.
Our final result will be marginalised posterior bounds.
As we will discuss below, such a procedure provides the most stringent constraints to date on this type of cosmological models.

The paper is organised as follows. In Section~\ref{pertue} we introduce the full system
of linear perturbation equations for the McDE scenario, in Section~\ref{an_sol} we derive
analytical solutions for the background critical points that determine a viable cosmological
expansion history, and in Section~\ref{num_sol} we compute the full numerical solutions of the
system. In Section~\ref{analysis} we present the datasets adopted for our analysis and we describe
the procedure for the direct comparison with observational data. In Section~\ref{results} we discuss
the main results of our work and provide observational constraints on the McDE parameters. Finally,
in Section~\ref{conclusions} we conclude. Furthermore, in Appendix~\ref{appendix} we
consider the effect of the uncoupled baryonic
component
% also included in the system making it more realistic},
showing that this does not significantly affect our main results.

\section{Linear perturbation equations}
\label{pertue}

The McDE model, proposed and investigated by
\cite{Baldi_2012a,Piloyan:2013mla,Baldi_2013}, is
 characterised by the existence of two different species of CDM particles with opposite couplings to
the same classical DE scalar field. The system is therefore
described by the following Lagrangian:
\begin{align}
S=\int d^{4}x\sqrt{-g}\biggl[\frac{M_{Pl}^{2}}{2}R-\frac{1}{2}{\phi}^{;\alpha}\phi_{;\alpha}-
\underset{\pm}{\sum}m_{\pm}e^{\pm
\sqrt{\frac{2}{3}}\frac{\beta}{M_{Pl}}\phi}\bar{\psi}_{\pm}\psi_{\pm}-V(\phi)+\mathcal{L}_{r}\biggl]
\,,\label{eq:Action-1}
\end{align}
where $\phi$ is the dark energy scalar field, $\psi_{\pm}$ represent the two CDM fields, $\beta $ is
a dimensionless parameter defining the strength of the interaction, and $\mathcal{L}_{r}$ is the
radiation Lagrangian.
In eq.~(\ref{eq:Action-1}) we have discarded the uncoupled baryonic component as its contribution does
not significantly alter the results of our analysis. However, in the Appendix~\ref{appendix}
we will drop this assumption and properly quantify the effect of baryons on our results. As we will
see, these have a rather small
impact, due to the low baryonic density observed today, which makes our simplified setup fully
justified.

We restrict our attention to a spatially flat FLRW metric. Perturbation equations corresponding to the model of eq.~(\ref{eq:Action-1})
on sub-horizon scales for each species of dark matter have been derived
in \cite{Baldi_2012a} and read:
\begin{align}
\ddot{\delta }_{-} & =-2H(1+\beta\frac{\dot{\phi}}{\sqrt{6}H})\dot{\delta }_{-}+4\pi G(\rho_{-}\Gamma_{A}\delta_{-}+\rho_{+}\Gamma_{R}\delta_{+})\,,\label{eq:pert_1-2-1}\\
\ddot{\delta }_{+} & =-2H(1-\beta\frac{\dot{\phi}}{\sqrt{6}H})\dot{\delta }_{+}+4\pi G(\rho_{-}\delta_{-}\Gamma_{R}+\rho_{+}\delta_{+}\Gamma_{A})\,,
\end{align}
where $\rho_{\pm}$ are the energy densities of the two CDM species,  $\delta _{\pm}$ their respective density contrasts, and where the $\Gamma $ factors
 \begin{align}
\Gamma_{R} & =1-\frac{4}{3}\beta^{2}\,,\label{eq:Gamma}\\
\Gamma_{A} & =1+\frac{4}{3}\beta^{2}\,
\end{align}
encode the effects of repulsive ($R$) and attractive ($A$) fifth-forces.

It is convenient to rewrite these equations employing the $e$-folding $N$
as the time variable, $N\equiv \ln a$, and to introduce the following dimensionless quantities:
\begin{align}
x^{2}\equiv \frac{\dot{\phi}^{2}}{6M_{{\rm Pl}}^{2}H^{2}}\,, & \qquad y^{2}\equiv \frac{V}{3M_{{\rm Pl}}^{2}H^{2}}\,,\label{eq:Cont_Sc-1}\\
z_{\pm}^{2}\equiv \frac{\rho_{\pm}}{3M_{{\rm Pl}}^{2}H^{2}}\,, & \qquad r^{2}\equiv \frac{\rho_{r}}{3M_{{\rm Pl}}^{2}H^{2}}\,.\label{eq:Cont_Sc_last-1}
\end{align}
The linear perturbations equations with respect to $N$ then read
\begin{align}
\delta_{-}^{''}+\left[ 2(1+\beta x)-\frac{1}{2}(3-3y^{2}+3x^{2}+r^{2})\right] \delta'_{-} & =\frac{3}{2}(z_{-}^{2}\Gamma_{A}\delta_{-}+z_{+}^{2}\Gamma_{R}\delta_{+})\,,\label{eq:pert_1-2-1-1}\\
\delta_{+}^{''}+\left[ 2(1-\beta x)-\frac{1}{2}(3-3y^{2}+3x^{2}+r^{2})\right] \delta'_{+} & =\frac{3}{2}(z_{-}^{2}\Gamma_{R}\delta_{-}+z_{+}^{2}\delta_{+}\Gamma_{A})\,,\label{eq:pert_min}
\end{align}
where we have made use of the background equation
\begin{equation}
\frac{H'}{H}=-\frac{1}{2}(3-3y^{2}+3x^{2}+r^{2})\,.
\end{equation}
The background behavior was studied and compared to observations in
\cite{Piloyan:2013mla}, where it was shown that the background
evolution on a spatially flat FLRW metric is characterized by several
critical-points defined as solutions for which $x,y,z_{\pm}$ are constant. These are summarised in Table~\ref{tbl_crt-2}. Among
these, two critical points {(point 2 and point 5)} represent accelerated stable solutions,
and two are metastable (saddle points) matter-dominated solutions {(point 3 and point 4)}: viable
cosmologies should connect {one of the two matter eras} to one of the two accelerated
regimes. As points 3 and 4 are metastable points,
their occurrence strongly depends on the initial conditions; the occurrence of the stable points, on the contrary,
depends only on the values of the parameters $\alpha,\beta$.
Therefore, by choosing the parameters in the appropriate range we will obtain either point 2 (in which dark energy dominates)
or 5 (in which dark energy and dark matter coexist) as final state.
The metastable matter eras will instead generally both occur for the same parameters, one after the other.
 However, point 3 will always be the last point before  dark energy domination, since the matter density dilutes more slowly
for point 3 than for point 4 (i.e.~the equation of state is smaller for point 3, see Table \ref{tbl_crt-2}).
Therefore, if the initial conditions are set far enough in the past we expect the phase relative to point 4 to end very early and not to affect the late-time observations (supernovae and growth rate) considered in this work.
%to which supernova and growth rate observations considered in this work are referred.
In the following
we will therefore always assume that we can neglect the possible point 4 matter era.

In the next Sections we will first solve the perturbation equations
analytically on the background critical points,
and then numerically along the full trajectory.

%One could introduce also an uncoupled baryon component in the system. This
%would avoid the local gravity constraints on the coupling $\beta$, leaving cosmology
%as the only tool to constrain it. In the Appendix we will show that this has a rather small
%impact, due to the low amount of baryons observed today, so in the rest of the paper
%for simplicity we will neglect the baryon component.

\section{Analytical solutions of the perturbation equations}
\label{an_sol}

\begin{table}[t]
{
\renewcommand{\arraystretch}{1.8}
\begin{tabular}{|c|c|c|c|c|c|c|c|}
\hline
Point  & $x$  & $y$  & $z_{+}$  & $z_{-}$  & \,$\Omega_{{\rm DE}}$\,  & $w_{eff}$  & $\mu$\tabularnewline
\hline
\hline
1  & $\pm1$  & 0  & 0  & 0  & 1  & 1  & 0\tabularnewline
\hline
2  & $\frac{\alpha}{3}$  & $\frac{1}{3}\sqrt{9-\alpha^{2}}$  & 0  & 0  & 1  & \,\,$-1+\frac{2\alpha^{2}}{9}$\,\,  & 0\tabularnewline
\hline
3  & 0  & 0  & $\frac{1}{\sqrt{2}}$  & $\frac{1}{\sqrt{2}}$  & 0  & 0  & $0$\tabularnewline
\hline
4  & $-\frac{2\beta}{3}$  & 0  & $\sqrt{1-\frac{4\beta^{2}}{9}}$  & 0  & $\frac{4\beta^{2}}{9}$  & $\frac{4\beta^{2}}{9}$  & 1\tabularnewline
\hline
5  & \,\, $\frac{3}{2(\alpha+\beta)}$ \,\,  & \,\, $\frac{\sqrt{9+4\alpha\beta+4\beta^{2}}}{2|\alpha+\beta|}$\,\,  & \,\,$\frac{\sqrt{-9+2\alpha\beta+2\alpha^{2}}}{\sqrt{2}|\alpha+\beta|}$
\,\,  & 0  & \,\, $\frac{9+2\alpha\beta+2\beta^{2}}{2(\alpha+\beta)^{2}}$ \,\, & \,\,$\frac{-\beta}{(\alpha+\beta)}$\,\,  & $1$\tabularnewline
\hline
\end{tabular}
}
\caption{Critical points for background equations provided in \cite{Piloyan:2013mla}.
Only the physical solutions for $x,y,z_{+},z_{-}$ are selected. Only
points 2 and 5 can have accelerated expansion ($w_{{\rm eff}}<-1/3$). }
\label{tbl_crt-2}
\end{table}

 On the particular background solutions corresponding to the critical points of Table~\ref{tbl_crt-2}, the perturbation equations become constant-coefficient
linear equations and, therefore, exactly solvable. Thus,
the analytical solutions of perturbations on the background critical points will
approximate the full solutions in the matter era (point 3)
and final accelerated stage (point 2 or point 5), respectively, as shown in the following sections.
 We define the total matter perturbation as:

\begin{equation}
\delta=\frac{\Omega_{-}\delta_{-}+\Omega_{+}\delta_{+}}{\Omega_{-}+\Omega_{+}} \,,
\end{equation}
and, correspondingly, the total and partial growth rates as:
\begin{equation}
f=\frac{\delta^{'}}{\delta}\,,\qquad f_{\pm}=\frac{\delta_{\pm}^{'}}{\delta_{\pm}} \,.
\label{eq:delta_tot}
\end{equation}

We consider only one quadrant of the parameter plane, $\left\{ \alpha > 0,\beta > 0\right\} $,
since due to the symmetry of the system
it is sufficient to solve equations with positive values of parameters.
 Indeed, the sign of the coupling is completely irrelevant and all our constraints will refer to its absolute value $|\beta |$.
The analytical solutions of eq.~(\ref{eq:pert_1-2-1-1}) are presented
in Table \ref{tbl_crt-1} for each of the five critical points. Only
the growing solutions of the dark matter components and total density
perturbation, $f_{\pm}$, $f$, respectively, are selected.
 For point 3 the perturbations $\delta_{+}$ and $\delta_{-}$ have opposite signs and compensate each other so that the total growth function $f$ is always unity. Nonetheless, near the transition between point 3 and point 2 (or point 5) the symmetry between the background densities of the two CDM species, that is ensured by the matter-dominated attractor, starts to be violated, which gives rise to the solution $f=f_{-}+3$ in this region. This can be justified if we write $f$ in more detail, as:

\begin{align}
f=\frac{\delta^{'}}{\delta}=\left[\frac{\Omega_{-}\delta_{-}+\Omega_{+}\delta_{+}}{\Omega_{-}+\Omega_{+}}\right]^{'}
/\left[\frac{\Omega_{-}\delta_{-}+\Omega_{+}\delta_{+}}{\Omega_{-}+\Omega_{+}}\right] \,\nonumber\\
=\frac{\Omega_{-}\delta_{-}^{'}+\Omega_{+}\delta_{+}^{'}}
{\Omega_{-}\delta_{-}+\Omega_{+}\delta_{+}}+\frac{\Omega_{-}^{'}\delta_{-}+\Omega_{+}^{'}\delta_{+}
}
{\Omega_{-}\delta_{-}+\Omega_{+}\delta_{+}}-
\frac{\Omega_{-}^{'}+\Omega_{+}^{'}}
{\Omega_{-}+\Omega_{+}} \,.
\label{eq:f_t_exp}
\end{align}
At point 3, the first term on the second line of eq.~(\ref{eq:f_t_exp}) is unity as long as $\Omega_{-}\delta_{-}$ and $\Omega_{+}\delta_{+}$ exactly compensate each other. However, as soon as $\Omega_{-}\neq\Omega_{+}$ (i.e.~when the system is about to move out of point 3) it behaves as $f_{-}$. In this case the second term is also different from zero.
In fact, analytical calculations show that $\Omega_{-}-\Omega_{+}\sim e^{3N}$ when the system is about to go out of point 3, which leads to:
\begin{equation}
\frac{\Omega_{-}^{'}\delta_{-}+\Omega_{+}^{'}\delta_{+}}{\Omega_{-}\delta_{-}+\Omega_{+}\delta_{+}}\rightarrow 3 \,,
\end{equation}
as $\delta_{-}\approx-\delta_{+}$.
Finally, the last term is zero at each critical point.
It is important to mention that the growth function $f$ passes through a singularity as the total perturbation $\delta$ goes through zero. However, the observable is $\delta'/\delta_{0}$ (see eq.~(\ref{obsf})), which is never singular.

\begin{figure}
\begin{centering}
\includegraphics[width=0.4\columnwidth]{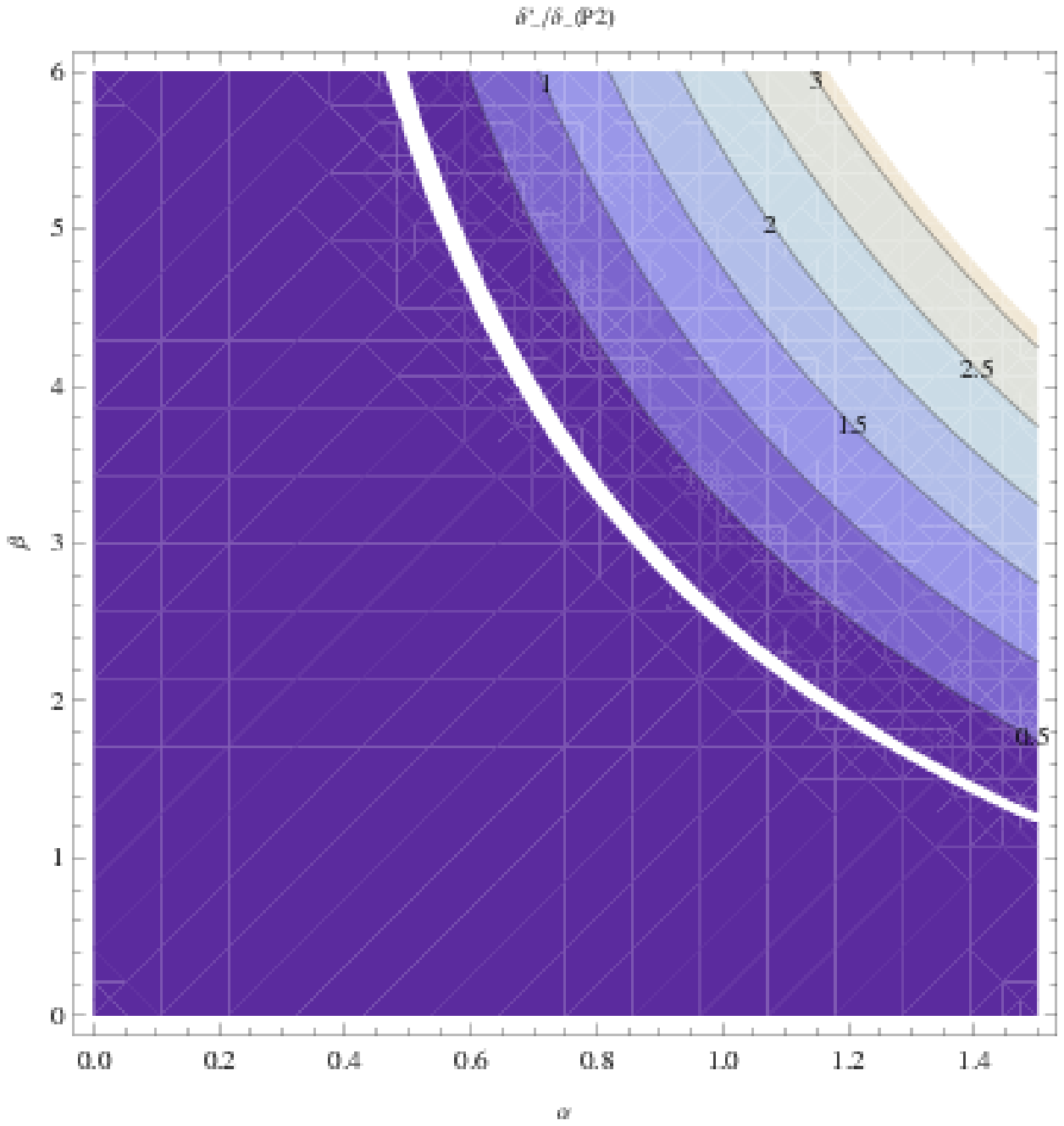}
\includegraphics[width=0.4\columnwidth]{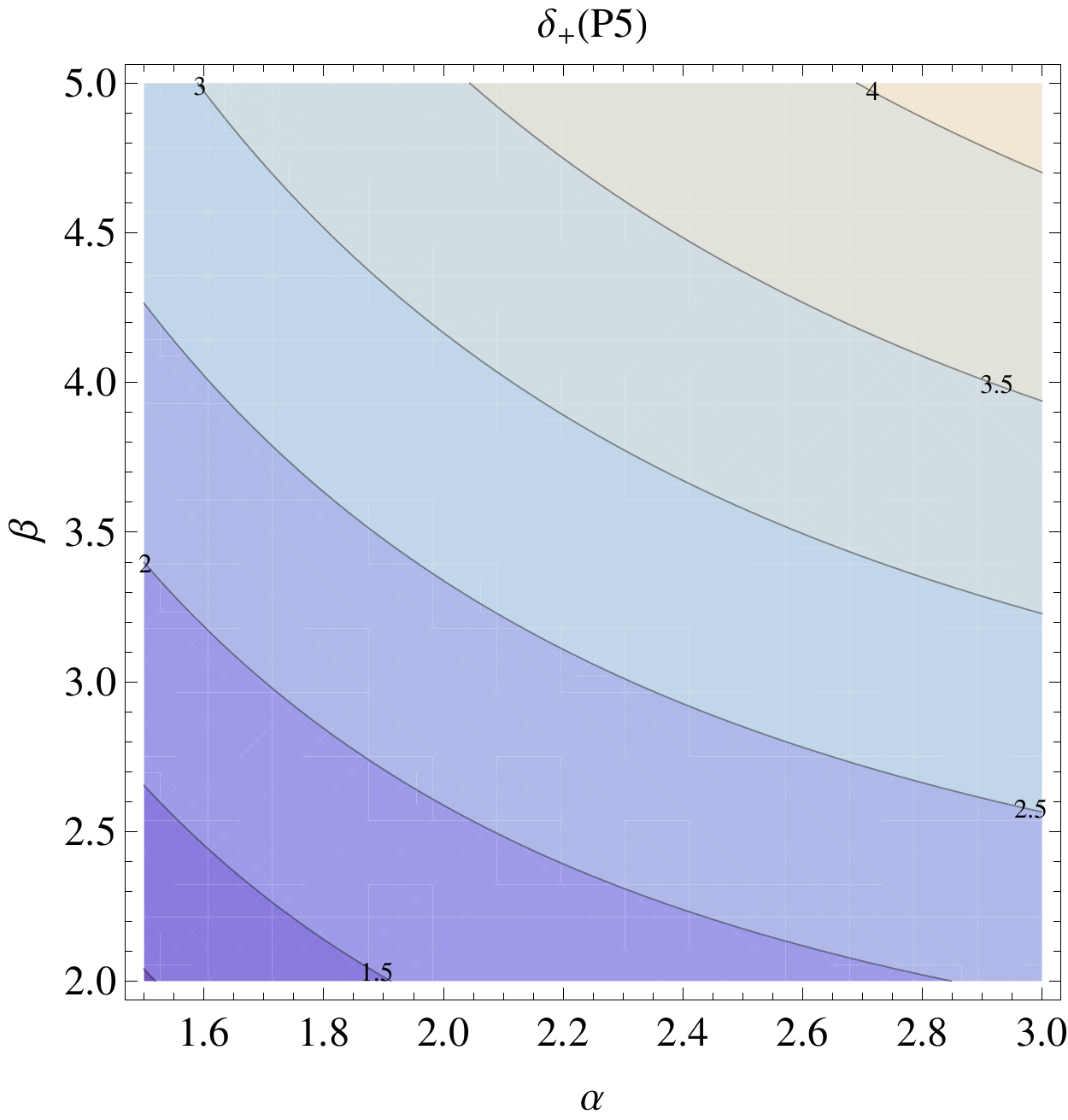}
\caption{ {\em Left panel:} Contour plots of the analytical solution for $f_{+}$ for
parameters in the stability region of point 2. $f_{-}$ is always zero in this region. {\em Right panel:} The same plot but for the stability region of point 5. }
\label{cp-p2_m}
\end{centering}
\end{figure}

 For a more complete picture, we present the contour plots of the growth functions for point 2 and point 5 in the left and right panels of Fig.~\ref{cp-p2_m}, respectively.
From these contour plots we conclude that larger values of the McDE characteristic parameters correspond
to larger growth rates; moreover, for a significant portion of the parameter space, the growth rate on the background critical point 2 is zero.

\begin{table}[t]
{
\renewcommand{\arraystretch}{1.8}
\begin{tabular}{|c|c|c|c|}
\hline
Point & $f_{+}$  & $f_{-}$  & $f$ \tabularnewline
\hline
\hline
1  & max$[-\frac{1}{2}-2\beta\: ,\:0]$  & $-\frac{1}{2}+2\beta$ or $0$  & $f_{-}$ \tabularnewline
\hline
2  & max$[\frac{1}{3}(-6+\alpha^{2}+2\alpha\beta)\: ,\:0]$  & \,\,max$[\frac{1}{3}(-6+\alpha^{2}-2\alpha\beta)\: ,\:0]$\,\,  & \,\,$f_{+}$\,\, \tabularnewline
\hline
3  & max$[\frac{1}{4}\left(-1+\sqrt{1+32\beta^{2}}\right), 1]$   & max$[\frac{1}{4}\left(-1+\sqrt{1+32\beta^{2}}\right) , 1] $ & { $f_{-}+3$}\tabularnewline
\hline
4  & $\frac{1}{12}\left(-3-4\beta^{2}+\sqrt{225-216\beta^{2}-112\beta^{4}}\right)$  &max$ [f_{+} ,-\frac{1}{2}+2\beta^{2}, 0 ] $& $f_{-}$ \tabularnewline
\hline
5  & \,\, $\frac{1}{4}(-1-\frac{3\beta}{(\alpha+\beta)}+\Delta)$\,\,
 & \,\,max$[f_{+},-5+\frac{9\alpha}{2(\alpha+\beta)}
, 0 ]$ & $f_{-}$\tabularnewline
\hline
\end{tabular}
}
\caption{Growth functions for each species and total
growth function at the background critical points.
Here $\Delta=\sqrt{\frac{4\alpha\beta(5+8\beta^{2})+\alpha^{2}(25+32\beta^{2})-4(27+35\beta^{2})}{(\alpha+\beta)^{2}}}$.}
\label{tbl_crt-1}
\end{table}

\section{Numerical solutions of the perturbation equations}
\label{num_sol}

\begin{figure}
\begin{centering}
\includegraphics[width=0.5\columnwidth]{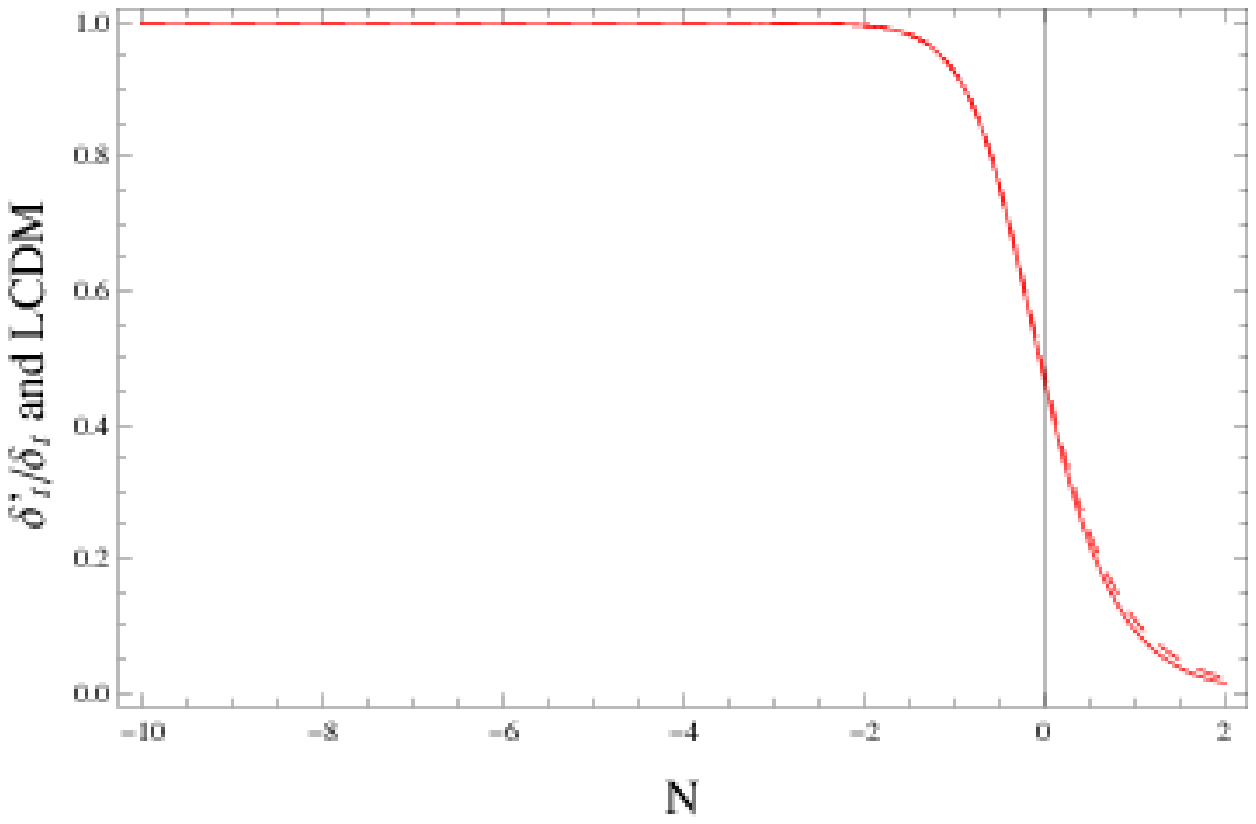}
\caption{Plot of $f$ (solid curve) when
$\alpha=0.1$ and $\beta=0$, and of $f_{\Lambda\,CDM}=\Omega_{m}^{0.54}$ (dashed curve). }
\label{f_f_L}
\end{centering}
\end{figure}

In this section we will illustrate the numerical solutions of the
perturbed system of equations (\ref{eq:pert_1-2-1-1}) and (\ref{eq:pert_min}) defined in Section
\ref{pertue} together with the background equations. Our numerical results show a very good
agreement with the  analysis of Section \ref{an_sol}. We will restrict the
range of the model's parameters to the stability and acceleration regions of
the background critical points of Table~\ref{tbl_crt-2} \citep[see Ref.][]{Piloyan:2013mla}.
As a first test of our numerical integration, we display in Fig.~\ref{f_f_L}
the total growth function of McDE with $\alpha=0.1$
and $\beta=0$ (i.e.~for the uncoupled case, solid curve) and the growth function $f_{\Lambda CDM}$ for the standard $\Lambda$CDM model given by $f_{\Lambda CDM}=\Omega_{m}^{\gamma}$
with $\gamma=0.54$ (dashed curve). The two curves show very good agreement, as expected for a standard Quintessence model with a shallow self-interaction potential (indeed the limiting case of $\alpha =0$, $\beta = 0$ exactly corresponds to a $\Lambda $CDM cosmology).

\begin{figure}
\begin{centering}
\includegraphics[width=0.4\columnwidth]{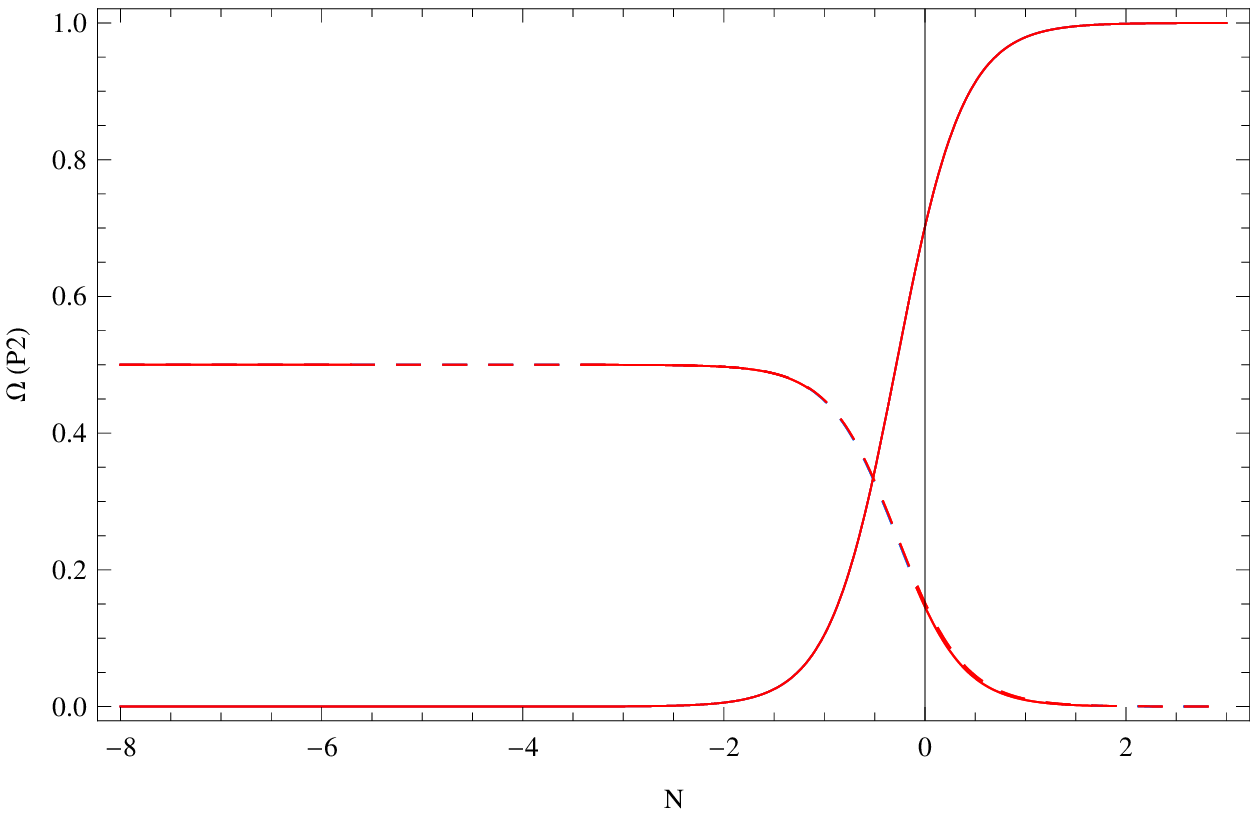}
\includegraphics[width=0.4\columnwidth]{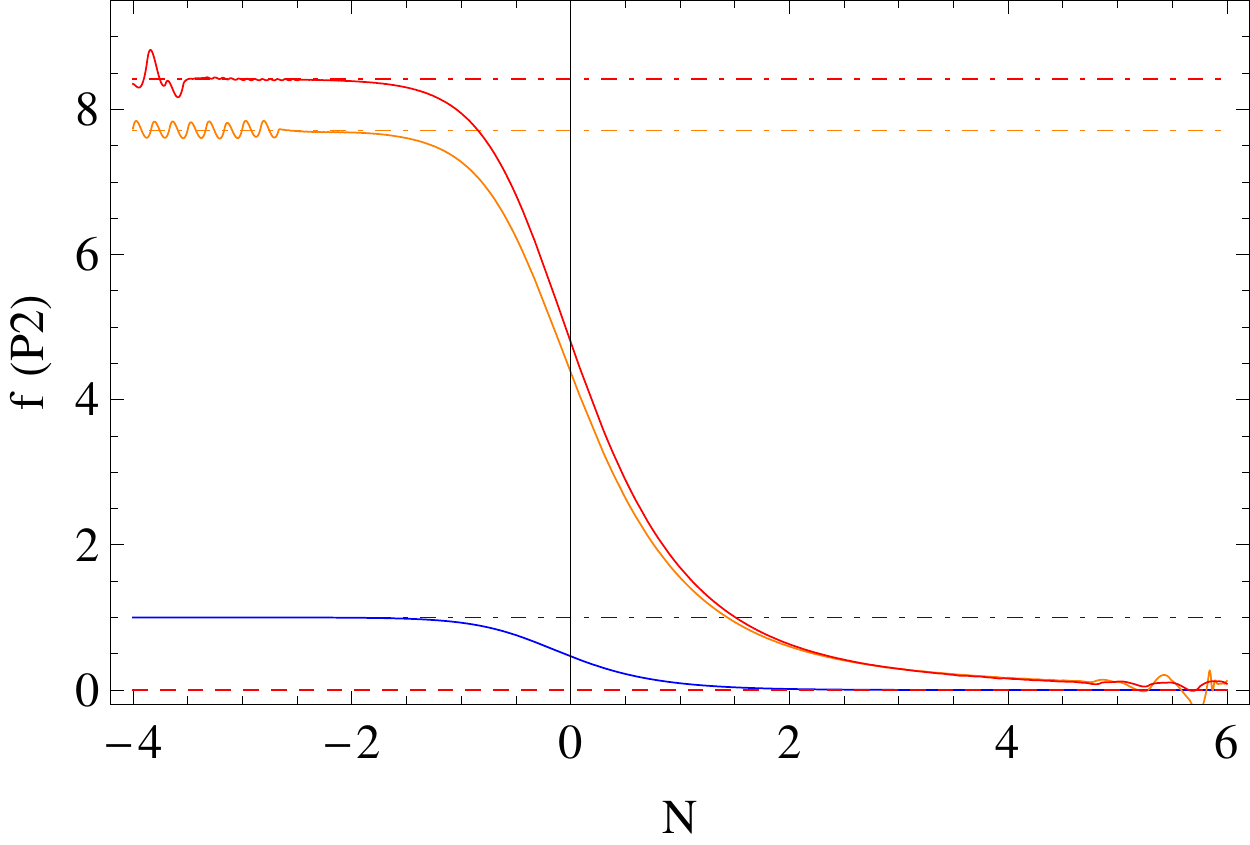}
\caption{The left panel is the plot of $\Omega_{i}$ when $\alpha=0.1$ and $\beta=0.5$. In this panel, solid curves correspond to $\Omega_{DE}$ and dashed curves correspond to $\Omega_{-}$ and $\Omega_{+}$.
The right panel is the plot of $f$ for $\alpha=0.1$ and $\beta=$0.5 (blue curve), 3.5 (orange curve), 4 (red curve). The horizontal lines
are the analytical  values for the matter era (dot-dashed) and the accelerated point 2 (dashed).
}
\label{p_2f_m}
\end{centering}
\end{figure}

\begin{figure}
\begin{centering}
\includegraphics[width=0.4\columnwidth]{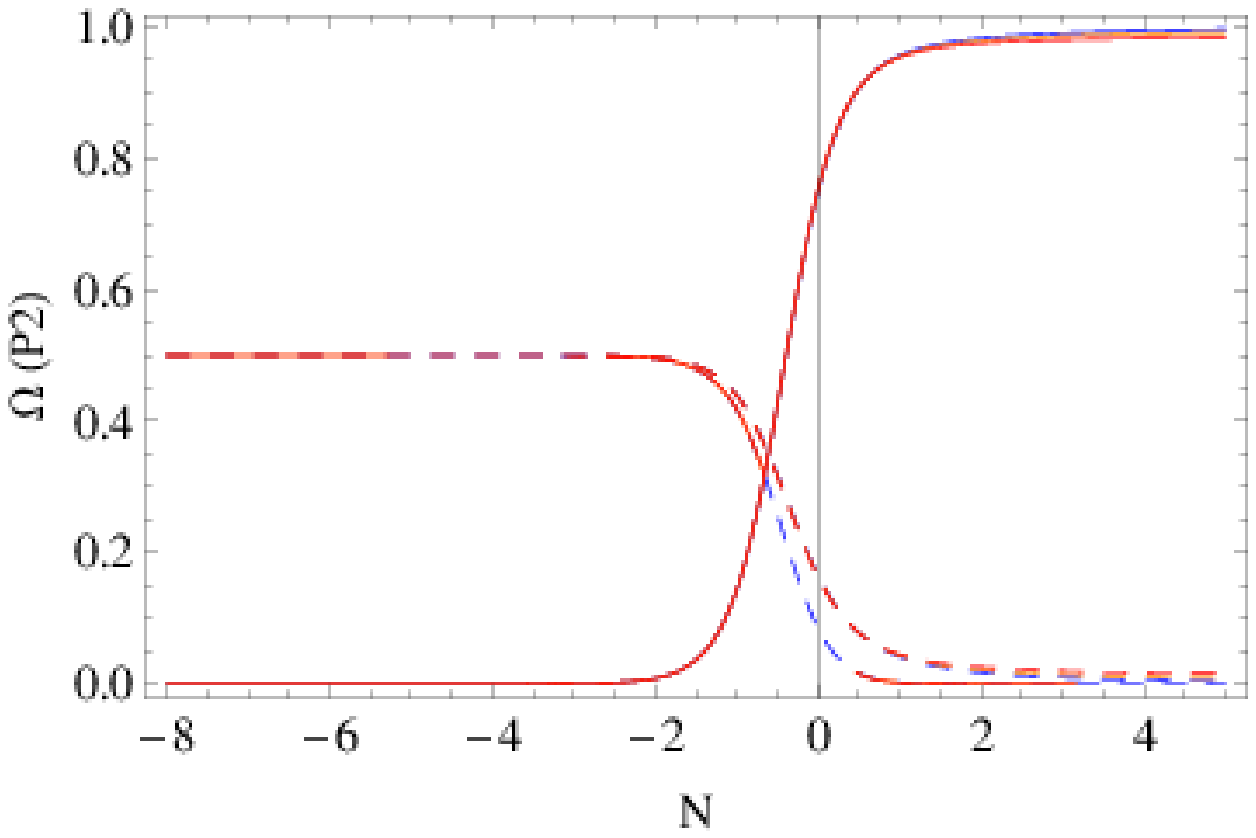}
\includegraphics[width=0.4\columnwidth]{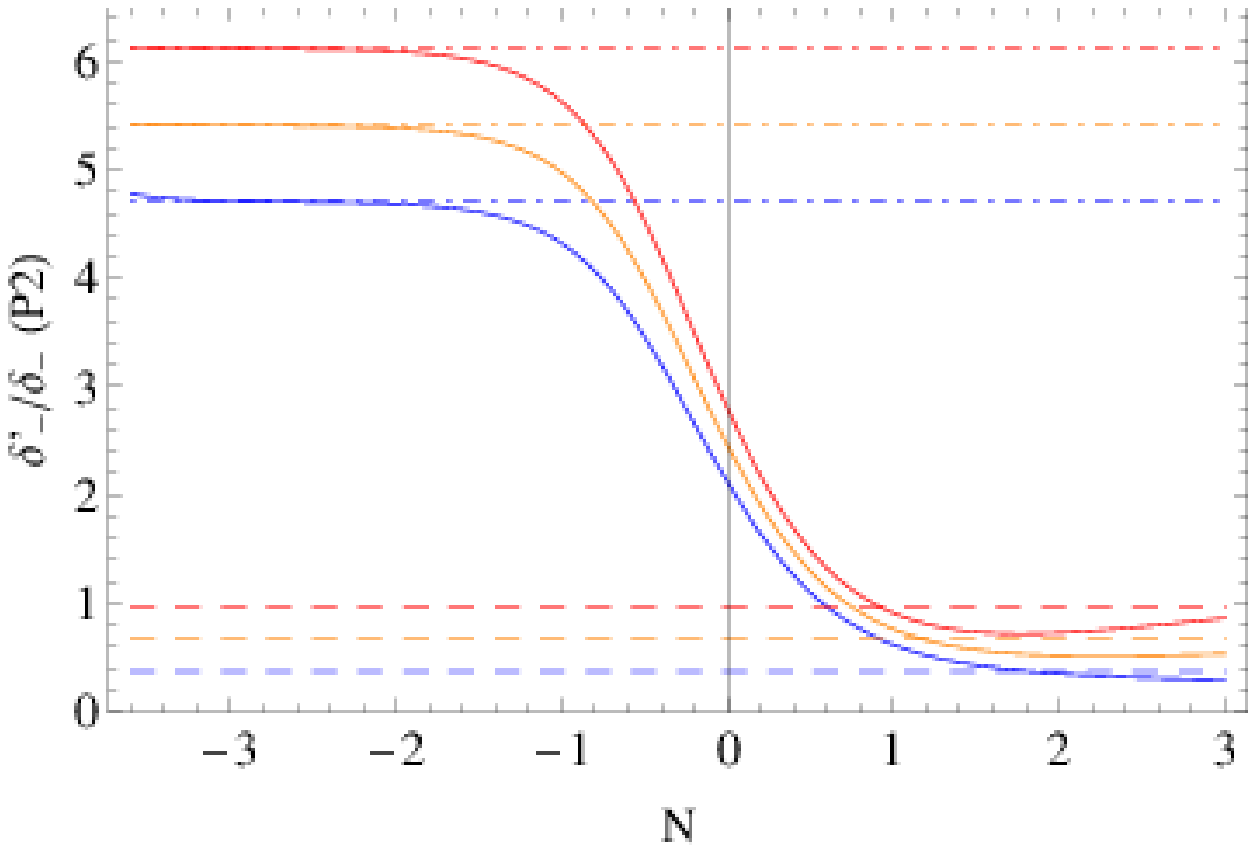}
\includegraphics[width=0.4\columnwidth]{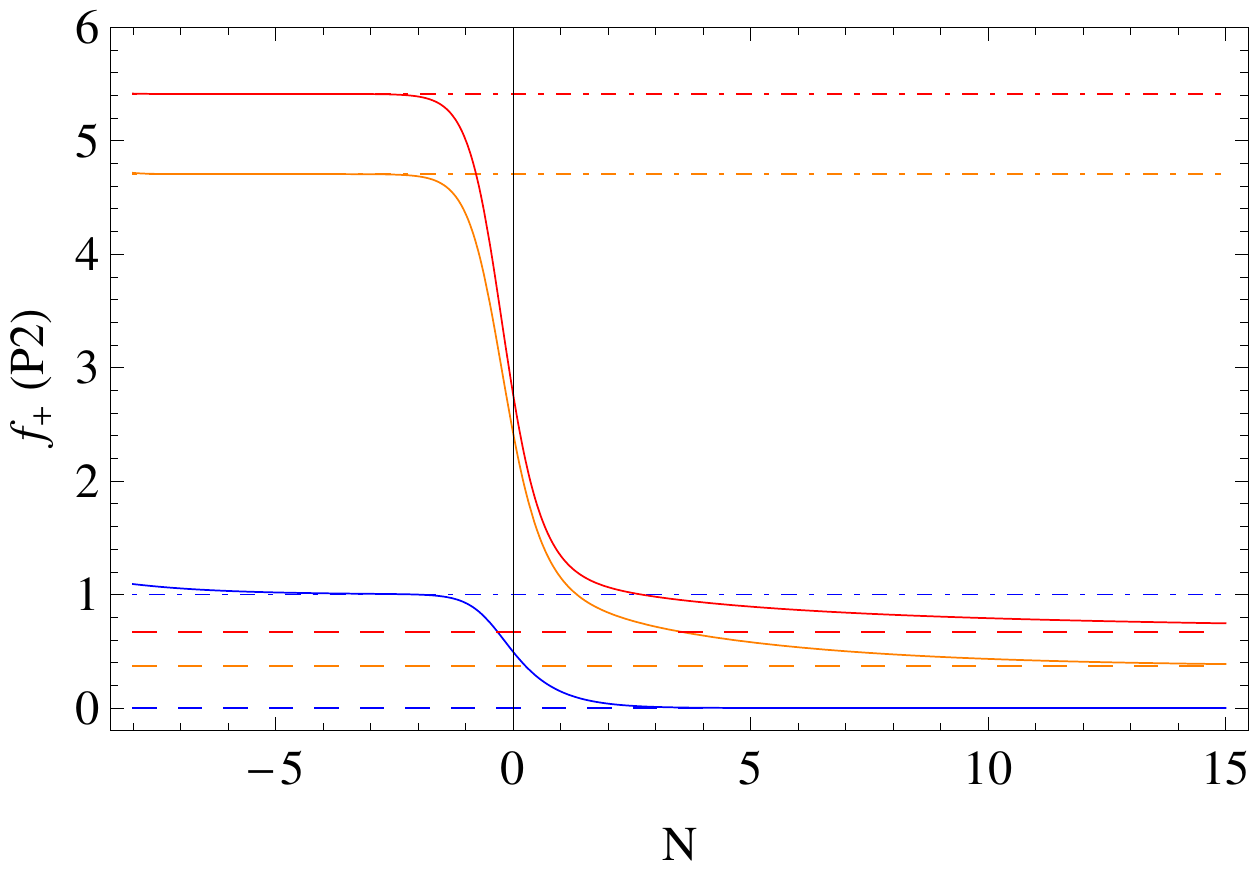}
\includegraphics[width=0.4\columnwidth]{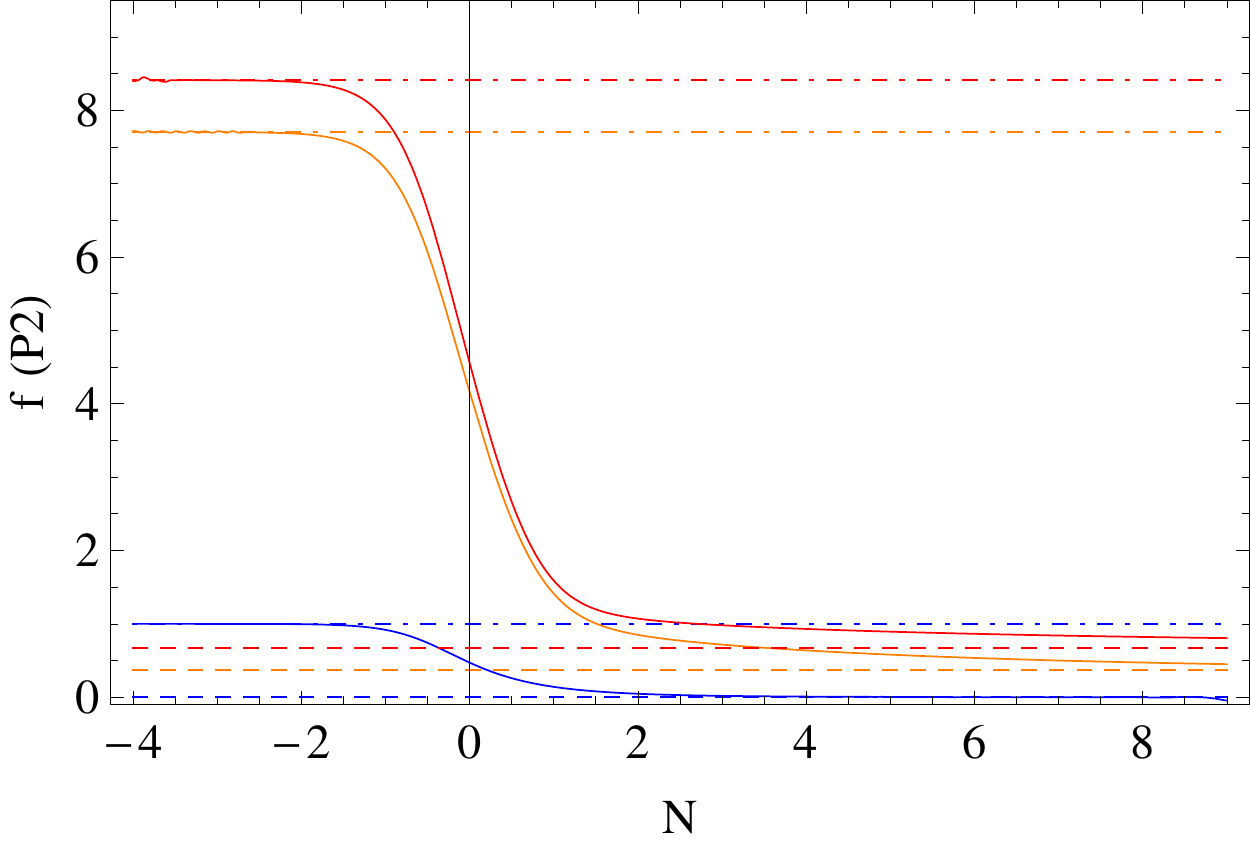}
\caption{The top-left panel is the plot
of $\Omega_{i}$ when $\alpha=0.9$ and $\beta=3$. In this panel, solid curves correspond to
$\Omega_{DE}$ and dashed curves correspond to $\Omega_{-}$ and $\Omega_{+}$.
The other panels are plots of $f_{-},f_{+},f$ for $\alpha=0.9$
and $\beta=$0.5 (blue solid curve), 3.5 (orange solid curve), 4 (red
solid curve). The dashed horizontal lines
are the analytical values for the matter era (dot-dashed) and the accelerated point 2 (dashed).
}
\label{p_2f_pp}
\end{centering}
\end{figure}

 As mentioned earlier, we set initial conditions far enough into the past such that the point 4 matter era has
already decayed away, leaving only point 3.
As we can see in Table \ref{tbl_crt-1}, on point 3 there are two possible growth rates,
unity and $\frac{1}{4}\left(-1+\sqrt{1+32\beta^{2}}\right)$.
Let us now define the  initial adiabaticity parameter:

\begin{equation}
A_{{\rm ic}}=\frac{\Omega_{-}\delta_{-i}}{\Omega_{+}\delta_{+i}}=\frac{1-\mu}{1+\mu}\frac{\delta_{-i}}{\delta_{+i}} \,,
\label{eq:adiab}
\end{equation}
where $\mu$ is the asymmetry parameter \citep[see Ref.][]{Baldi_2012a} defined as
\begin{equation}
\mu=\frac{\Omega_{+}-\Omega_{-}}{\Omega_{+}+\Omega_{-}}\,, \label{eq:Myu}
\end{equation}
and the contrasts $\delta_{\pm i}$ are evaluated at the initial time.
We see that $A_{{\rm ic}}=1$ implies adiabatic initial conditions. We find that if one starts
with  $A_{{\rm ic}}=1$ at very high redshifts,
then initially the growth rate equals  unity.
 However, this trajectory is unstable for $|\beta | \geq \beta _{\rm G} = \sqrt{3}/2$ such that
soon the growth rate moves to the second value $\frac{1}{4}\left(-1+\sqrt{1+32\beta^{2}}\right)$.
If $|\beta | < \beta _{\rm G}$, instead, the growth rate remains stably at unity.
That is, adiabatic fluctuations are unstable, as already
found in Ref.~\citep[][]{Baldi_2012a}, for $|\beta |\geq \beta _{\rm G}$. If instead one starts with  $A_{{\rm ic}}$ substantially
different from unity, then the growth rate goes directly to max$[1,\frac{1}{4}\left(-1+\sqrt{1+32\beta^{2}}\right)]$.

It is instructive to find the transition point analytically.
The full perturbation solutions of point 3 are the following:
\begin{eqnarray}
\delta_{+}&=&\frac{\delta_{+i}}{2}\left[(1+A_{\rm ic})e^{f_{1}(N-N_{i})}+(1-A_{\rm ic})e^{f_{2}(N-N_{i})}\right] \,,
\label{eq:cr_3p}\\
\delta_{-}&=&\frac{\delta_{+i}}{2}\left[(1+A_{\rm ic})e^{f_{1}(N-N_{i})}-(1-A_{\rm ic})e^{f_{2}(N-N_{i})}\right] \,,
\label{eq:cr_3m}
\end{eqnarray}
{where $A_{\rm ic}$ and $\delta_{+i}$ are the initial adiabaticity  and initial perturbation for the CDM
species with positive coupling at $N_{i}$, respectively, and}
\begin{equation}
f_{1}=1 \,,
\quad
f_{2}=\frac{1}{4}(-1+\sqrt{1+32\beta^{2}}) \,.
\end{equation}
We can now find the point at which the second term starts to dominate,
i.e.~the transition time $N_{0}$ at which the two terms in eqs.~(\ref{eq:cr_3p}) and (\ref{eq:cr_3m}) are equal. This is given by:
\begin{equation}
N_{0}-N_{i}=\frac{1}{f_{2}-f_{1}}\ln \left|\frac{1+A_{\rm ic}}{1-A_{\rm ic}}\right|\,.
\end{equation}
{
Therefore,
as expected, the transition point  $N_{0}$ is very close to the initial time $N_{i}$, unless $A_{{\rm ic}}$ is
extremely close to unity.
For instance, when $\beta=1$ and the initial adiabaticity is $A_{\rm ic}=2$ the transition point
occurs $0.5$ e-foldings after the initial time and if $A_{\rm ic}=1.01$ it occurs after $2.5$ e-foldings.
 The above derivation explicitly shows that adiabatic initial conditions will rapidly evolve into a non-adiabatic
 state for coupling values $|\beta |\geq \beta _{\rm G}$, while for $|\beta |<\beta _{\rm G}$ the initial adiabaticity is preserved.
As a consequence, possible observational effects associated with the transition between adiabatic and non-adiabatic initial conditions
would be relevant only if the transition occurred in the redshift range covered by observational data, which is a condition that requires a high level of fine-tuning
of the model parameters.
 Therefore, in the following we will restrict to the case of non-adiabatic initial conditions and we will assume $A_{{\rm ic}}=2$, without loss of
generality.

The total growth rate $f$ is presented
in the right panel of Fig.~\ref{p_2f_m} for some values of parameters that lie within
the stable range of point 2, comparing with the exact solutions on
the critical points. We also present the evolution
of background fractional densities for the same values of parameters in the left panel of Fig.~\ref{p_2f_m}.
For the same values of $\beta$ but different $\alpha=0.9$ we plot the different growth rates $f_{-},f_{+},f$ in the last three panels of  Fig.~\ref{p_2f_pp}. Again, in the first panel of Fig.~\ref{p_2f_pp}, we illustrate the corresponding evolution
of background fractional densities $\Omega_{i}$. We stress here that plots for fractional densities are made for one set of parameters only as the other cases do not differ at the background level. In these two figures, parameters are chosen in the stable range of point 2.
 Similarly, for parameters within the stable region of point 5, we illustrate the
growth rate and the background fractional density evolution in Fig.~\ref{p_5f_m}.

The numerical integration of equations (\ref{eq:pert_1-2-1-1}) and (\ref{eq:pert_min})
together
 with the background equations shows a very interesting effect:
during  the observationally relevant range $z\approx 1$, the growth rate  gets strongly enhanced
when $\beta$ grows larger than $\beta _{\rm G}$, before
being driven back to a small value when dark energy fully dominates.
In Fig.~\ref{betabe} we illustrate this behavior plotting the quantity $\delta'/\delta_{0}=f(z)G(z)$
(where $G(z)$ is the growth factor normalized to unity today), since this is the observational quantity we will compare our
model to in the next section.
% As the figure shows, at progressively lower redshifts the growth rate becomes larger and hardly compatible with observations.
As can be seen, the
McDE behavior suddenly deviates from the $\Lambda$CDM case as $|\beta |$
grows larger than $\beta _{\rm G}$.
This leads
us to expect that for $|\beta |$ larger than  $\beta _{\rm G}$ the model becomes
rapidly inconsistent with observations
 and that growth rate data can place tight constraints on the coupling value,
as indeed we are going to find
in the next Section.

\begin{figure}
\begin{centering}
\includegraphics[width=0.4\columnwidth]{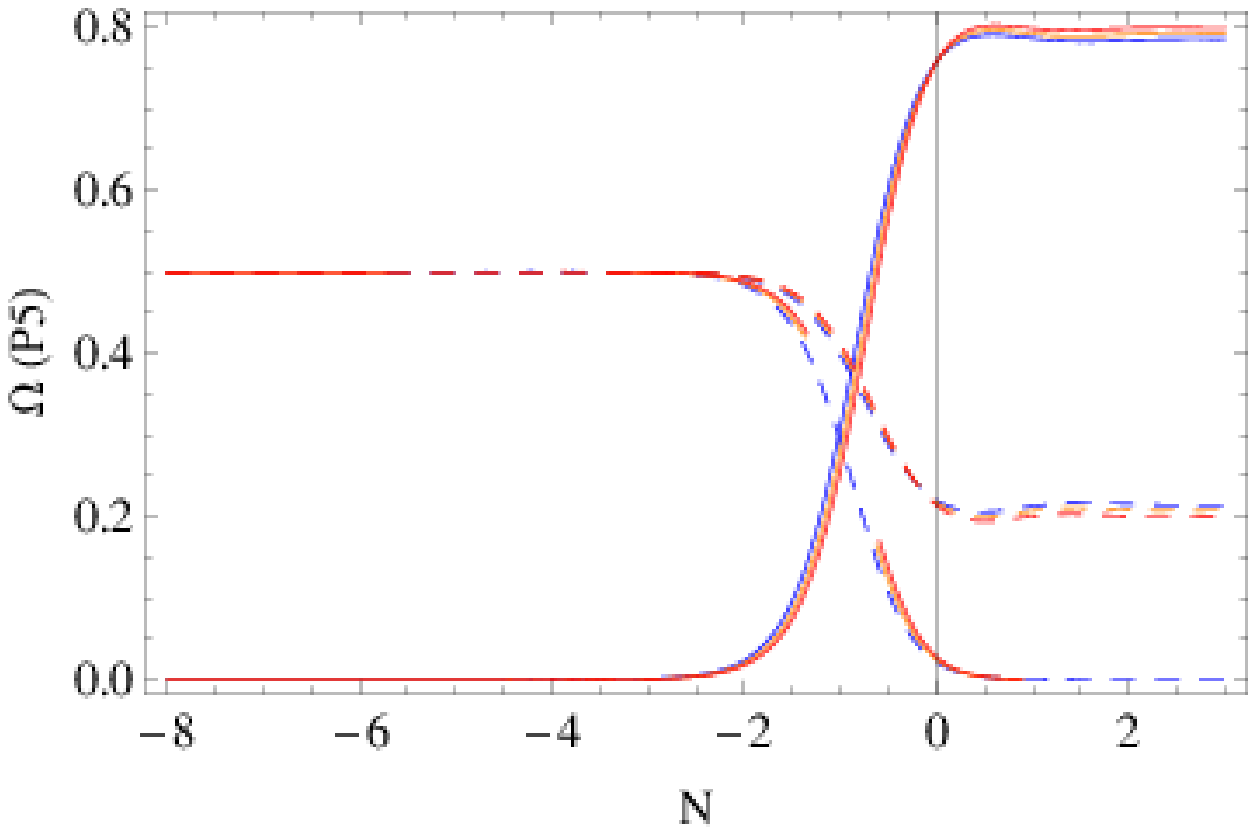}
\includegraphics[width=0.4\columnwidth]{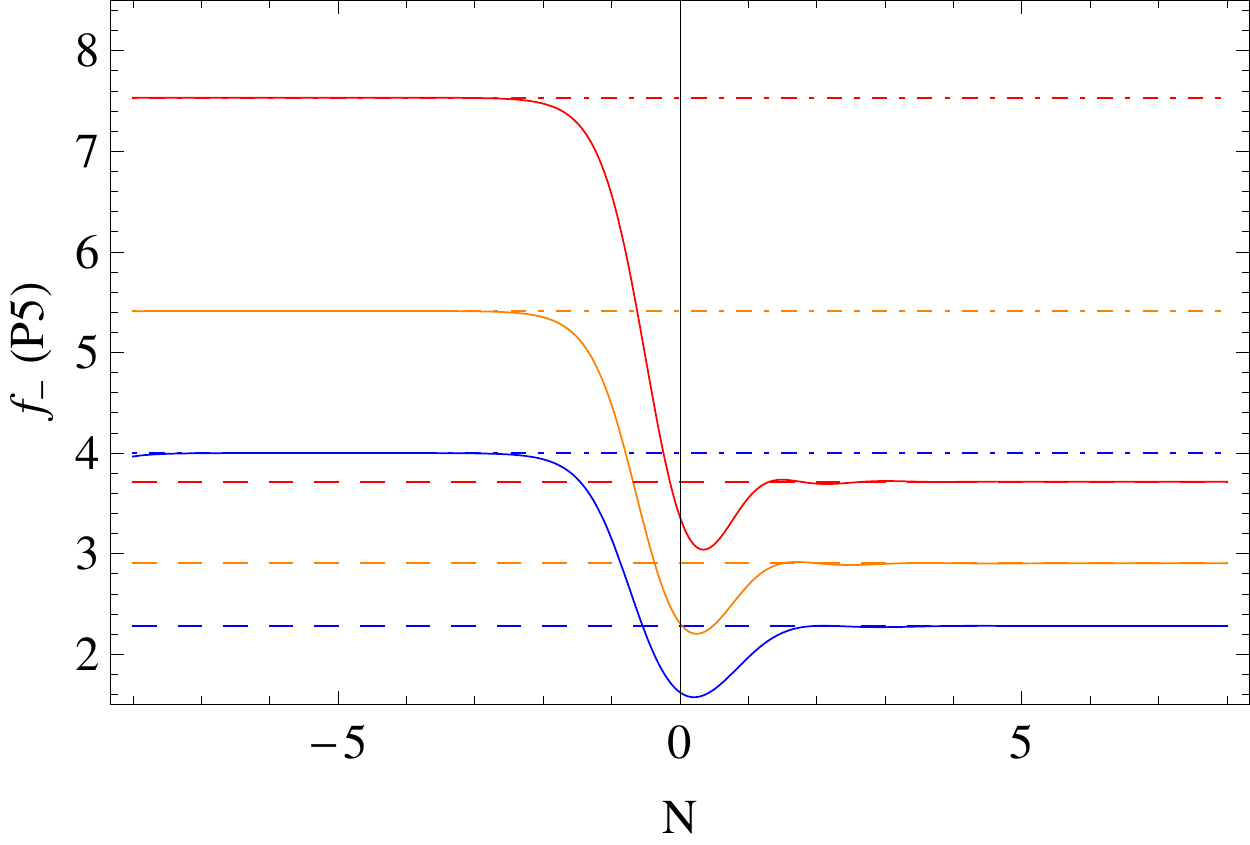}
\includegraphics[width=0.4\columnwidth]{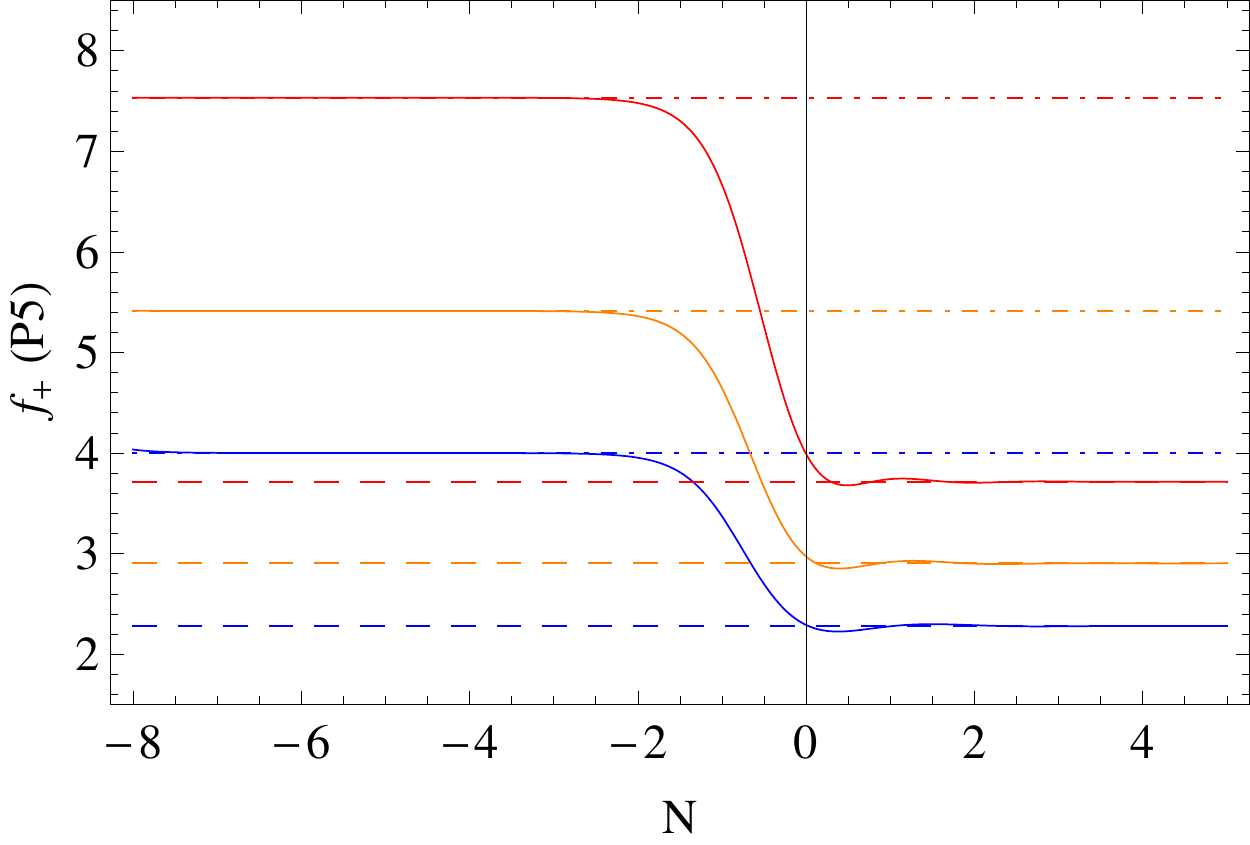}
\includegraphics[width=0.4\columnwidth]{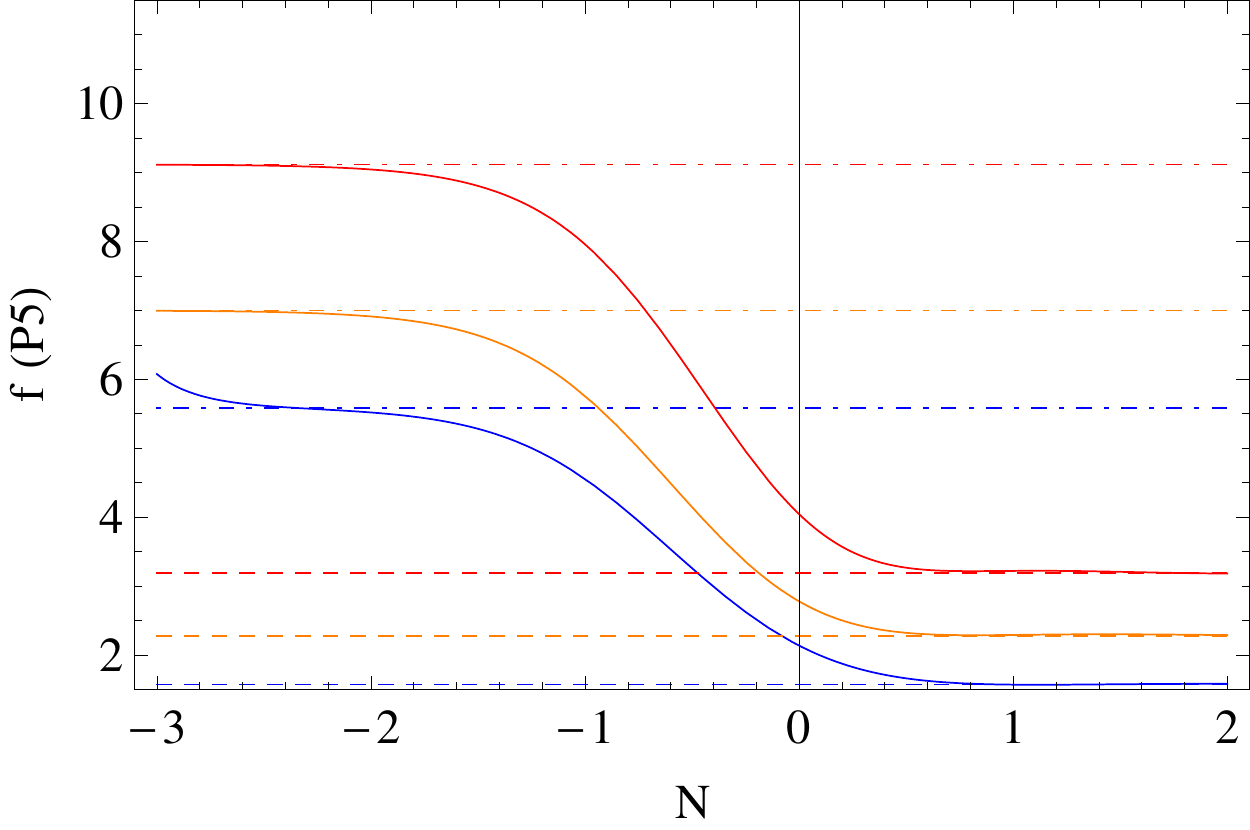}
\caption{ As Fig.~\ref{p_2f_pp} but for the stable region of point 5.}
\label{p_5f_m}
\end{centering}
\end{figure}

%\begin{figure}
%\begin{centering}
%\includegraphics[width=0.5\columnwidth]{Fig/p5_Om_alf_2_c_35_4_45_De07_mu_0}
%\caption{$\Omega_{i}$ for $\alpha=2$ and $\beta=$2 (blue),
%3 (orange), 4 (red), where solid curves
%correspond to $\Omega_{DE}$ and dashed curves correspond to $\Omega_{-}$
%and $\Omega_{+}$. Parameters are chosen in the stable range of point
%5.}
%\label{p_5om}
%\end{centering}
%\end{figure}

\begin{figure}
\begin{centering}
\includegraphics[width=0.48\columnwidth]{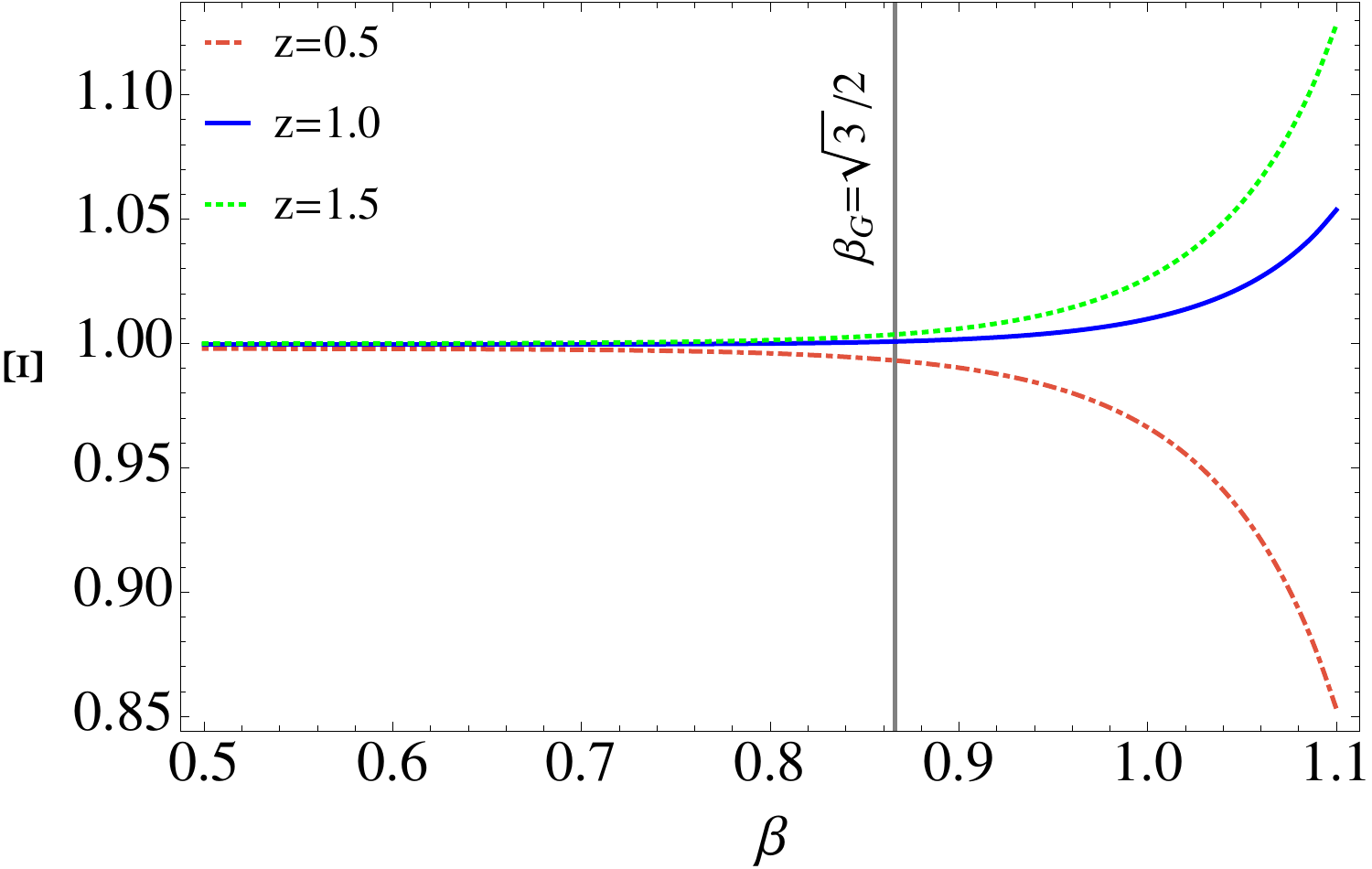}
\quad
\includegraphics[width=0.48\columnwidth]{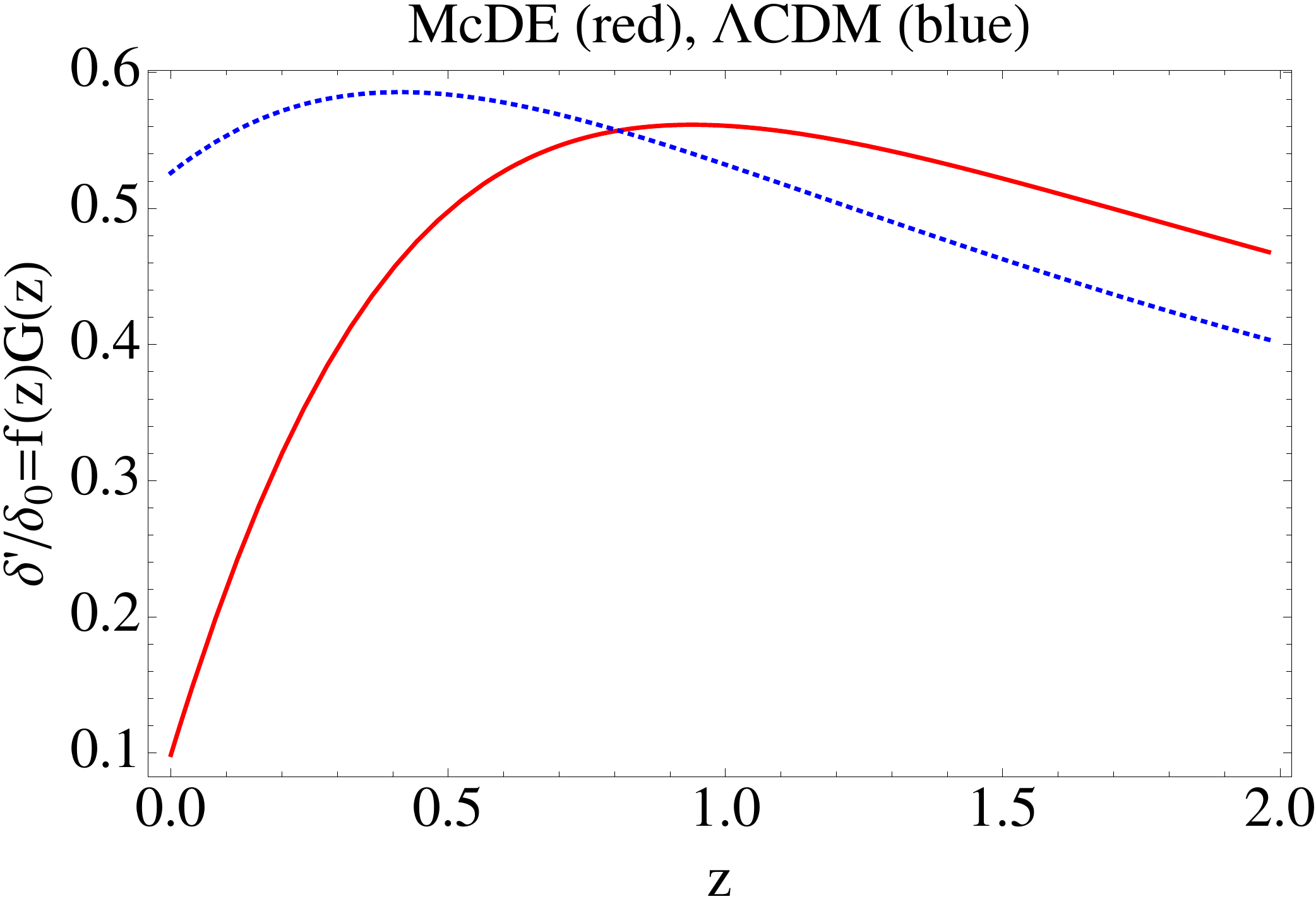}
\caption{{\em Left Panel:} The ratio $\Xi\equiv \dfrac{f(z)G(z)|_{{\rm McDE}}}{f(z)G(z)|_{{\rm \Lambda CDM}}}$
as a function of the coupling $\beta$ at $z=\left\{0.5\,,1\,,1.5\right\}$ and for parameters
$\alpha=0.1$ and $\Omega_{{\rm DE},0}=0.692$. As can be seen, the
McDE behavior suddenly deviates from the $\Lambda$CDM case as $|\beta |$
grows larger than $\beta _{\rm G}$.
%, with the effect becoming more severe at lower redshifts.
{\em Right Panel:} evolution with respect to redshift of $\delta'/\delta_{0}=f(z)G(z)$ for the McDE model with $\beta=1.1$, $\alpha=0.1$ and $\Omega_{{\rm DE},0}=0.692$ (red solid line) and for the $\Lambda$CDM model with the same dark-energy content (blue dotted line).
The relative trend of the two curves at different redshifts explains the opposite behavior between the $z=0.5$ and the $z \geq 1$ curves shown in the left panel.
}
\label{betabe}
\end{centering}
\end{figure}

\section{Comparison to observations}
\label{analysis}

In our previous work Ref.~\cite{Piloyan:2013mla} we employed the Union2.1
Compilation~\cite{Suzuki_etal_2012} of Type Ia supernovae to constrain the background
behavior of the McDE model. We found that supernova data can constrain
the slope of the self-interaction potential $\alpha$, which is found
to be bound to values $\le1.5$ at the $3\sigma$ confidence level.
On the other hand, we found a flat posterior likelihood for the initial
asymmetry parameter, $\mu_{in}$, which is therefore completely unconstrained
by the data, and we derived an extremely loose bound $|\beta |\lesssim  83$ (at the
$2\sigma$ confidence level) on the coupling parameter. This showed
how efficient is the McDE model in mimicking the $\Lambda$CDM model
at the background level. Here we will extend the previous analysis
by confronting the McDE model with present growth rate
data as well as forecasted future data from upcoming wide-field surveys: as we will see the study of linear perturbations in the McDE
model will allow us to put much tighter constraints on the coupling
parameter.
We proceed by describing the different data sets entering our investigation, and the likelihood estimator adopted for the comparison with the theoretical predictions.

\subsection{SN data}
\label{supernovae}

At the background level, we will make use of two different SN datasets. The first is the Union2.1 Compilation~\cite{Suzuki_etal_2012}
of 580 Type Ia SNe in the redshift range $z=0.015-1.414$. More precisely,
we use the magnitude vs.~redshift table (without systematic errors)
publicly available at the Supernova Cosmology Project \href{http://supernova.lbl.gov/Union/}{webpage}.
The second dataset corresponds instead to the forecasted sample of two years of observations
by the Large Synoptic Survey Telescope and features a total of $10^{5}$
supernovae in the redshift range $z=0.1-1.0$ with the redshift distribution
as given in \cite{Abell:2009aa}. We will refer to this dataset at
the ``LSST 100k'' catalog.

The predicted theoretical magnitudes are related to the luminosity
distance $d_{L}$ by:
\begin{equation}
m(z)=5\log_{10}\frac{d_{L}(z)}{10\,\textrm{pc}}\,,
\end{equation}
which is computed under the assumption of spatial flatness:
\begin{equation}
d_{L}(z)=(1+z)\int_{0}^{z}\frac{\rd\tilde{z}}{H(\tilde{z})}\,.
\end{equation}
The luminosity distance $d_{L}(z)$ is obtained by integrating numerically
the background McDE equations as explained in Ref.~\cite{Piloyan:2013mla}.
The $\chi^{2}$ function, on which the likelihood analysis will be
based, is then:
\begin{equation}
\chi_{SNIa}^{'2}=\sum_{i}\frac{[m_{i}-m(z_{i})+\xi]^{2}}{\sigma_{i}^{2}}\,,
\end{equation}
where the index $i$ labels the elements of the supernova dataset.
The parameter $\xi$ is an unknown offset sum of the supernova absolute
magnitudes, of $k$-corrections and other possible systematics. As
usual, we marginalize the likelihood $L'_{SNIa}=\exp(-\chi_{SNIa}^{'2}/2)$
over $\xi$, such that $L_{SNIa}=\int\rd\xi\, L'_{SNIa}$, leading to a new
marginalized $\chi^{2}$ function:
\begin{equation}
\chi_{SNIa}^{2}=S_{2}-\frac{S_{1}^{2}}{S_{0}}\,,
\end{equation}
where we neglected a cosmology-independent normalizing constant, and
the auxiliary quantities $S_{n}$ are defined as:
\begin{equation}
S_{n}\equiv\sum_{i}\frac{\left[m_{i}-m(z_{i})\right]^{n}}{\sigma_{i}^{2}}\,.
\end{equation}
As $\xi$ is degenerate with $\log_{10}H_{0}$, we are effectively
marginalizing also over the Hubble constant.

\subsection{$f\sigma_{8}(z)$ data}
\label{fs8}

At the linear perturbation level, we will build the growth-rate likelihood using two different datasets.
The first contains the latest data \citep[see][]{Macaulay:2013swa}
from 6dFGS~\cite{Beutler:2012px}, LRG~\cite{Samushia:2011cs},
BOSS~\cite{Tojeiro:2012rp}, WiggleZ~\cite{Blake:2012pj} and VIPERS~\cite{delaTorre:2013rpa}.
The second dataset approximates instead the forecasted accuracy of a
future Euclid-like mission and it has been obtained in Ref.~\cite{2013LAFogli}.

These different growth-rate data are  given
as a set of values $d_{i}$ where $i=\left\{ \right.$6dFGS, LRG, BOSS, WiggleZ, VIPERS, Euclid$\left. \right\}$ and where
\begin{equation} \label{obsf}
d=f\sigma_{8}(z)=f(z)\sigma_{8}G(z)=\sigma_{8}\delta'/\delta_{0}\,.
\end{equation}
Let us denote our theoretical estimates as $t_{i}=\delta'_{i}/\delta_{0}$,
where $\delta$ indicates the total density perturbation.
We can then build
a $\chi^{2}$ function that reads:
\begin{equation}
\chi_{f\sigma_{8}}^{'2}=\left(d_{i}-\sigma_{8}t_{i}\right)C_{ij}^{-1}\left(d_{j}-\sigma_{8}t_{j}\right)\,,
\end{equation}
where $C_{ij}$ is the covariance matrix of the data. Since we do
not know $\sigma_{8}$ and cannot use the standard estimates because
they have been obtained assuming the standard $\Lambda $CDM model, we need to marginalize
the likelihood $L'_{f\sigma_{8}}=\exp(-\chi_{f\sigma_{8}}^{'2}/2)$
over $\sigma_{8}$, such that $L_{f\sigma_{8}}=\int\rd\sigma_{8}\, L'_{f\sigma_{8}}$,
leading to a new marginalized $\chi^{2}$ function:
\begin{equation}
\chi_{f\sigma_{8}}^{2}=S_{20}-\frac{S_{11}^{2}}{S_{02}}\,,
\end{equation}
where we neglected a cosmology-independent normalizing constant, and
the auxiliary quantities $S_{nm}$ are defined as:
\begin{align}
S_{11} & =d_{i}C_{ij}^{-1}t_{j}\,,\\
S_{20} & =d_{i}C_{ij}^{-1}d_{j}\,,\\
S_{02} & =t_{i}C_{ij}^{-1}t_{j}\,.
\end{align}
We are effectively marginalizing also over the initial value of $\delta_{0}$
 as the latter is degenerate with $\sigma_{8}$.

\subsection{Full likelihood}

The full likelihood is based on the total $\chi^{2}$:
\begin{equation}
\chi_{{\rm tot}}^{2}=\chi_{SNIa}^{2}+\chi_{f\sigma_{8}}^{2}\,,
\end{equation}
which depends on the three main parameters $\Omega_{{\rm DE},0},\alpha,|\beta |$
plus other parameters specifying the initial conditions, $A_{{\rm ic}},\mu_{{\rm in}},\delta_{\pm,{\rm in}},\delta'_{\pm,{\rm in}}$.
As discussed above, we will solve the perturbation equations with
non-adiabatic initial conditions at very early times: $A_{{\rm ic}}=2$. This choice is conservative for coupling values $|\beta |\geq \beta _{\rm G}$
since adiabatic initial conditions would affect the observable quantities only in the highly fine-tuned case where the departure from adiabaticity occurs at very recent times, while for $|\beta |<\beta _{\rm G}$ any choice of $A_{{\rm ic}}$ is equivalent since adiabaticity is conserved.
This together
with the fact that we are effectively marginalizing over $\delta_{\pm,{\rm in}}$ implies
that the likelihood $L_{{\rm tot}}=\exp(-\chi_{{\rm tot}}^{2}/2)$
depends very weakly on the initial conditions parameters, which we have
therefore fixed for convenience to $A_{{\rm ic}}=2$, $\mu_{{\rm in}}=0$,
$\delta'_{+,{\rm in}}=\delta_{+,{\rm in}}=\exp N_{{\rm in}}$, $\delta_{-,{\rm in}}=A_{{\rm ic}}\frac{1+\mu}{1-\mu}\delta_{+,{\rm in}}$,
$\delta'_{-,{\rm in}}=\delta_{-,{\rm in}}$\,,
 consistently with the evolution of perturbations during matter domination.

%%%%%%%%%%%%%%%%%%%%%%%
\section{Results}
\label{results}
%%%%%%%%%%%%%%%%%%%%%%%

\begin{figure}
\begin{centering}
\includegraphics[width=0.48\columnwidth]{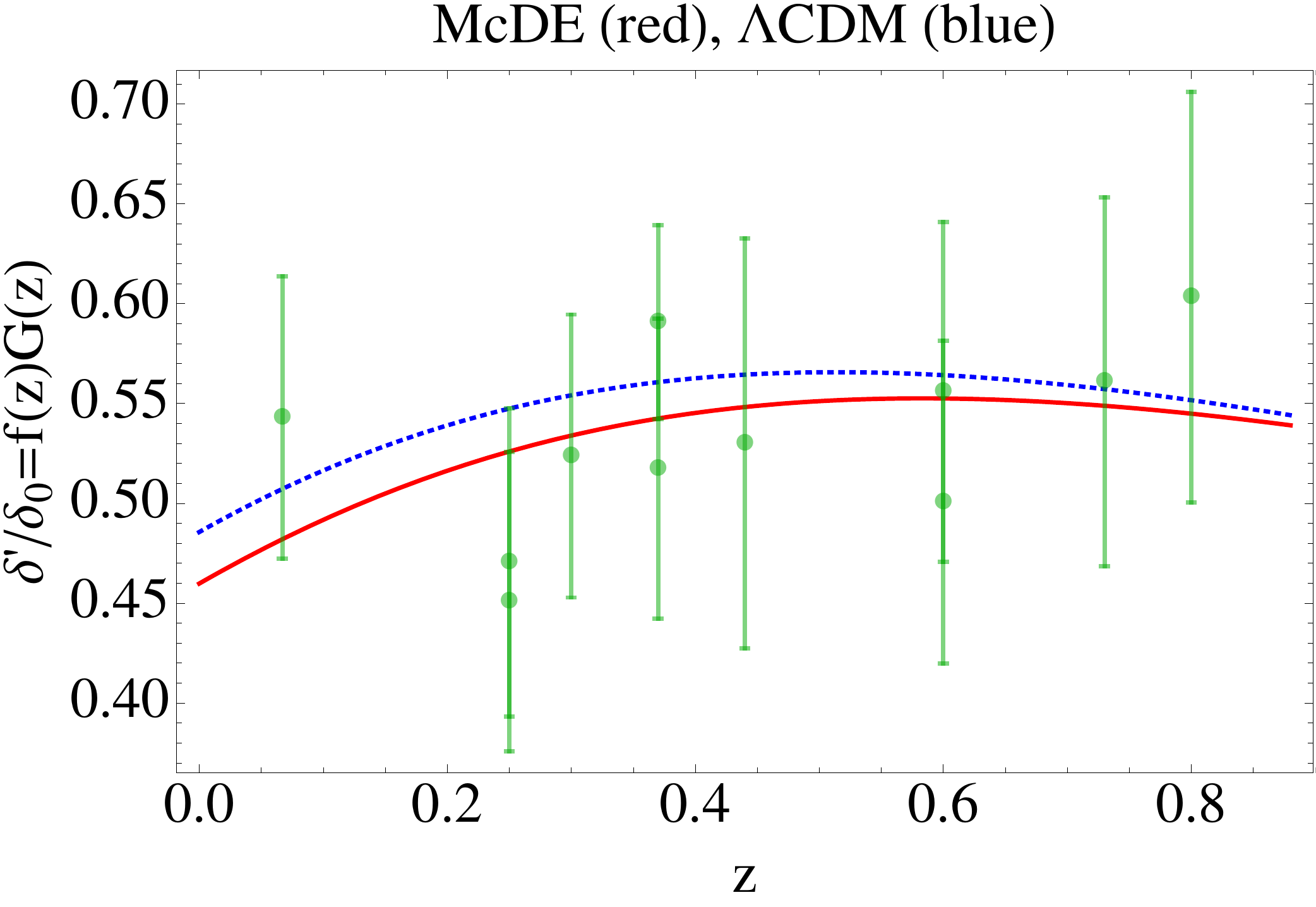}
\quad
\includegraphics[width=0.48\columnwidth]{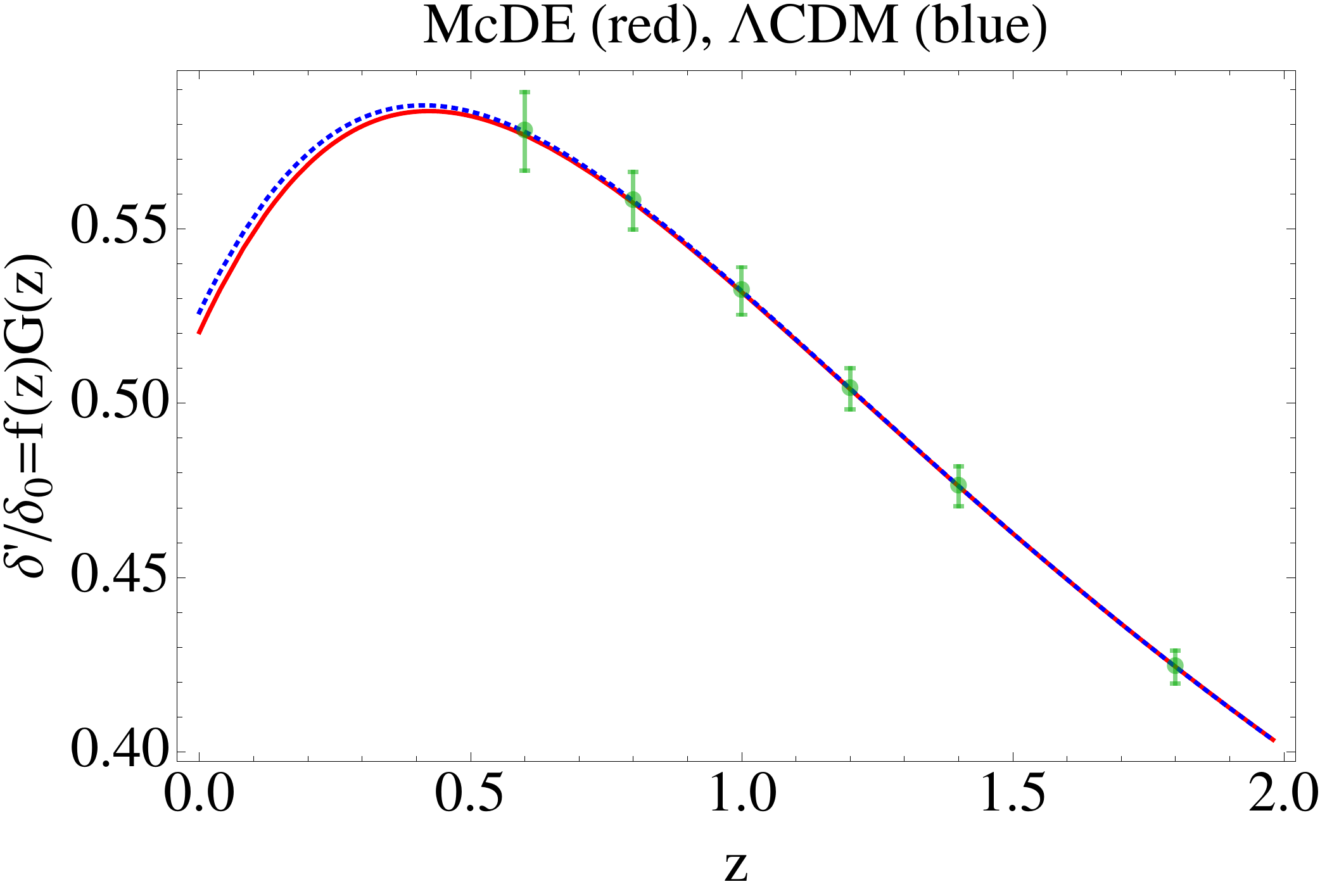}
\caption{Best-fit McDE model (second and fourth columns of Table \ref{tab:1D} for left and right panel, respectively) together with $f\sigma_{8}$
data points for present (left panel) and future (right panel) data.
The $\Lambda$CDM curve is displayed for comparison and
is relative to a $\Lambda$CDM model with the same $\Omega_{{\rm DE},0}$.
As the likelihood is marginalized over $\sigma_{8}$, a possible vertical
shift is inconsequential.}
\label{data}
\end{centering}
\end{figure}

%\begin{figure}
%\begin{centering}
%\includegraphics[width=0.55\columnwidth]{Figs-results/data-E}
%\caption{ Best-fit McDE model (see {\rop forth} column of Table \ref{tab:1D}) together with $f\sigma_{8}$
%data used. The $\Lambda$CDM curve is displayed for comparison and
%is relative to a $\Lambda$CDM model with the same $\Omega_{{\rm DE},0}$.}
%\label{data-E}
%\end{centering}
%\end{figure}

\begin{table}
\begin{centering}
{
\renewcommand{\arraystretch}{1.8}
\begin{tabular}{|lcccc|cc|}
\hline
Parameter  &
{\begin{minipage}{45pt}
\centering
Best Fit\\
{\small (SN+$f\sigma_{8}$)}
\end{minipage}}  & 95\% c.l.&  {\begin{minipage}{45pt} \centering Best Fit\\
{\small (LSST+Euclid)} \end{minipage}}  & 95\% c.l.&  {\begin{minipage}{45pt} \centering Best Fit\\
{\small (SN)} \end{minipage}}  & 95\% c.l. \tabularnewline
\hline
\hline
$\Omega_{{\rm DE},0}$  & 0.734 & $[0.684,0.824]$ & 0.692  & $[0.688,0.698]$ & 0.719 & $[0.680,0.765]$\tabularnewline
$\alpha$  & 0.66  & $[0,1.36]$  & 0.12  & $[0,0.54]$ & 0.62 & $[0,1.01]$ \tabularnewline
$\beta$  & 0.79  & $[0,0.88]$   & 0.03  & $[0,0.85]$ & 6.4 & $[0,83]$\tabularnewline
\hline
%$\mu_{{\rm in}}$  & unconstrained  & unconstrained  \\
%$\delta'_{{\rm in}}$  & unconstrained  & unconstrained  \\
% &  & \tabularnewline
\end{tabular}
}
\caption{ Best-fit values and 95\% confidence intervals for the parameters
of the model discussed in this paper (for point 2 and point 5 together) when using the combined Union2.1
supernova dataset and the latest $f\sigma_{8}$ data (2$^{st}$ and 3$^{nd}$ columns), or using the combined LSST
100k supernova dataset and the forecasted Euclid-like $f\sigma_{8}$
data (4$^{rd}$ and 5$^{th}$ columns), while the 6$^{th}$ and 7$^{th}$ columns report the results from ~\cite[][]{Piloyan:2013mla} when using only background observables and analyzing only point 2.}
\label{tab:1D}
\end{centering}
\end{table}

\begin{figure}
\begin{centering}
\includegraphics[width=0.45\columnwidth]{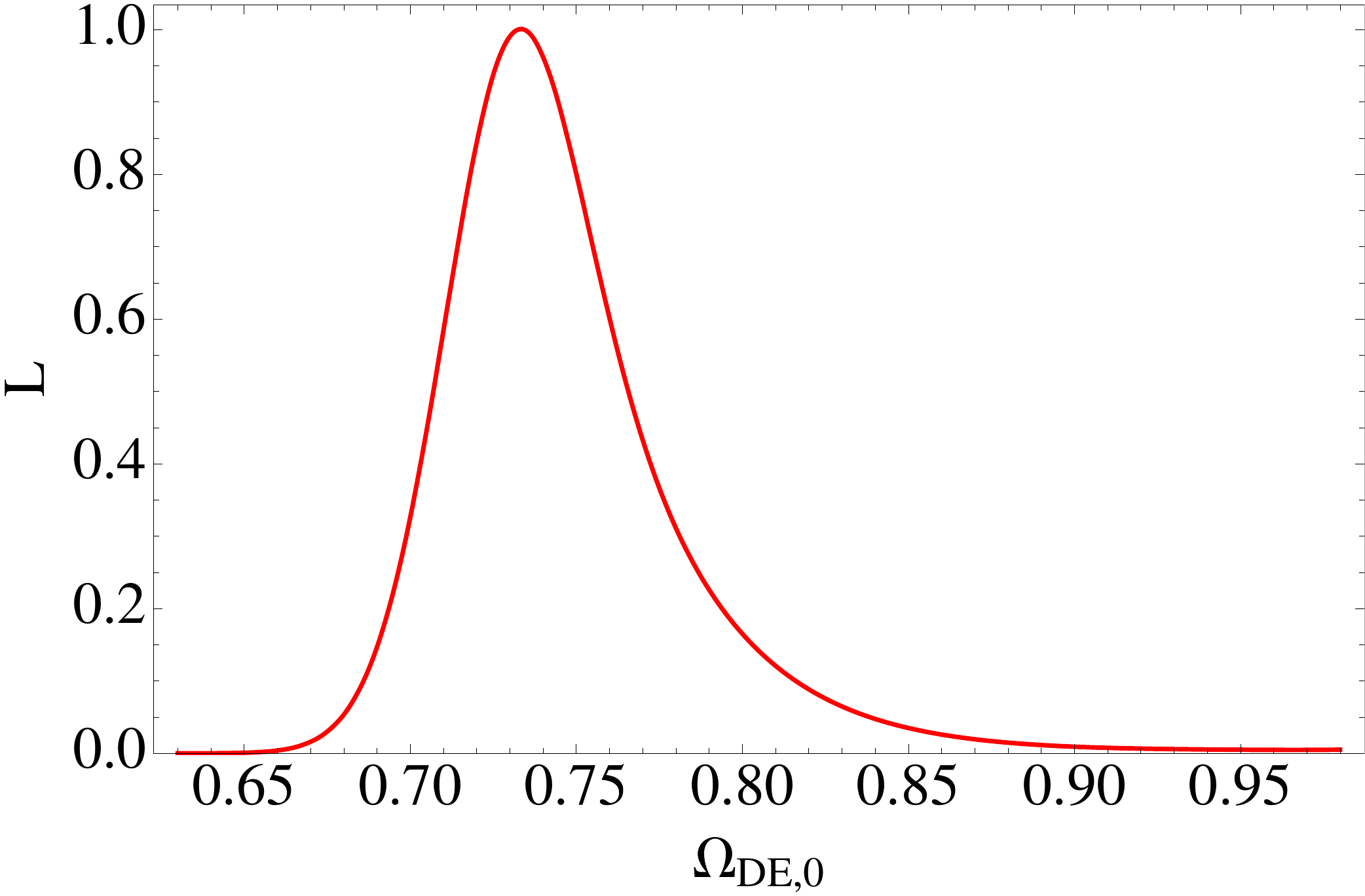}\qquad{}
\includegraphics[width=0.45\columnwidth]{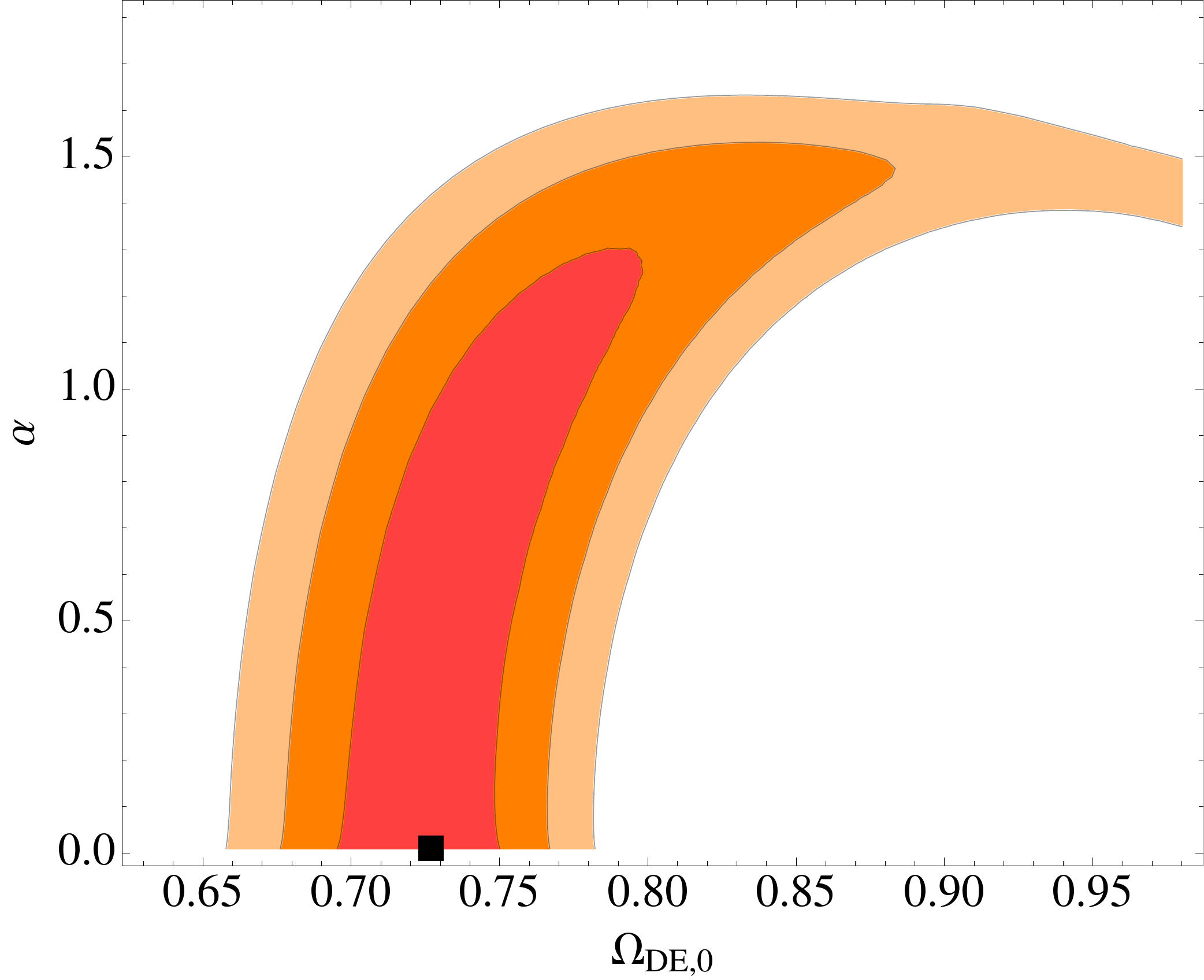}\\
 \ \\
 \ \\
 \includegraphics[width=0.45\columnwidth]{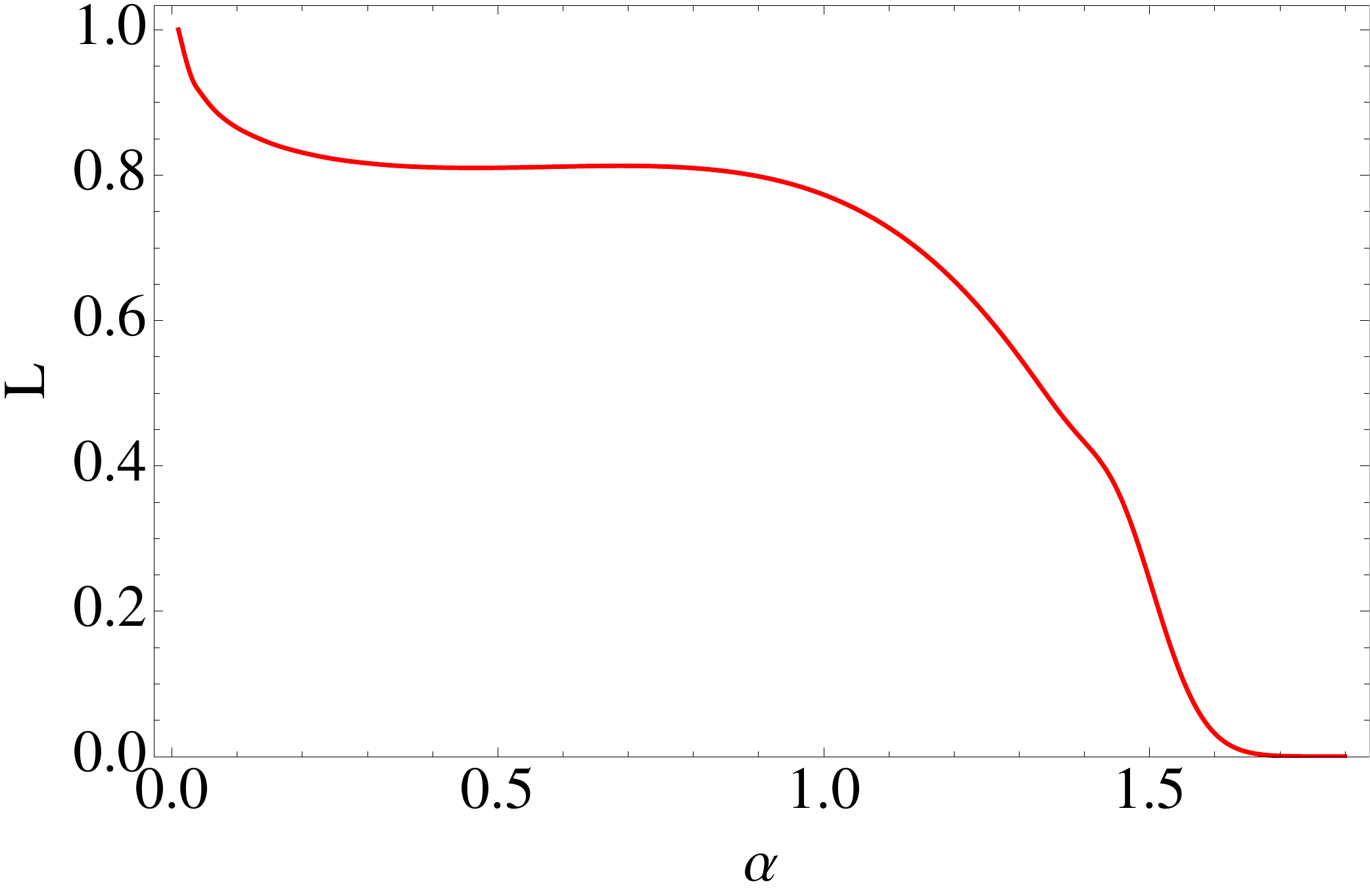}\qquad{}
\includegraphics[width=0.45\columnwidth]{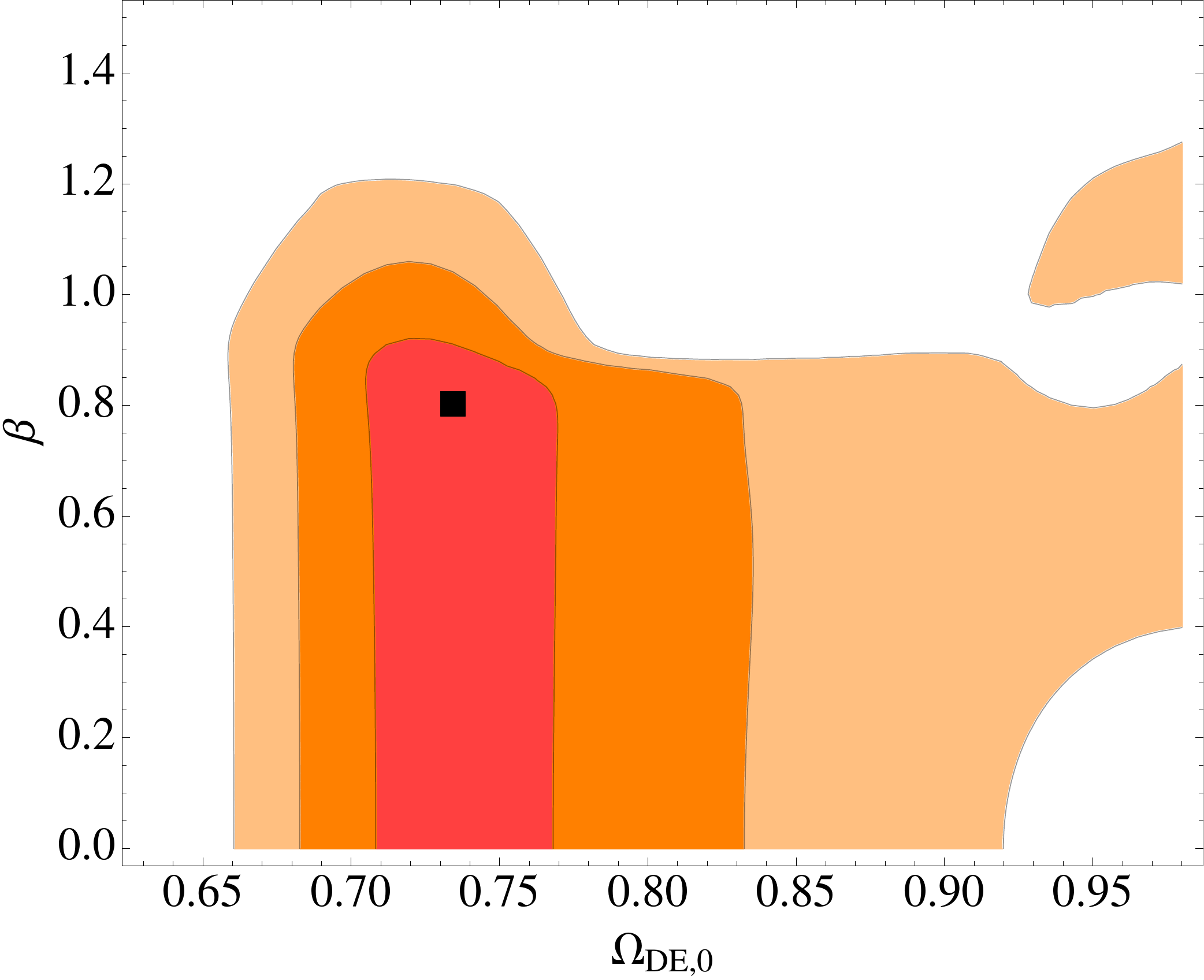}\\
 \ \\
 \ \\
 \includegraphics[width=0.45\columnwidth]{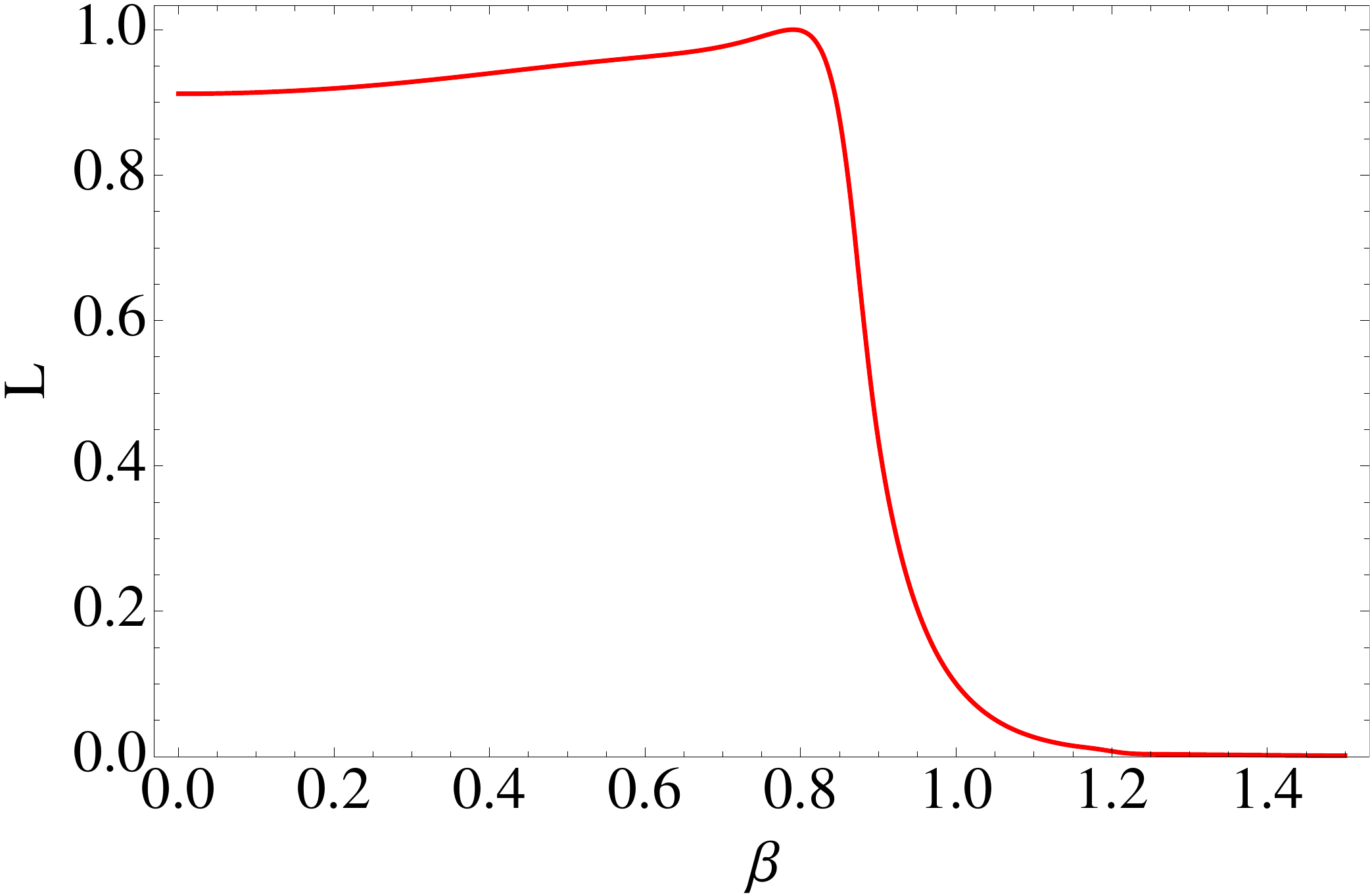}\qquad{}
\includegraphics[width=0.45\columnwidth]{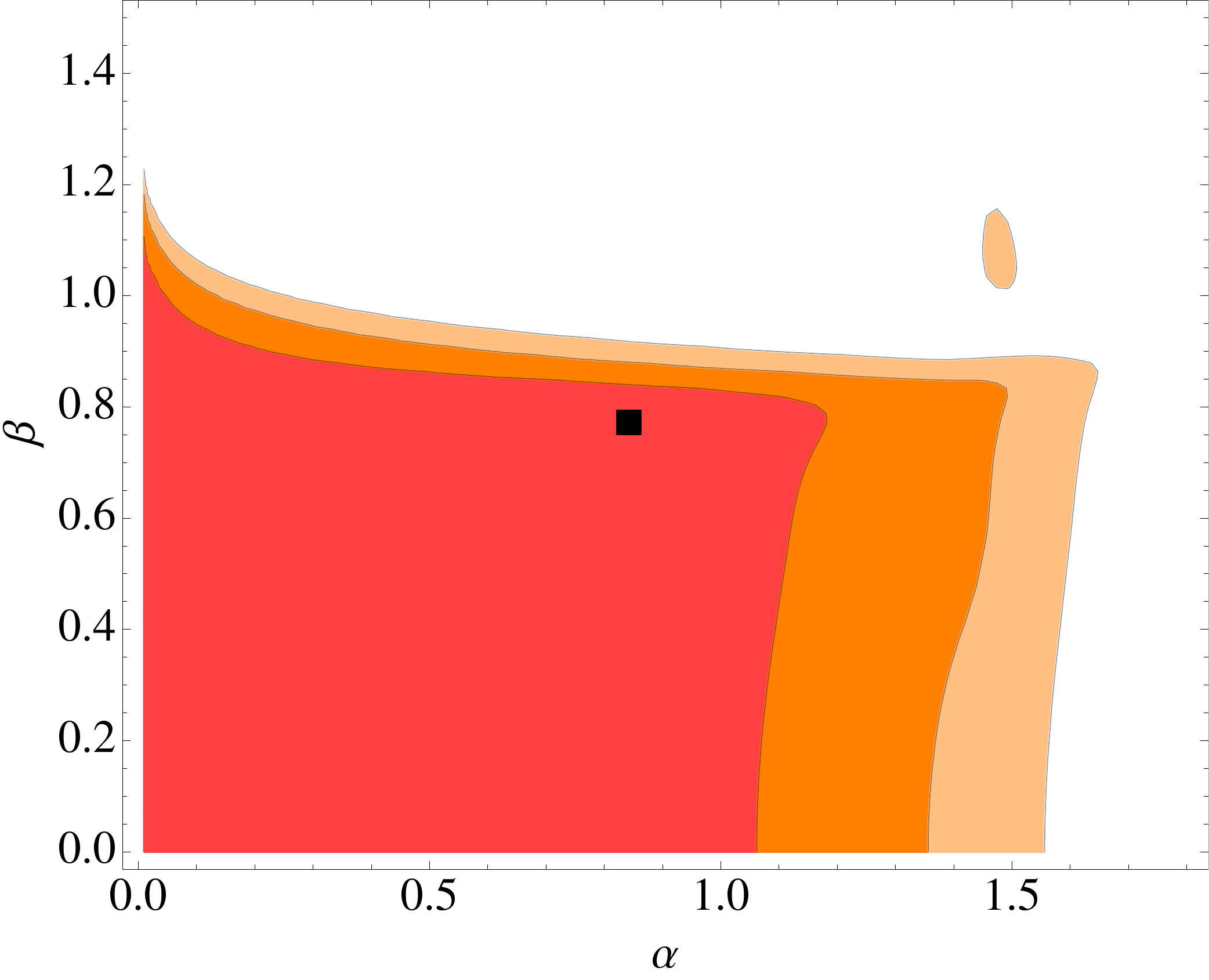}
\caption{{\em On the Left:} 1-dimensional marginalized posterior distributions
(for point 2 and point 5 together) on the parameters $\{\Omega_{{\rm DE},0},\alpha,\beta\}$
when fitting the model of this paper to the Union2.1 SN Compilation
(see Section \ref{supernovae}) and the latest $f\sigma_{8}$ data
(see Section \ref{fs8}). See second and third columns of Table~\ref{tab:1D} for best-fit values
with 95\% confidence intervals. {\em On the Right:} $1\sigma$,
$2\sigma$ and $3\sigma$ confidence-level contours for the relevant
2-dimensional marginalized posterior distributions. The black squares
mark the best-fit values. The degeneracy between $\Omega_{{\rm DE},0}$
and the parameters $\alpha,\beta$ makes values of $\Omega_{{\rm DE},0}\simeq0.90$
possible at the $3\sigma$ level.}
\label{posts}
\end{centering}
\end{figure}

The results obtained with the Union2.1 supernova dataset (see Section
\ref{supernovae}) and the latest $f\sigma_{8}$ data (see Section
\ref{fs8})
are shown in Fig.~\ref{data} (left panel), Fig.~\ref{posts} and in the 2$^{nd}$ and 3$^{rd}$ columns of Table~\ref{tab:1D}.
The best-fit values reported in the latter are relative to the full 3-dimensional likelihood.
 The last two columns of Table~\ref{tab:1D} report the constraints obtained in \cite{Piloyan:2013mla}
when using only background data.
As we can see, the effect of including growth rate data in the analysis is dramatic:
the 2$\sigma$ confidence region for the coupling $|\beta |$ shrinks from $|\beta |\lesssim 83$ when using supernovae
only to $|\beta |\lesssim 0.88$ when present growth rate data is included. % in the analysis.
%, or $|\beta |\lesssim 0.85$ for present and future growth rate data, respectively.
This is indeed expected, since one of the main motivations
for the introduction of the McDE model was to explore a case in which
the $\Lambda$CDM expansion is followed closely while the perturbations
deviate significantly and show new effects.

%As can be seen from Fig.~\ref{posts}, the likelihood depends non-trivially
%on $\beta$. The reason for the features at $\beta\sim1$ is actually
%rather simple. As explained in Section~\ref{fs8}, we marginalize
%the likelihood over the value of $\sigma_{8}$. This means that a
%possible vertical displacement in the growth rate is inconsequential
%as far as the likelihood function is concerned. Coming back to the
%features in $\beta$ of Fig.~\ref{posts}, if for example we move
%on the line $\alpha=1$ for increasing $\beta$, we find that $f$
%agrees well with the $\Lambda$CDM growth rate for $\beta\lesssim0.8$,
%then for larger $\beta$ values it starts deviating more and more.
%However, at $\beta\sim1$ the shape of the McDE growth rate is similar
%to the $\Lambda$CDM one (even though quite larger) and, because of
%the marginalization on $\sigma_{8}$, it has a comparably good $\chi^{2}$.
%This explains the sudden increase in the likelihood function at $\beta\sim1$.

\begin{figure}
\begin{centering}
\includegraphics[width=0.45\columnwidth]{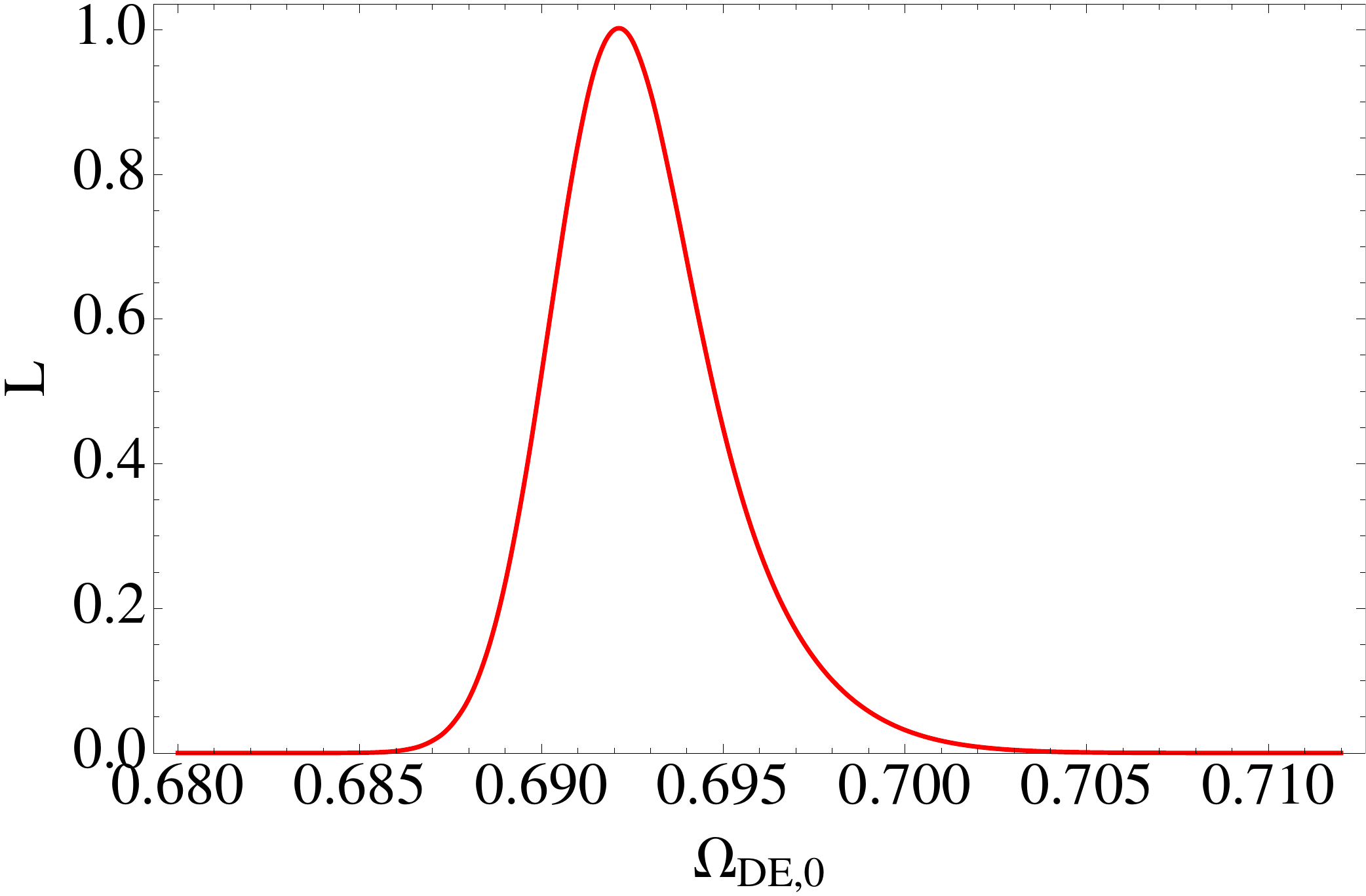}\qquad{}
\includegraphics[width=0.45\columnwidth]{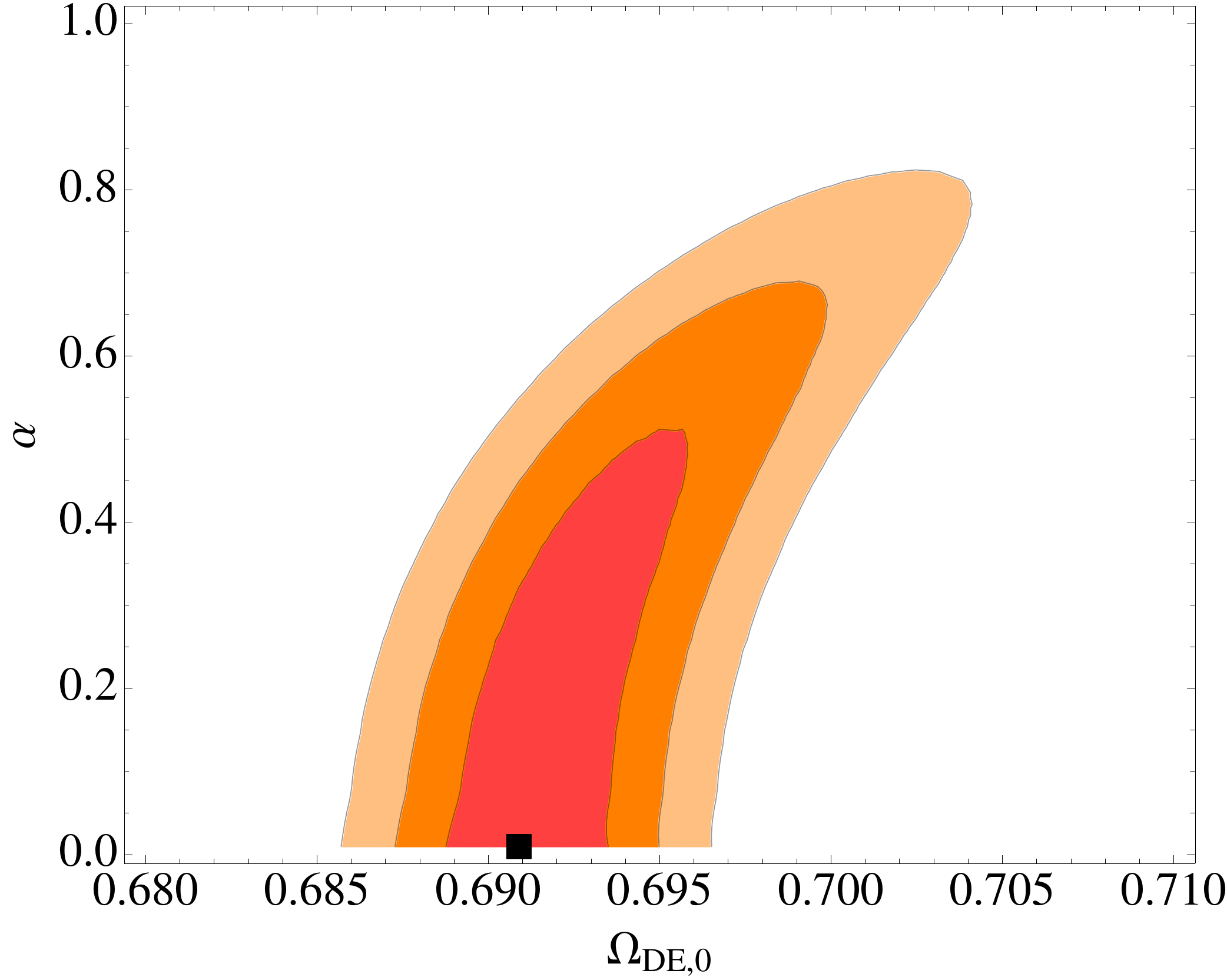}\\
 \ \\
 \ \\
 \includegraphics[width=0.45\columnwidth]{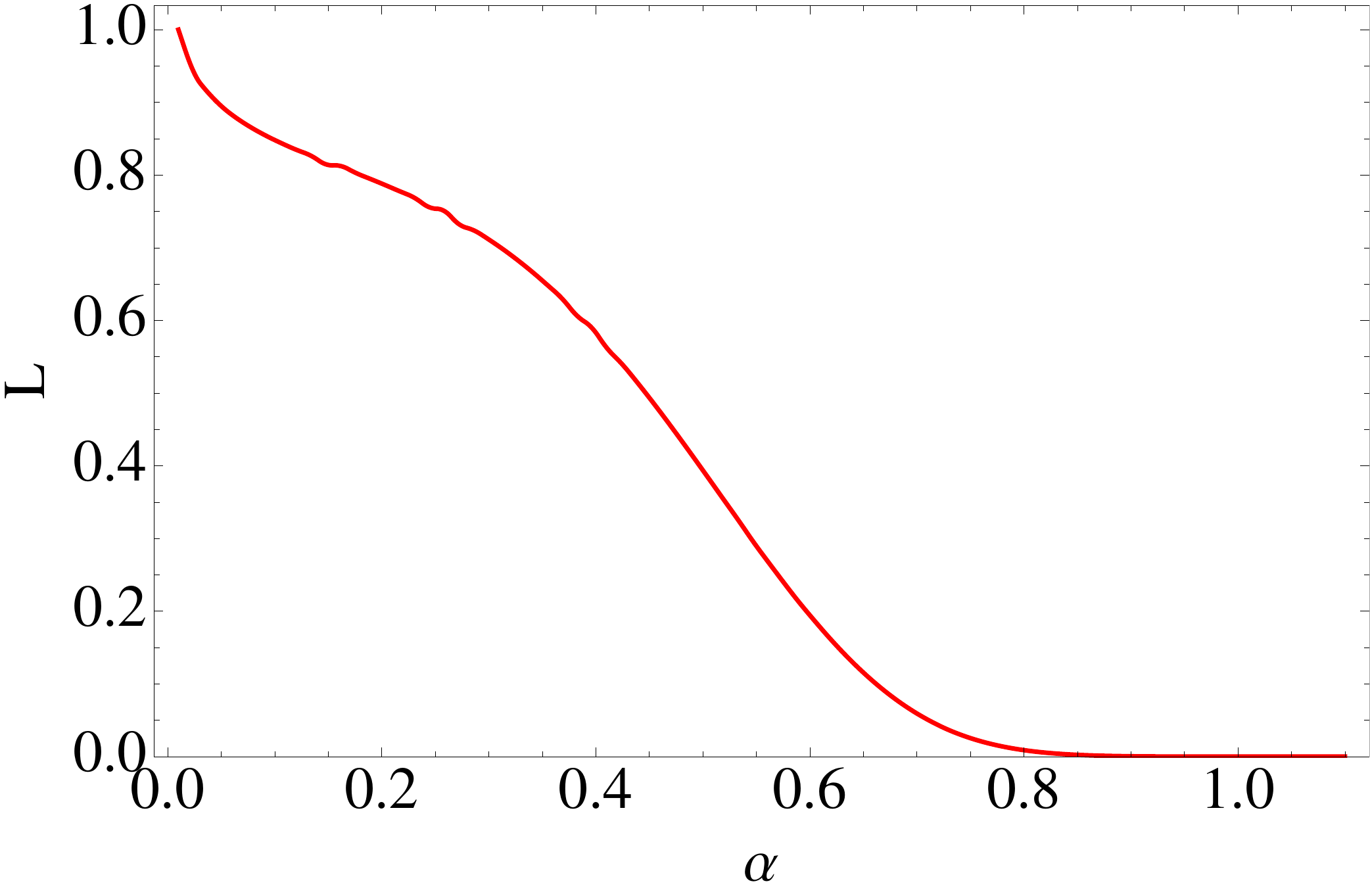}\qquad{}
\includegraphics[width=0.45\columnwidth]{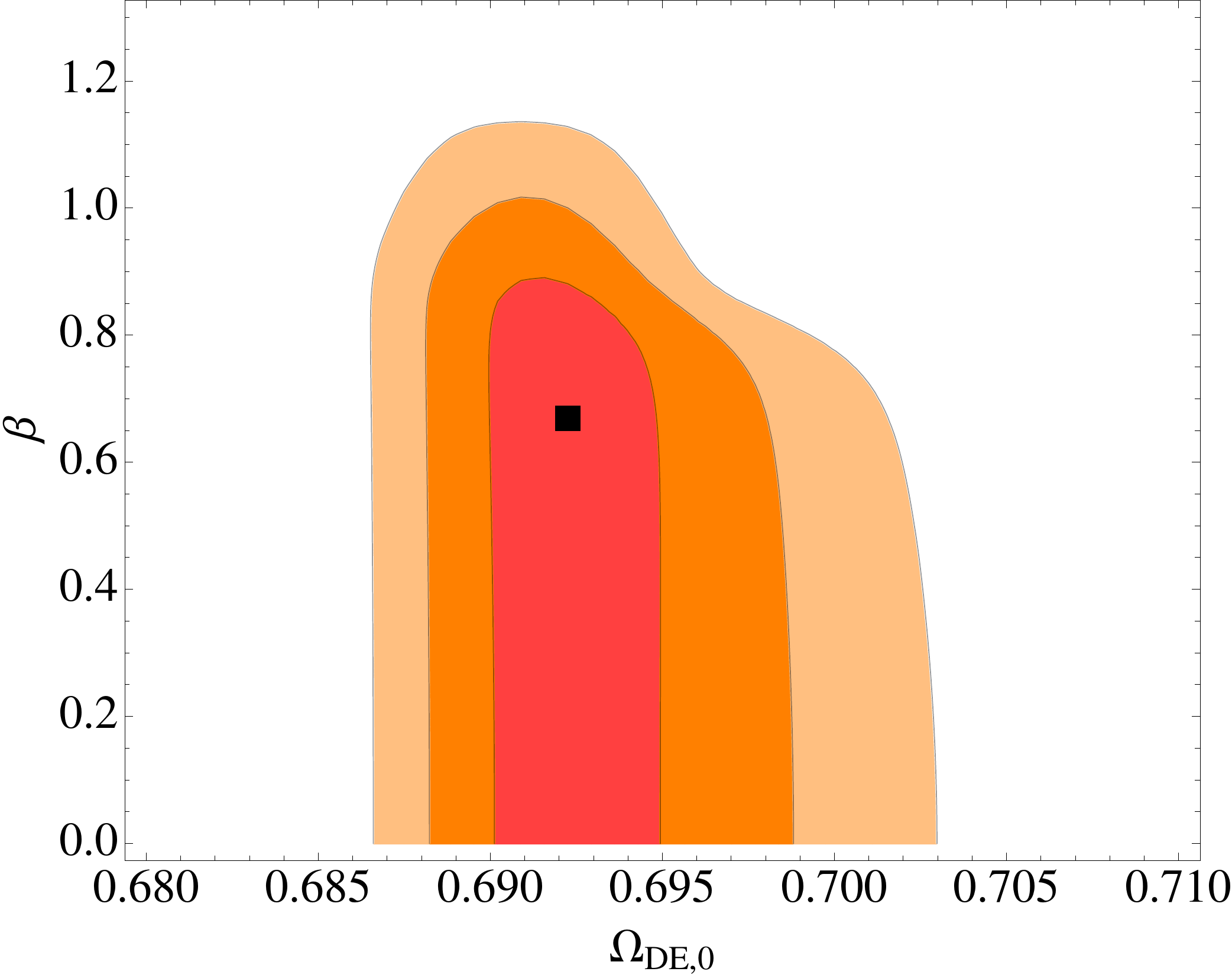}\\
 \ \\
 \ \\
 \includegraphics[width=0.45\columnwidth]{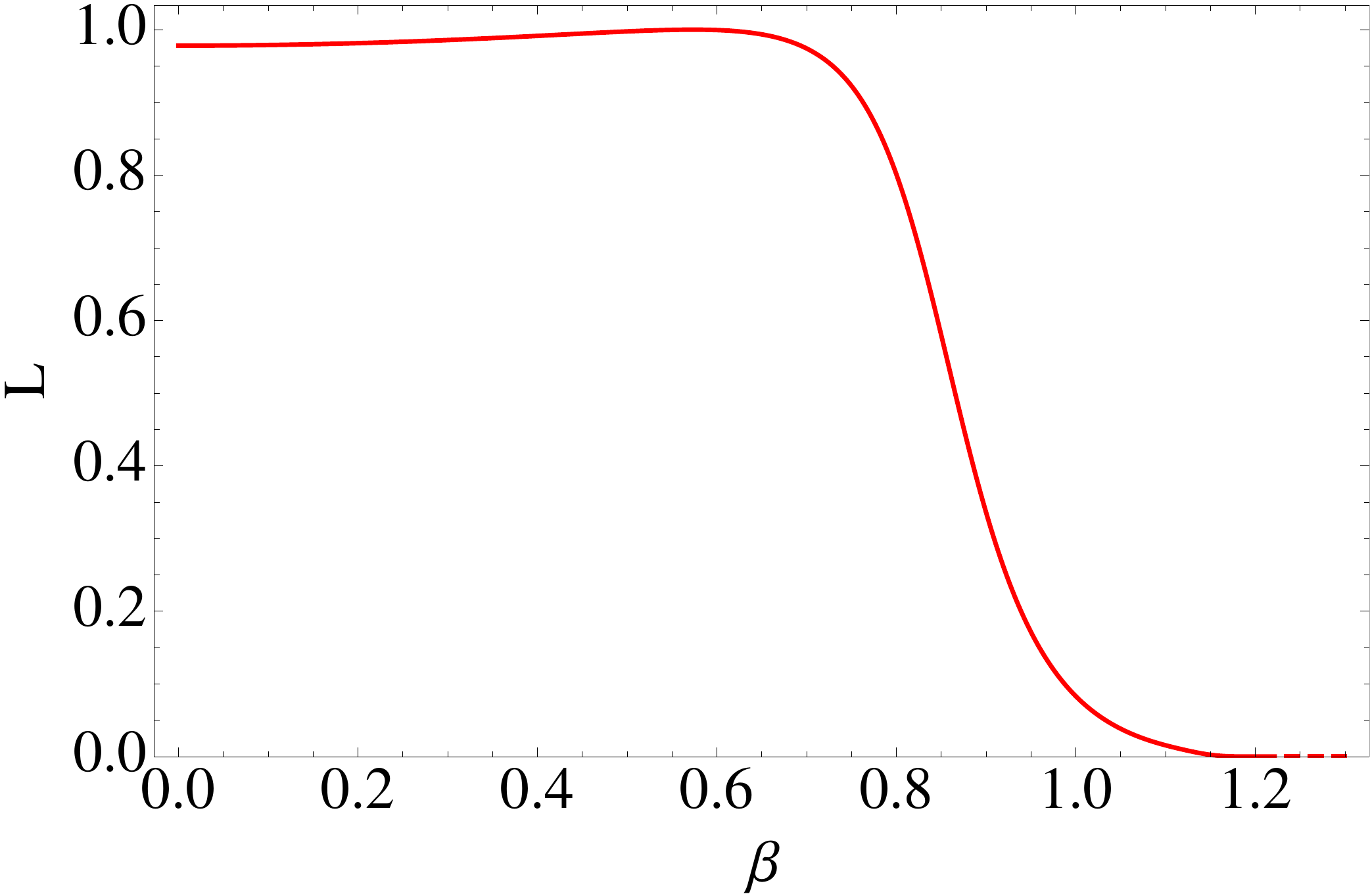}\qquad{}
\includegraphics[width=0.45\columnwidth]{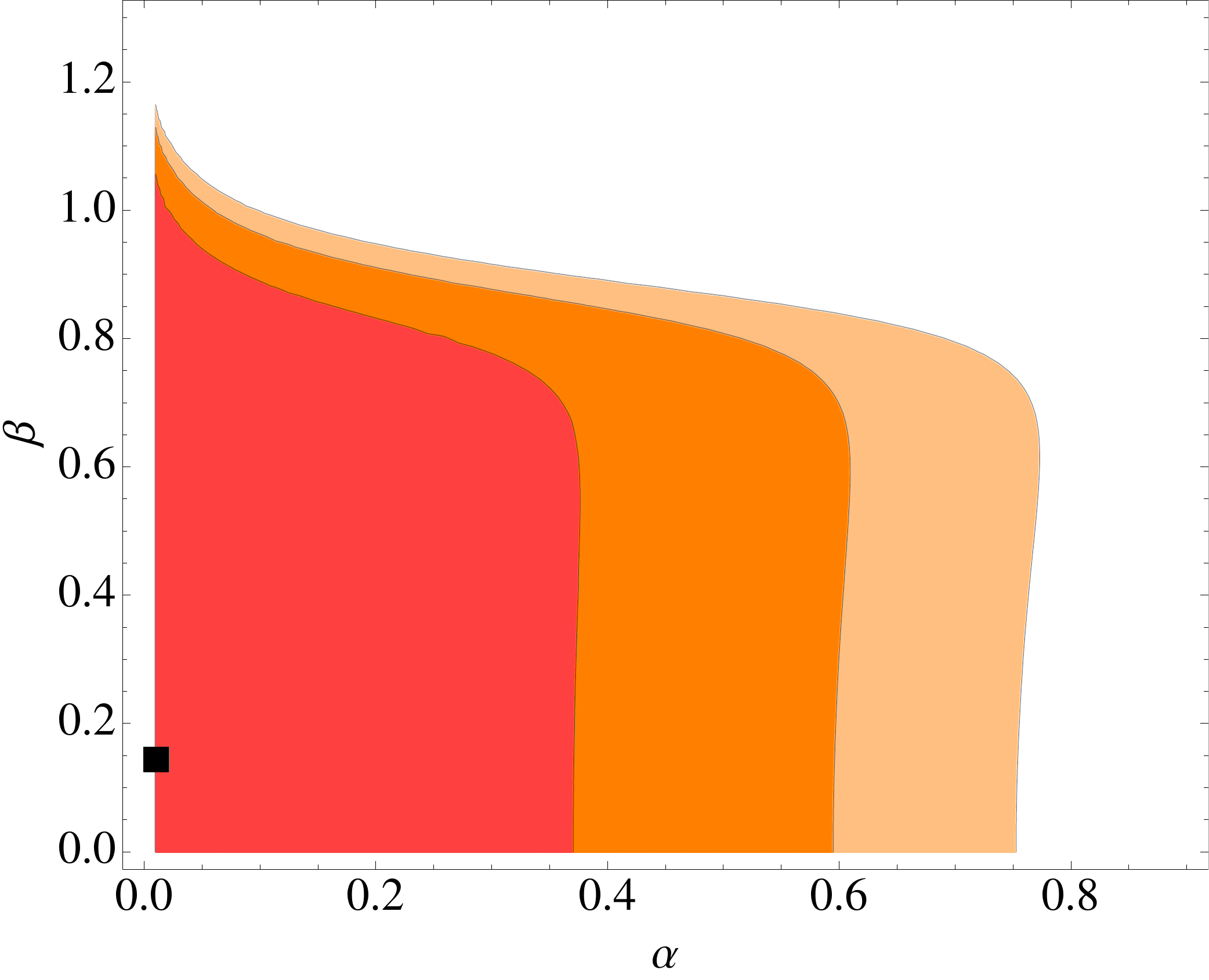}
\caption{{\em On the Left:} 1-dimensional marginalized posterior distributions
(for point 2 and point 5 together) on the parameters $\{\Omega_{{\rm DE},0},\alpha,\beta\}$
when fitting the model of this paper to the LSST 100k SN dataset (see
Section \ref{supernovae}) and the forecasted Euclid-like $f\sigma_{8}$
data (see Section \ref{fs8}). See 4$^{th}$ and 5$^{th}$ columns of Table~\ref{tab:1D} for best-fit
values with 95\% confidence intervals. {\em On the Right:} $1\sigma$,
$2\sigma$ and $3\sigma$ confidence-level contours for the relevant
2-dimensional marginalized posterior distributions. The black squares
mark the best-fit values.}
\label{posts-E}
\end{centering}
\end{figure}

The results of the combined LSST 100k supernova dataset (see Section
\ref{supernovae}) and the forecasted Euclid-like $f\sigma_{8}$ data
(see Section \ref{fs8}) are shown in Fig.~\ref{data} (right panel), Fig.~\ref{posts-E} and in the 4$^{th}$ and 5$^{th}$ columns of
Table \ref{tab:1D}.
These two forecasted catalogs are relative to a $\Lambda$CDM fiducial cosmology, using the best-fit cosmological parameters from the Planck Collaboration~\citep[][Table 5, last column]{Ade:2013zuv}.
As we can see in the plots of Fig.~\ref{posts-E}, with such kind of data
we expect
to improve constraints on all parameters with respect to present-day data. The 95\% limit on
$\alpha$
is reduced by more than a factor of 2. The 95\% constraint on $|\beta |$ is instead only marginally improved.
This shows how McDE models might be difficult to constrain even with the exquisite
quality of future Euclid-like data.
The constraints on $|\beta |$ are not strongly improved
because the deviation of the McDE growth rate with respect to the
$\Lambda$CDM one does not scale linearly with $\beta$, but it has
a step-like behavior, as displayed in Fig.~\ref{betabe}, which shows indeed that the McDE
growth rate mimics the $\Lambda$CDM model for $|\beta |\lesssim \beta _{\rm G}$,
but %strongly
departs from it for larger $|\beta |$ values. It is therefore
very difficult to constrain the coupling parameter to values smaller
than $|\beta |\sim \beta _{\rm G}$. Nonlinear clustering data may prove necessary
to further constrain McDE models. In particular, a first analysis
of the nonlinear evolution of structures within a McDE scenario has been attempted in \cite{Baldi_2013}, highlighting for the first time
very specific effects like the halo fragmentation process and a peculiar shape of the distortion of the matter power spectrum at small scales.
A more detailed investigation of such effects with higher-resolution N-body simulations is presently ongoing, and will be discussed in an upcoming paper.

%With respect to present-day data we expect
%to improve the constraints on $\alpha$ and $\Omega_{{\rm DE},0}$,
%but not on the coupling parameter $\beta$. The constraints on $\beta$
%are not improved because the deviation of the McDE growth rate with
%respect to the $\Lambda$CDM one does not scale linearly with $\beta$,
%but it has a step-like behavior. Fig.~\ref{betabe} shows indeed
%that the McDE growth rate mimics the $\Lambda$CDM model for $\beta\lesssim2$,
%but strongly departs from it for $\beta\gtrsim3$. Therefore, as present-day
%observations already constrain $\beta$ to be $\lesssim2$, little
%or no progress is achieved by using better data. This shows how difficult
%it is to rule out a model such as McDE. Nonlinear clustering data
%may prove necessary to further constrain multi-coupled dark energy
%models.

\section{Conclusions}
\label{conclusions}

The Multi-coupled Dark Energy model has been recently proposed as a simple extension of the standard coupled Quintessence scenario with the
intriguing feature of showing an effective screening of the interaction between Dark Energy and Cold Dark Matter particles, without requiring additional free parameters.
As a consequence of such screening, the background evolution of the universe closely follows the standard $\Lambda $CDM expansion history even for very large values of the coupling constant.
This effect makes the Multi-coupled Dark Energy scenario an ideal benchmark to test the discriminating power of present and future multi-probe observational surveys since it maximises the degeneracy
with the standard cosmological model in all probes that test only the background cosmic evolution.
In a previous paper \citep[][]{Piloyan:2013mla} we have quantified such degeneracy by comparing the predicted expansion history of Multi-coupled Dark Energy models with real observational data
consisting of the recent Union2.1 Compilation~\cite{Suzuki_etal_2012} of Type Ia supernovae, confirming that the background expansion history has a very low constraining power with respect to this scenario.
The present paper represents the natural extension of the analysis performed in our previous work to the linear evolution of density perturbations,
%where more prominent features are expected to arise due to
which are expected to show new physics because of
the attractive and repulsive
fifth-forces acting between Cold Dark Matter particles, a consequence of their individual coupling to the Dark Energy field.

In order to compare the predicted behavior of linear density perturbations with both presently available and future forecasted data on the growth of structures, we have first derived the full system of perturbed dynamical equations at first order
for a generic Multi-coupled Dark Energy cosmology, and analytically solved such set of equations on the few particular background solutions that we had identified as phase-space critical points of the system in our previous work.
This has allowed us to obtain the exact solution for the linear growth rate in both the past matter-dominated epoch and future Dark Energy-dominated regime.

Then, we have numerically computed the full solution of linear perturbation equations along the whole expansion history of the universe, for a wide range of model parameters, and compared the numerical solutions with the analytical ones in the appropriate regimes, finding excellent matching between the two. With our numerical solver at hand we have then performed a likelihood analysis by sampling the model's parameter space and comparing the derived
evolution with recent observational data of the growth rate, including
data sets from 6dFGRS, LRG, BOSS, WiggleZ and VIPERS, as well as with future data consistent
with the forecasted accuracy of the Euclid satellite. Our analysis
has shown that -- as expected -- the growth of density perturbations can strongly constrain
Multi-coupled Dark Energy scenarios, putting tight bounds on the coupling constant which is constrained to
$|\beta | \lesssim 0.88$ and $|\beta |\lesssim 0.85$ at 95\% confidence level when considering present and future data sets, respectively.

Interestingly, we have also found that the evolution of linear density perturbations encoded by the growth rate shows a sharp deviation from the standard $\Lambda $CDM evolution when the coupling constant $|\beta |$ approaches and overcomes
a critical value $\beta _{\rm G}=\sqrt{3}/2$, corresponding to the coupling value that determines fifth-forces with the same strength as standard gravitational interactions. Nonetheless, as our 95\% confidence levels on the coupling directly show,
presently available data at the linear level are not yet capable of excluding a coupling value of $|\beta | = \beta _{\rm G}$, and therefore cannot rule out scalar interactions of gravitational strength in the context of a Multi-coupled Dark Energy framework.
The natural further extension of this analysis is then to investigate the effects of the Multi-coupled Dark Energy scenario in the nonlinear regime of structure formation by means of dedicated high-resolution N-body simulations, in order
to highlight possible characteristic footprints of the model that might allow to further tighten its viable parameter space. Such task is ongoing and will be discussed in an upcoming dedicated paper.

%\clearpage
\section*{Acknowledgements}
L.A. and V.M. acknowledge financial support from DFG through the project TRR33 "The Dark Universe".
M.B. is supported by the  Marie Curie Intra European Fellowship
``SIDUN"  within the 7th Framework  Programme of the European Commission. A.P. thanks DAAD for support.

\appendix

\section{Including a baryonic component}
\label{appendix}

\subsection{Equations}
\label{b-equations}

\begin{figure}
\begin{centering}
\includegraphics[width=0.4\columnwidth]{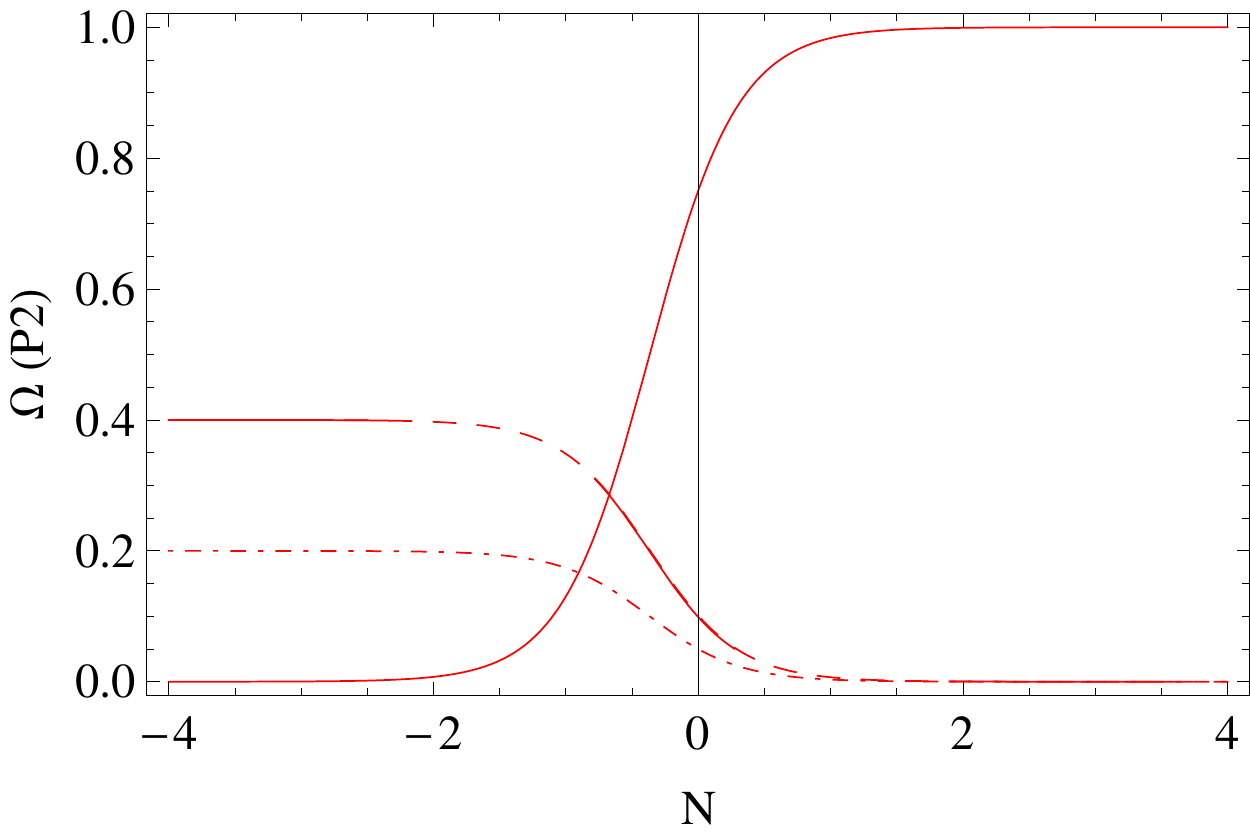}
\quad
\includegraphics[width=0.4\columnwidth]{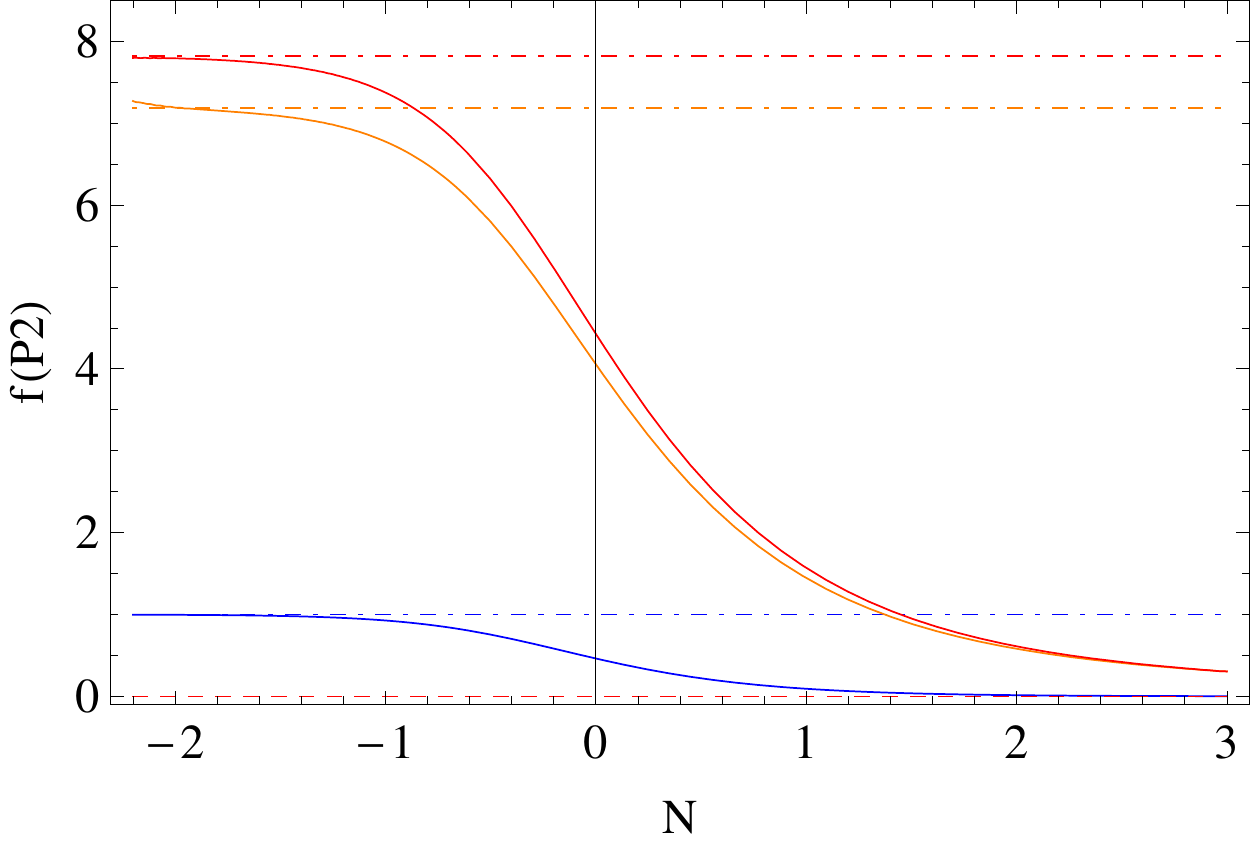}
\caption{ {\em Left panel:} The fractional densities $\Omega_{i}$ for $\alpha=0.1$ and $\beta=0.5$. Solid curves correspond
to $\Omega_{DE}$,dot-dashed curves correspond to $\Omega_{b}$, dashed
curves correspond to $\Omega_{-}$ and $\Omega_{+}$. Parameters are
chosen in the stable range of point 2. {\em Right panel:} The evolution of $f$ for $|\beta | = 0.5$ (blue), $3.5$ (orange), and $4$ (red). The dashed horizontal lines
are the analytical  values for the matter era and the accelerated point 2.}
\label{p_2_om-bar}
\end{centering}
\end{figure}

We will now consider the case in which, besides dark matter, there is also a baryonic component
which does \emph{not} couple to dark energy.
We will fix the baryonic content of the universe to the best-fit value $\Omega_{b,0}=0.048$ from the
 Planck Collaboration~\citep[][Table 5, last column]{Ade:2013zuv}.
The background equations will be modified as follows:
\begin{align}
2\frac{H'}{H} & =-(3-3y^{2}+3x^{2}+r^{2})\,,\label{eq:Frid_EQ-H-prime}\\
x' & =-x\frac{H'}{H}-3x+\alpha y^{2}+\beta(z_{2}^{2}-z_{1}^{2})\,,\label{eq:Frid_EQ-1}\\
y' & =-y\frac{H'}{H}-\alpha xy\,,\label{eq:Frid_EQ-2}\\
z_{+}' & =-z_{+}\frac{H'}{H}-\frac{3}{2}z_{+}+\beta xz_{+}\,,\label{eq:Frid_EQ-3}\\
z_{-}' & =-z_{-}\frac{H'}{H}-\frac{3}{2}z_{-}-\beta xz_{-}\,,\label{eq:Frid_EQ-4}\\
r' & =-r\frac{H'}{H}-2r\,,\label{eq:Frid_EQ-5} \\
z_{b}' &=-z_{b}\frac{H'}{H}-\frac{3}{2}z_{b}\,,\label{eq:barz}
\end{align}
where
\begin{equation}
z_{b}^{2}\equiv \frac{\rho_{b}}{3M_{{\rm Pl}}^{2}H^{2}} \,.
\end{equation}
The additional critical points, present when $\Omega_{b}\not=0$, are listed in Table
\ref{tbl_crt-bar}.
The linear perturbation equations, including baryons,  read now:
\begin{align}
\delta_{-}^{''}+(2(1+\beta x)-\frac{1}{2}(3-3y^{2}+3x^{2}+r^{2}))\delta'_{-} & =\frac{3}{2}(z_{-}^{2}\Gamma_{A}\delta_{-}+z_{+}^{2}\Gamma_{R}\delta_{+}+z_{b}^{2}\delta_{b})\,,\label{eq:pert_1-2-1-1-1}\\
\delta_{+}^{''}+(2(1-\beta x)-\frac{1}{2}(3-3y^{2}+3x^{2}+r^{2}))\delta'_{+} & =\frac{3}{2}(z_{-}^{2}\Gamma_{R}\delta_{-}+z_{+}^{2}\delta_{+}\Gamma_{A}+z_{b}^{2}\delta_{b})\,,\label{eq:pert_d} \\
\delta_{b}^{''}+(2-\frac{1}{2}(3-3y^{2}+3x^{2}+r^{2}))\delta'_{b}&=\frac{3}{2}(z_{-}^{2}\delta_{-}
+z_{+}^{2}\delta_{+}+z_{b}^{2}\delta_{b})\,.\label{eq:delta_bar}
\end{align}
The analytical solutions for the new background critical points are listed in Table~\ref{tbl_crt-f}.
 We plot the fractional densities for the new point 2 in the left panel of Fig.\ref{p_2_om-bar}, while in the right panel we display the numerical solutions for $f$ (solid curves) and its analytical predictions in matter domination (dot-dashed) and DE domination (dashed).
The matter point 3 is modified by the additional
non-zero $\Omega_{b}$. For instance, in the matter era the dot-dashed curve
 shows $\Omega_{b}=0.2$ and $\Omega_{\pm}=0.4$ (dashed curve) instead of $\Omega_{\pm}=0.5$. The growing solution found for the new critical point 6 in
the matter era provided in Table~\ref{tbl_crt-bar} is
$f=1 \,{\rm or}\, \frac{1}{4}\left(-1+\sqrt{1+64z_{cr}^{2}\beta^{2}}\right)+3$
where $z_{cr}$,  defined as $z_{cr}=\sqrt{\frac{1-z_{bI}^2}{2}}$, corresponds to the fractional density of each CDM species in the matter era, with $z_{bI}$ the initial baryonic fractional density. Note that the additional 3 in the expression $\frac{1}{4}\left(-1+\sqrt{1+64z_{cr}^{2}\beta^{2}}\right)+3$ appears for the same reason as explained in Section \ref{an_sol}.
This growing rate, in the previous case with no baryonic component (corresponding to $z_{cr}=\frac{\sqrt{2}}{2}$) goes back to the old solution (i.e.~the present point 6 in Table~\ref{tbl_crt-bar} goes back to the previous point 3 in Table~\ref{tbl_crt-2} in the appropriate limit).
One can
also see the differences between the numerical solutions with and without baryons by comparing Fig.~\ref{p_2_om-bar} with Fig.~\ref{p_2f_m}.

\begin{table}[t]
\begin{centering}
{
\renewcommand{\arraystretch}{1.8}
\begin{tabular}{|c|c|c|c|c|c|c|c|c|}
\hline
 Point& $x$  & $y$  & $z_{+}$  & $z_{-}$  &  $z_{b}$  &\,$\Omega_{{\rm DE}}$\,  & $w_{eff}$  & $\mu$\tabularnewline
\hline
\hline
6  & 0  & 0  & $\sqrt{\frac{1-z_{b}^{2}}{2}}$ & $\sqrt{\frac{1-z_{b}^{2}}{2}}$  &  $z_{b}$  & 0 & 0 & 0 \tabularnewline
\hline
7 & $\frac{3}{2\alpha}$  & $\frac{3}{2\alpha}$  & 0  & 0  & $\sqrt{1-\frac{9}{2\alpha^{2}}}$  & \,\,$\frac{9}{2\alpha^{2}}$\,\,  & 0 &0\tabularnewline
\hline
\end{tabular}
}
\caption{Additional critical points for background equations (\ref{eq:Frid_EQ-H-prime})-(\ref{eq:barz})
when  uncoupled baryons are included.
Only the physical solutions for $x,y,z_{+},z_{-},z_{b}$ are selected.}
\label{tbl_crt-bar}
\end{centering}
\end{table}

\begin{table}[t]
\begin{centering}
{
\renewcommand{\arraystretch}{1.8}
\begin{tabular}{|c|c|c|c|c|c|}
\hline
 Point & $f_{+}$ &$f_{-}$& $f$  & $f_{b}$\tabularnewline
\hline
\hline
6  & $f$  & $f$  &$\frac{1}{4}\left(-1+\sqrt{1+64z_{cr}^{2}\beta^{2}}\right)+3$ & 1 \tabularnewline
\hline
7 & -- &--  & $\frac{1}{4}\left(-1+\sqrt{25+\frac{108}{\alpha^{2}}}\right)$ & $\frac{1}{4}\left(-1+\sqrt{25+\frac{108}{\alpha^{2}}}\right)$\tabularnewline
\hline
\end{tabular}
}
\caption{Growth functions for additional critical points presented in Table \ref{tbl_crt-bar}.}
\label{tbl_crt-f}
\end{centering}
\end{table}

\subsection{Results}
\label{b-results}

\begin{table}
\begin{centering}
{
\renewcommand{\arraystretch}{1.8}
\begin{tabular}{|lcccc|cc|}
\hline
Parameter  &
{\begin{minipage}{45pt}
\centering
Best Fit\\
{\small (SN+$f\sigma_{8}$)}
\end{minipage}}  & 95\% c.l.&  {\begin{minipage}{45pt} \centering Best Fit\\
{\small (LSST+Euclid)} \end{minipage}}  & 95\% c.l.&  {\begin{minipage}{45pt} \centering Best Fit\\
{\small (SN)} \end{minipage}}  & 95\% c.l. \tabularnewline
\hline
\hline
$\Omega_{{\rm DE},0}$  & 0.734  & $[0.684,0.826]$   & 0.692  & $[0.688,0.698]$ & 0.719 & $[0.680,0.765]$\tabularnewline
$\alpha$  & 0.66  & $[0,1.37]$   & 0.12  & $[0,0.54]$ & 0.62 & $[0,1.01]$ \tabularnewline
$\beta$  & 0.88  & $[0,0.98]$   & 0.03  & $[0.,0.93]$ & 6.4 & $[0,83]$ \tabularnewline
\hline
%$\mu_{{\rm in}}$  & unconstrained  & unconstrained  \\
%$\delta'_{{\rm in}}$  & unconstrained  & unconstrained  \\
 %&  & \tabularnewline
\end{tabular}
}
\caption{ As Table~\ref{tab:1D} but for the case where the uncoupled baryonic fraction is also included.}
%Best-fit values with 95\% confidence intervals for the parameters
%of the model discussed in this paper with $\Omega_{b,0}=0.048$ when using the combined Union2.1
%supernova dataset and the latest $f\sigma_{8}$ data. We employed
%a flat prior on the parameters. See Fig.~\ref{posts-b} for a plot
%of the marginalized posterior distributions. 3 and 4 columns are best-fit values with 95\% confidence intervals for the parameters
%of the model discussed in this paper with $\Omega_{b,0}=0.048$ when using the combined LSST
%100k supernova dataset and the forecasted Euclid-like $f\sigma_{8}$
%data. We employed a flat prior on the parameters. See Fig.~\ref{posts-E-b}
%for a plot of the marginalized posterior distributions. See Section \ref{b-results} for more details, and (4 and 5 columns) Table~\ref{tab:1D} (relative to the case of $\Omega_{b,0}=0$) for comparison.}
\label{tab:1D-b}
\end{centering}
\end{table}

The results of the combined Union2.1 supernova dataset (see Section
\ref{supernovae}) and the latest $f\sigma_{8}$ data (see Section
\ref{fs8}) for the case of $\Omega_{b,0}=0.048$ are shown in Fig~\ref{posts-b} and in Table \ref{tab:1D-b} (2$^{nd}$ and 3$^{rd}$ columns).
The results of the combined LSST 100k supernova dataset (see Section
\ref{supernovae}) and the forecasted Euclid-like $f\sigma_{8}$ data
(see Section \ref{fs8}) for the case of $\Omega_{b,0}=0.048$ are shown in Fig.~\ref{posts-E-b} and in
Table \ref{tab:1D-b} (4$^{th}$ and 5$^{th}$ columns). The last two columns of Table~\ref{tab:1D-b} report the constraints from \cite{Piloyan:2013mla} obtained by using only background data.
In both cases, the constraints on the model parameters $\Omega_{{\rm DE},0},\alpha$ are basically
unchanged, while the constraints on the coupling $\beta$ are slightly weakened, but without substantial modification.
This is expected since in the limit where all matter is composed of uncoupled baryons, the value of $|\beta |$ becomes
obviously irrelevant.

\begin{figure}
\begin{centering}
\includegraphics[width=0.45\columnwidth]{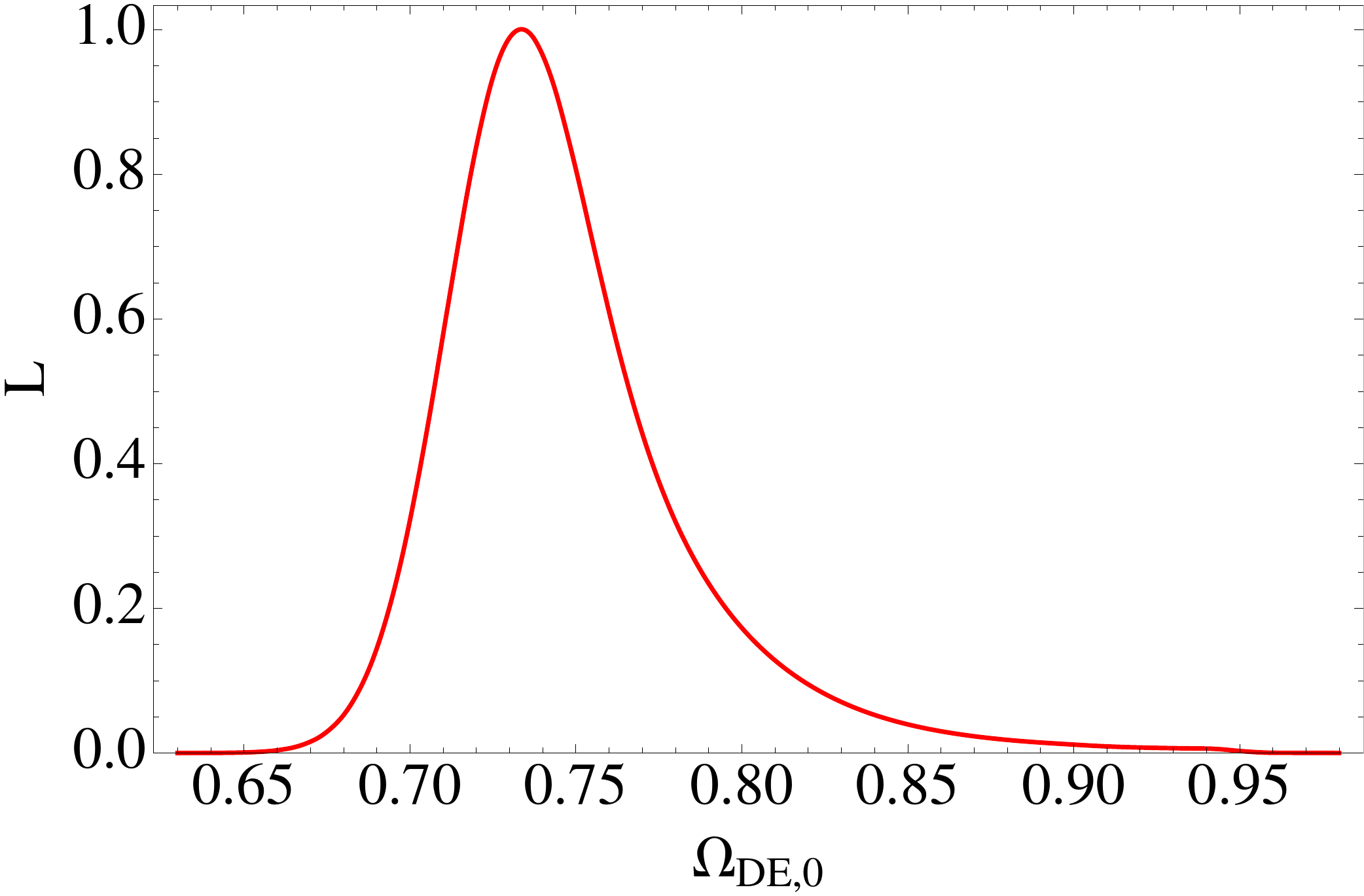}\qquad{}
\includegraphics[width=0.45\columnwidth]{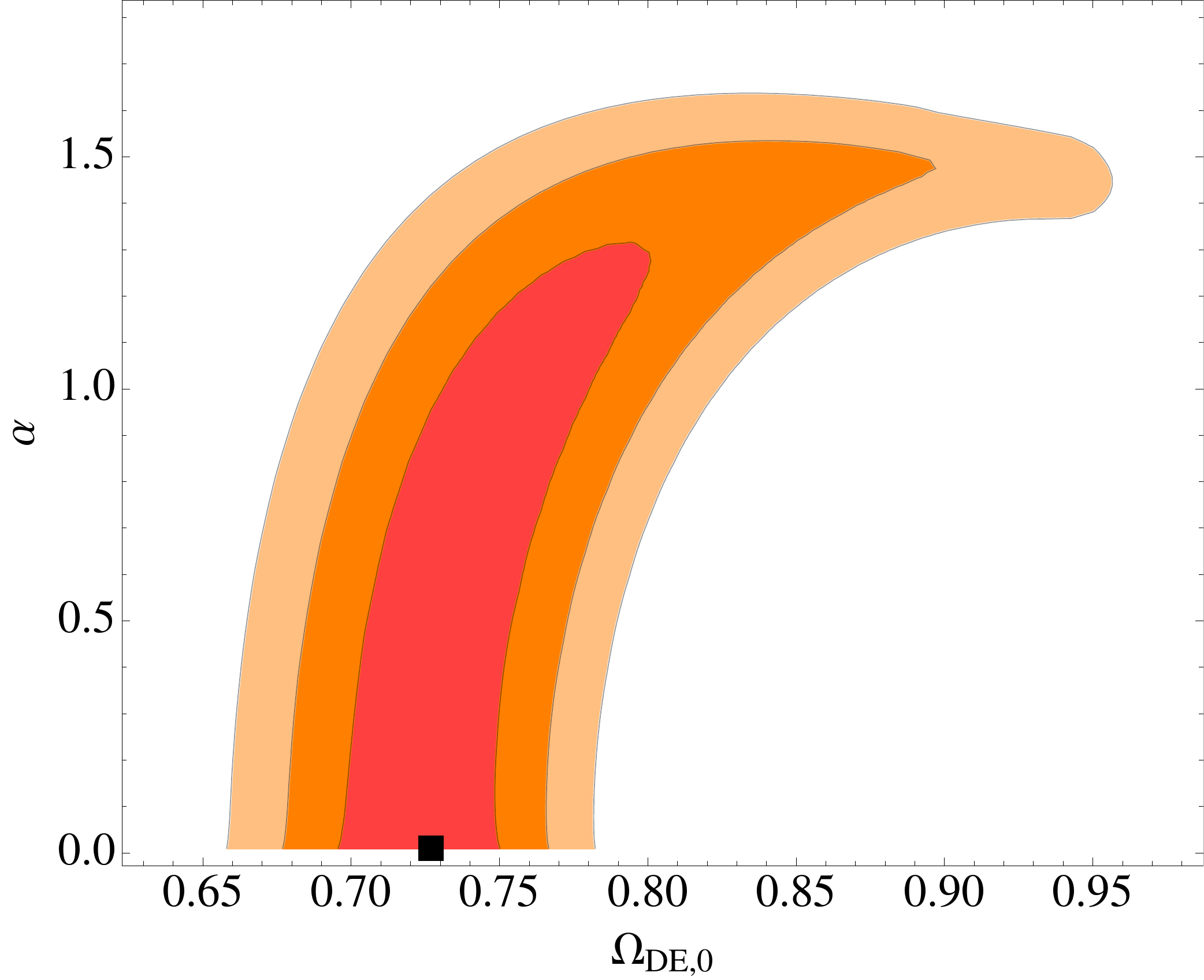}\\
 \ \\
 \ \\
 \includegraphics[width=0.45\columnwidth]{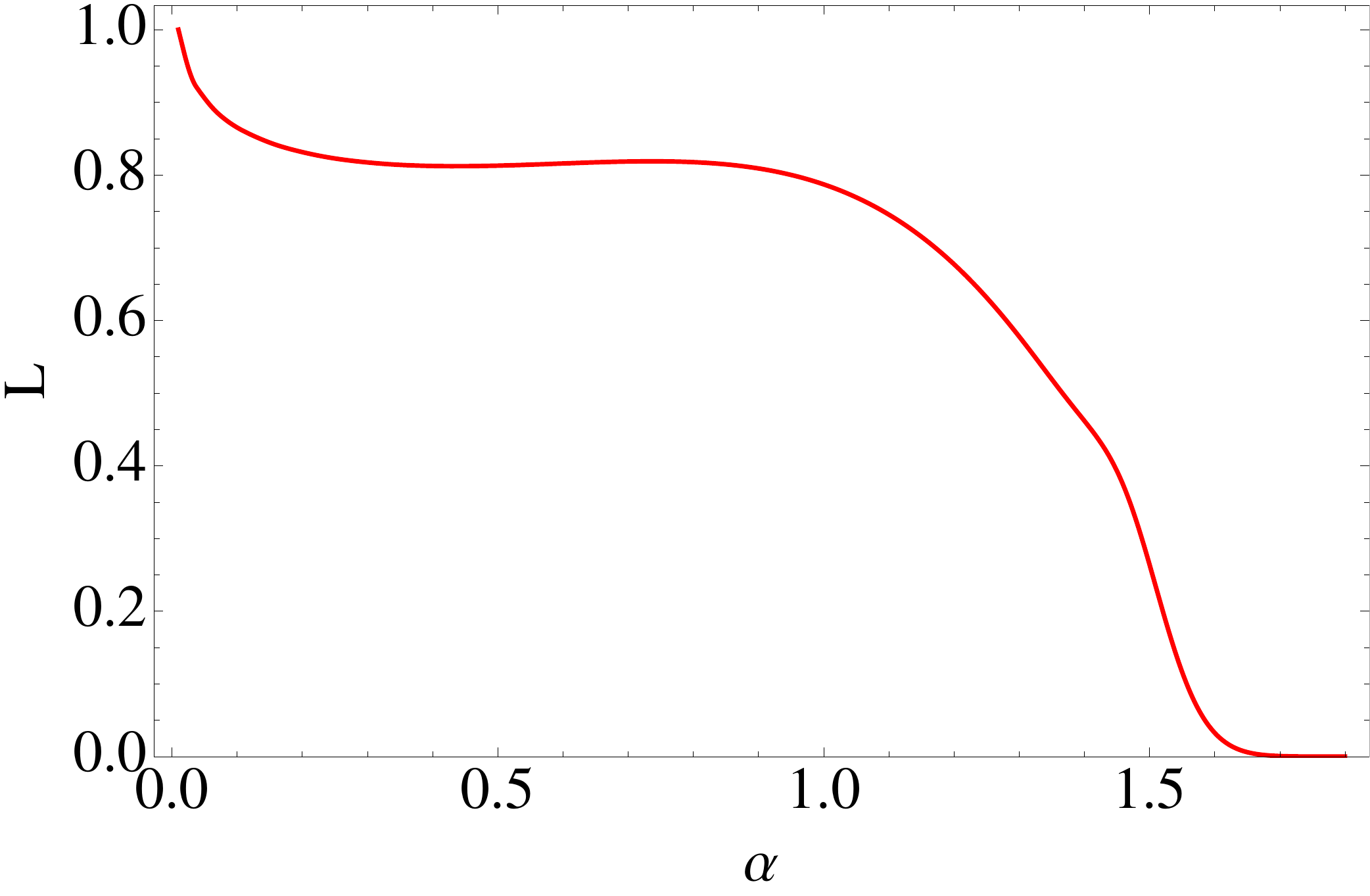}\qquad{}
\includegraphics[width=0.45\columnwidth]{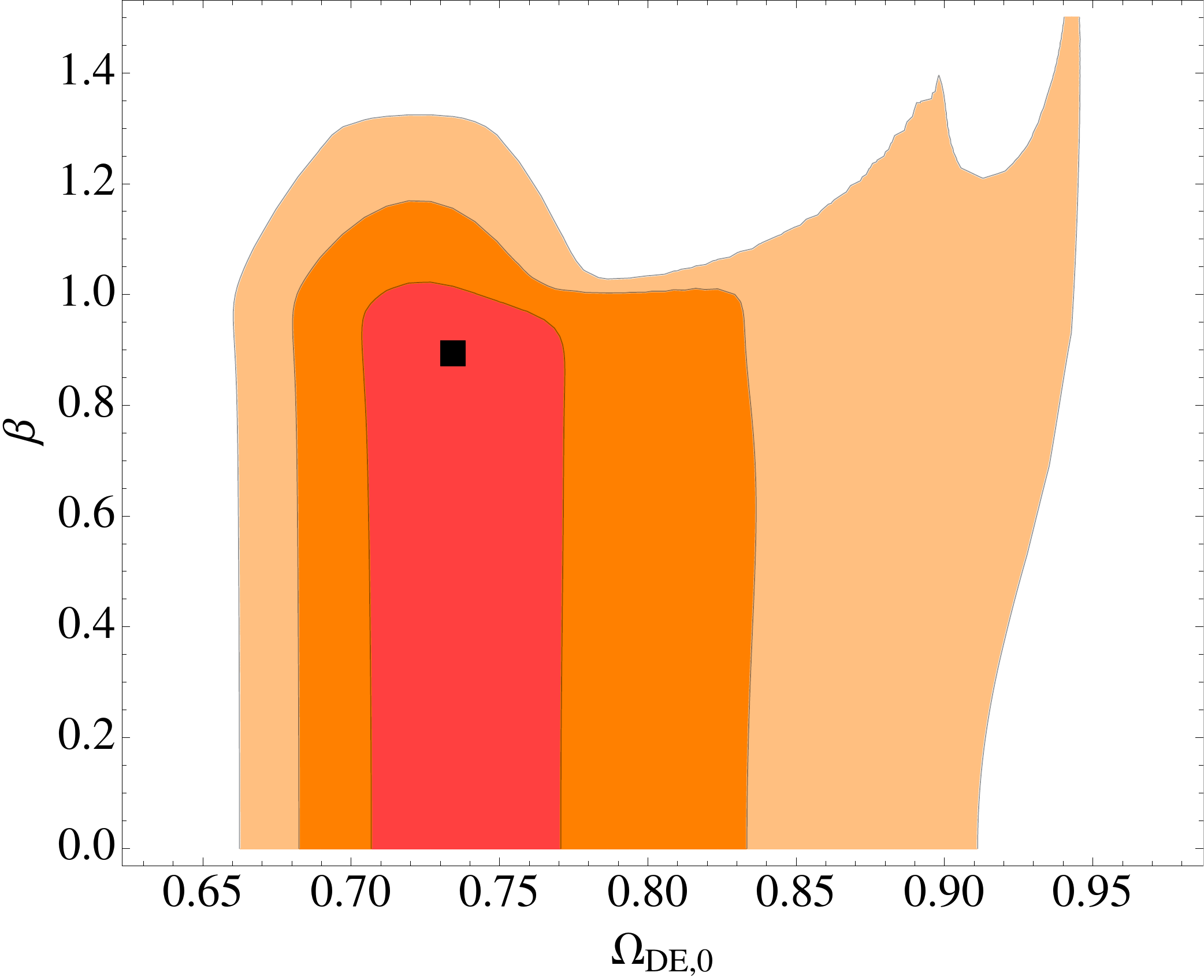}\\
 \ \\
 \ \\
 \includegraphics[width=0.45\columnwidth]{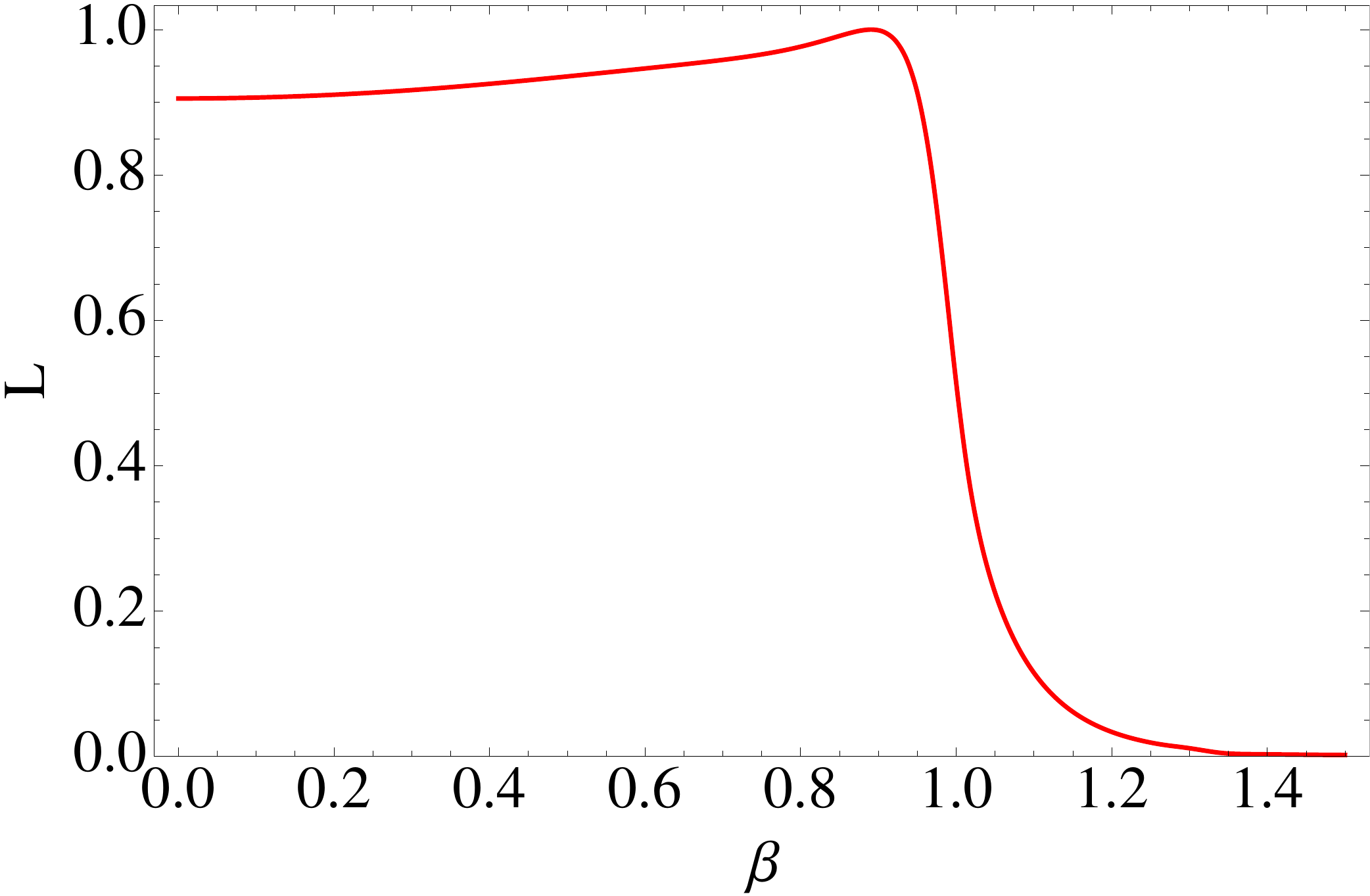}\qquad{}
\includegraphics[width=0.45\columnwidth]{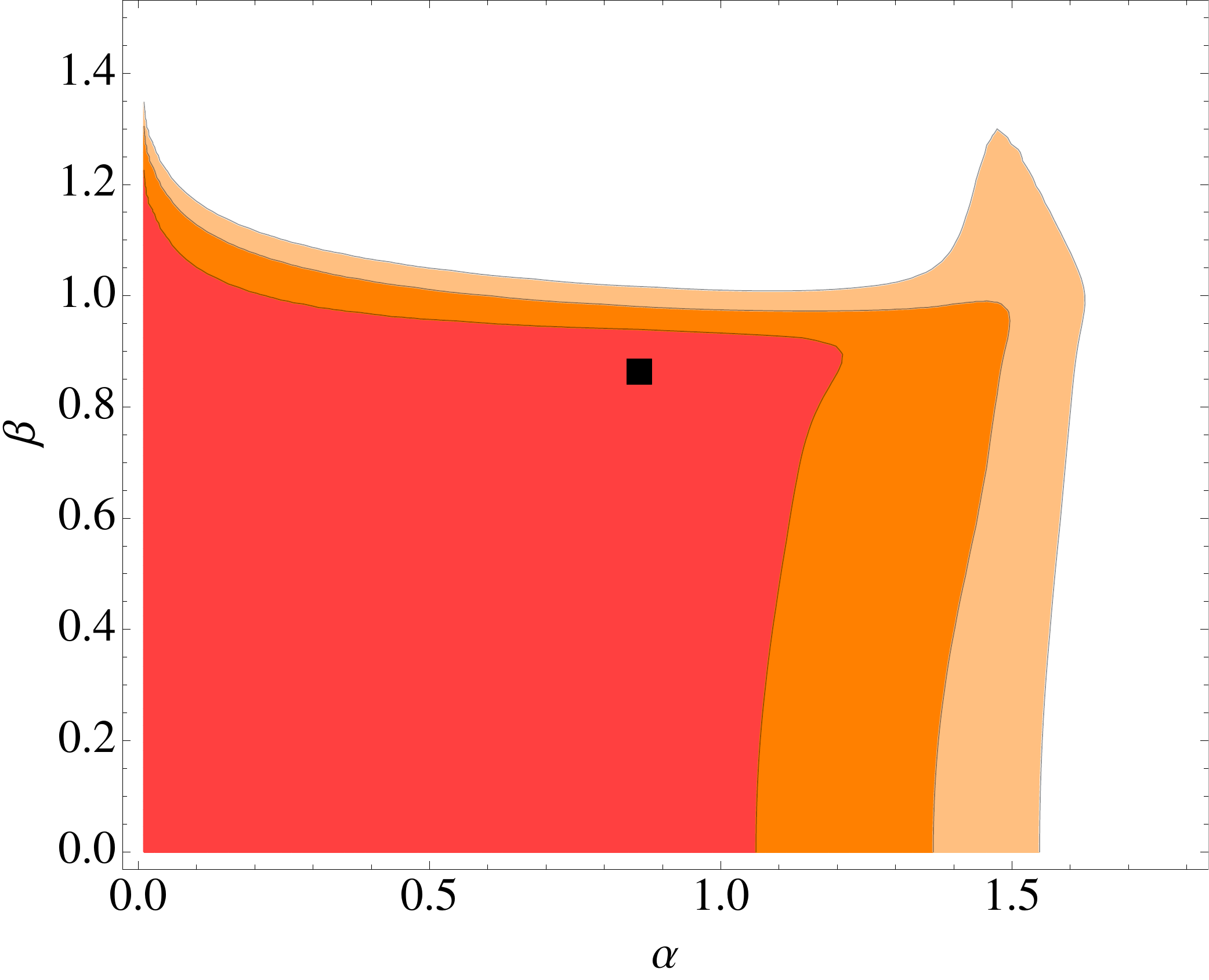}
\caption{{\em On the Left:} 1-dimensional marginalized posterior distributions
(for point 2 and point 5 together) on the parameters $\{\Omega_{{\rm DE},0},\alpha,\beta\}$
when fitting the model of this paper with $\Omega_{b,0}=0.048$ to the Union2.1 SN Compilation
(see Section \ref{supernovae}) and the latest $f\sigma_{8}$ data
(see Section \ref{fs8}). See Table~\ref{tab:1D-b} (2$^{nd}$ and 3$^{rd}$ columns) for best-fit values
with 95\% confidence intervals. {\em On the Right:} $1\sigma$,
$2\sigma$ and $3\sigma$ confidence-level contours for the relevant
2-dimensional marginalized posterior distributions. The black squares
mark the best-fit values. This plot should be compared to Fig.~\ref{posts} where the baryonic content has been neglected.}
\label{posts-b}
\end{centering}
\end{figure}

\begin{figure}
\begin{centering}
\includegraphics[width=0.45\columnwidth]{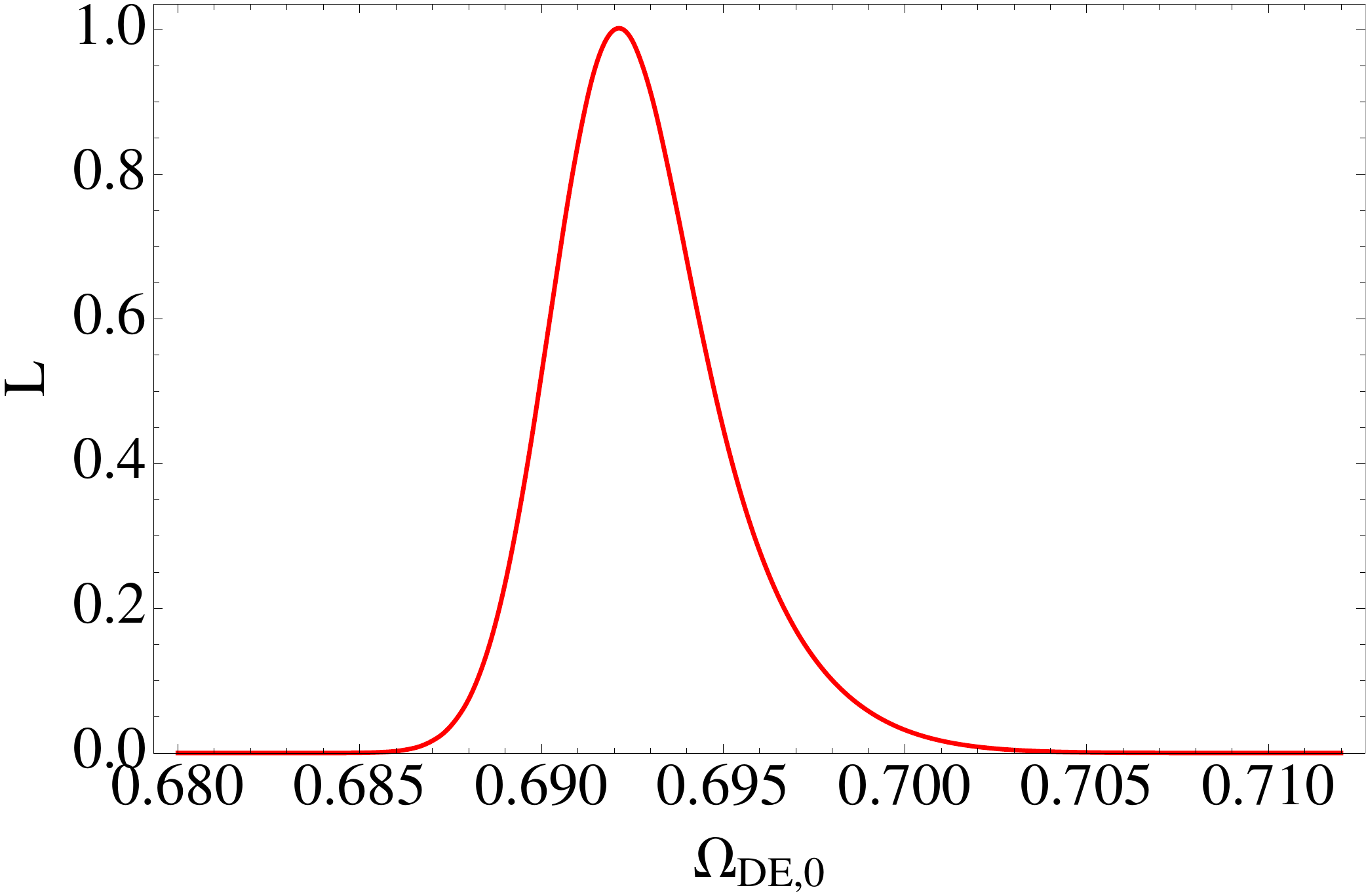}\qquad{}
\includegraphics[width=0.45\columnwidth]{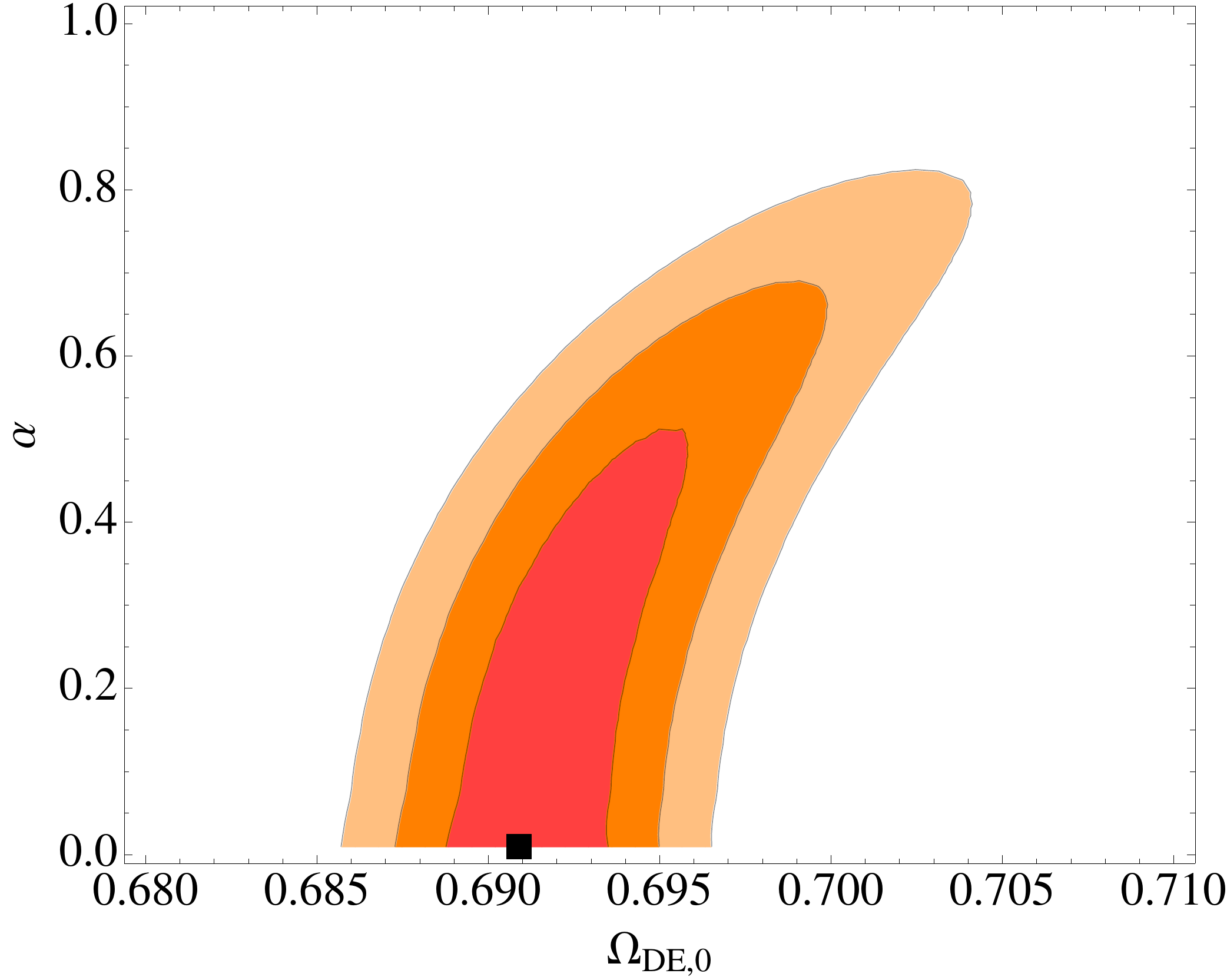}\\
 \ \\
 \ \\
 \includegraphics[width=0.45\columnwidth]{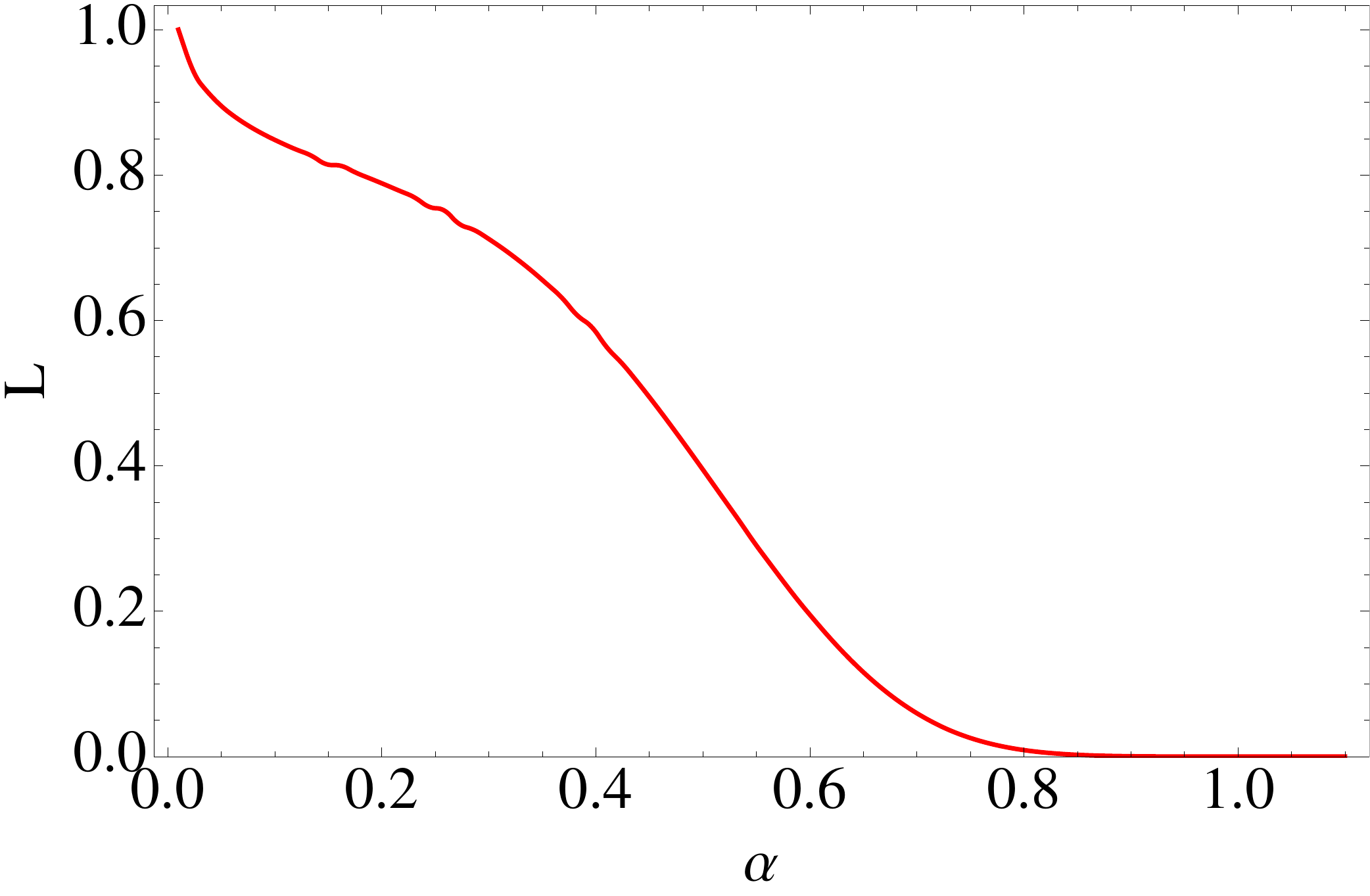}\qquad{}
\includegraphics[width=0.45\columnwidth]{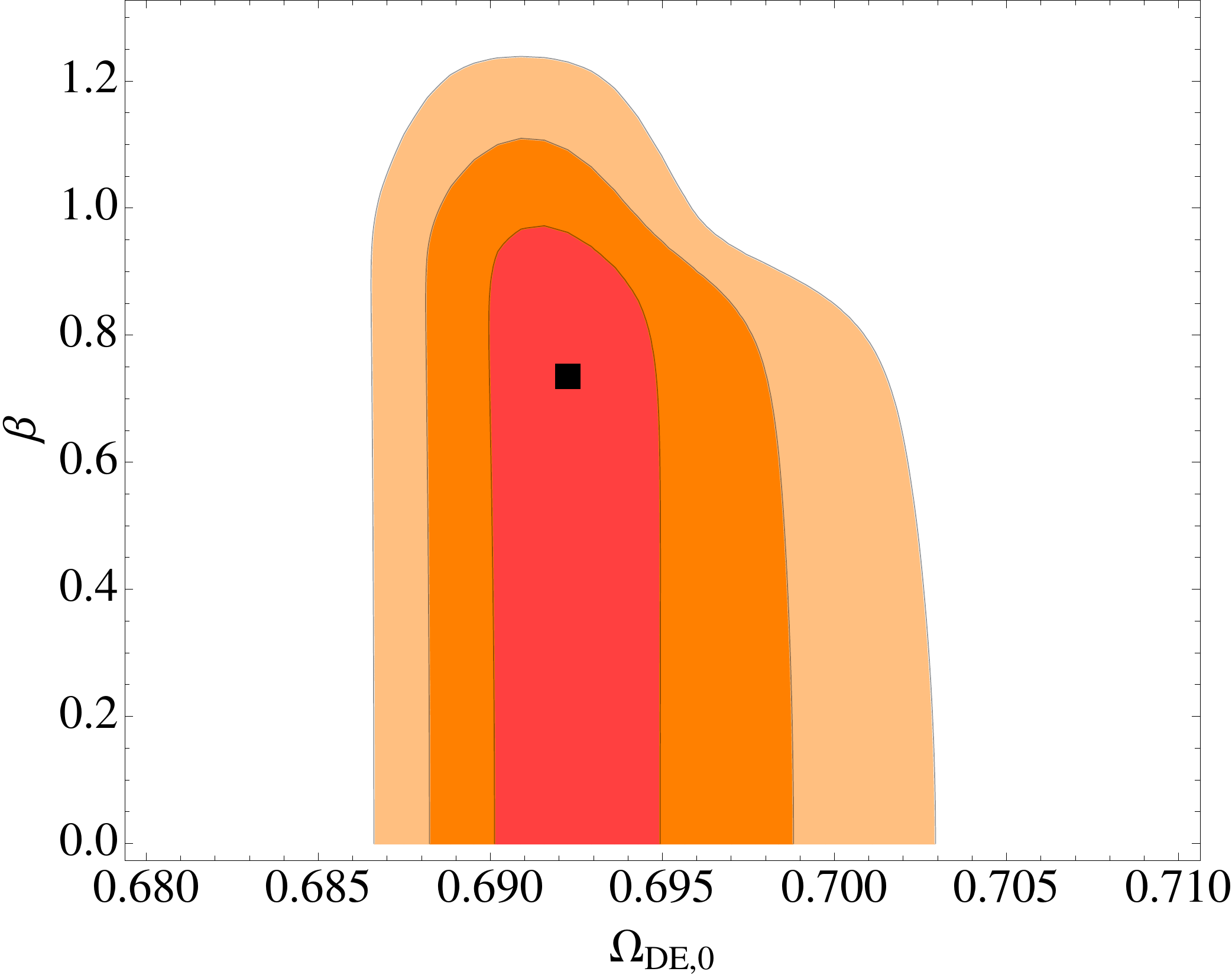}\\
 \ \\
 \ \\
 \includegraphics[width=0.45\columnwidth]{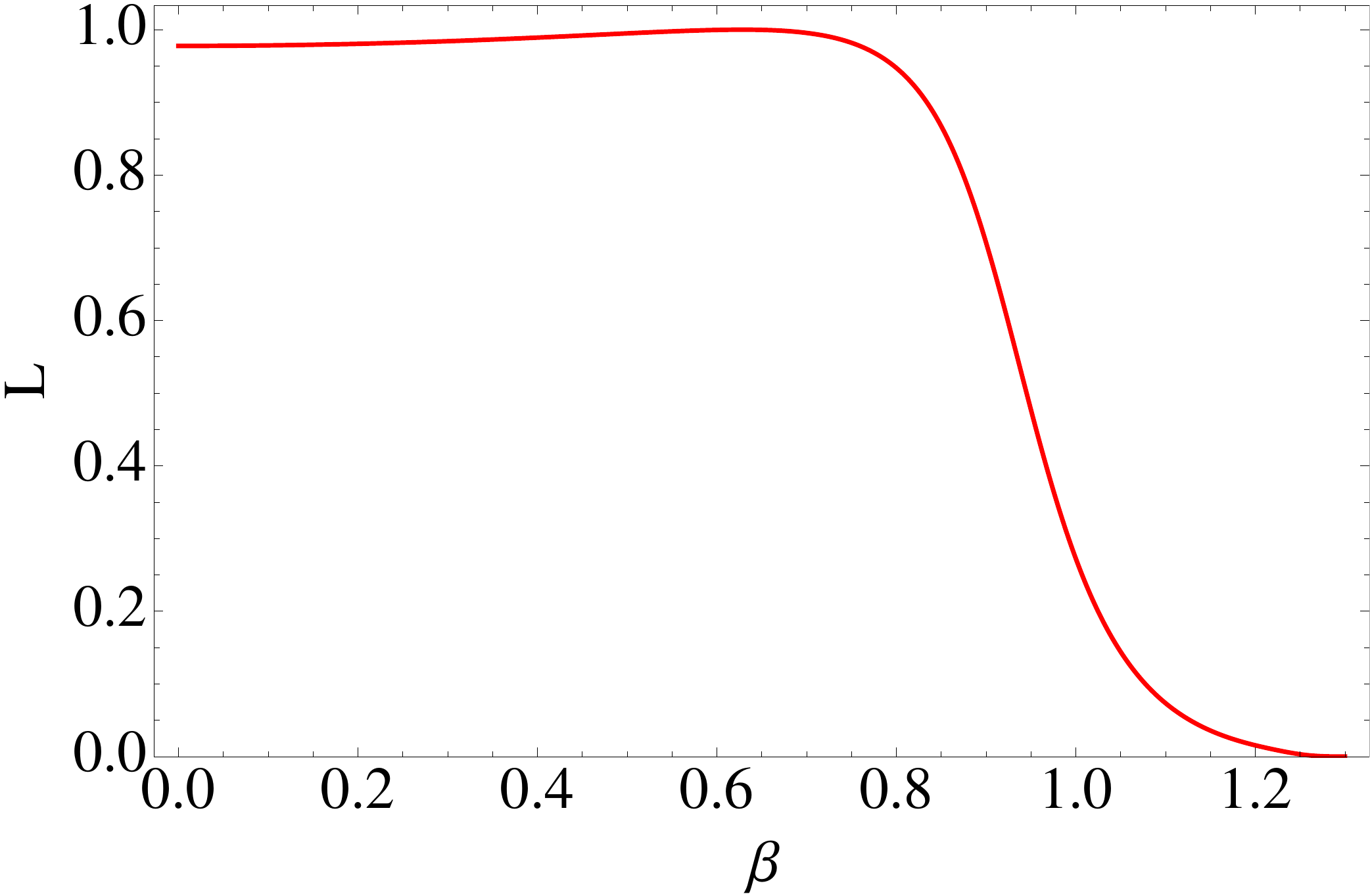}\qquad{}
\includegraphics[width=0.45\columnwidth]{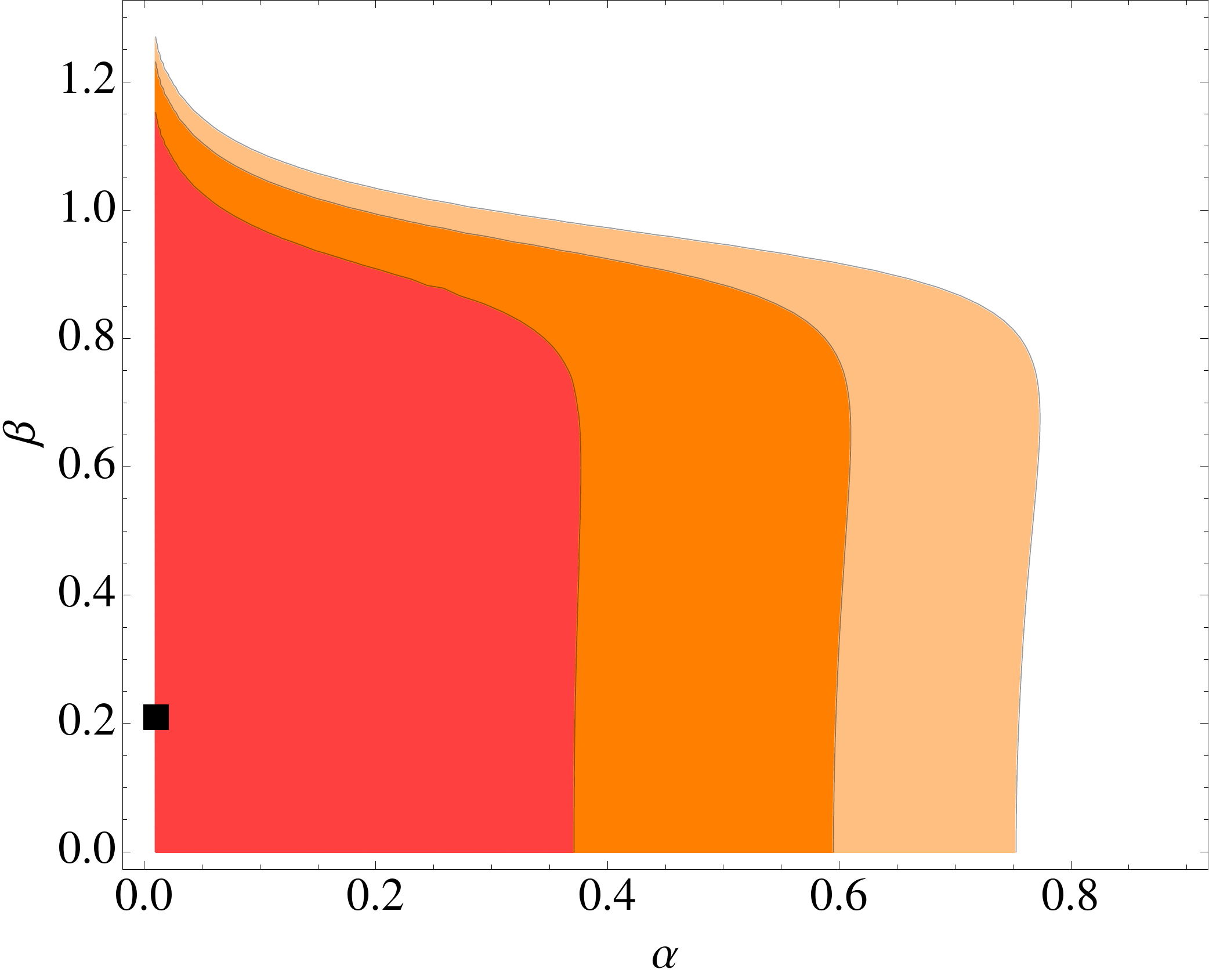}
\caption{{\em On the Left:} 1-dimensional marginalized posterior distributions
(for point 2 and point 5 together) on the parameters $\{\Omega_{{\rm DE},0},\alpha,\beta\}$
when fitting the model of this paper with $\Omega_{b,0}=0.048$ to the LSST 100k SN dataset (see
Section \ref{supernovae}) and the forecasted Euclid-like $f\sigma_{8}$
data (see Section \ref{fs8}). See Table~\ref{tab:1D-b} (4$^{th}$ and 5$^{th}$ columns) for best-fit
values with 95\% confidence intervals. {\em On the Right:} $1\sigma$,
$2\sigma$ and $3\sigma$ confidence-level contours for the relevant
2-dimensional marginalized posterior distributions. The black squares
mark the best-fit values.  This plot should be compared to Fig.~\ref{posts-E} where the baryonic content has been neglected.}
\label{posts-E-b}
\end{centering}
\end{figure}

\bibliographystyle{JHEP}
\bibliography{baldi_bibliography,bibliography,amendola}

\end{document}